\documentclass[letterpaper,12pt]{JHEP3}
\usepackage{amsmath,amssymb}
\usepackage{latexsym}
\usepackage{epsfig}
\usepackage{cite}
\usepackage[english]{babel}

\newcommand{\dd}{{\rm d}}

\newcommand{\eq}{\begin{equation}}
\newcommand{\feq}{\end{equation}}
\newcommand{\eqn}{\begin{eqnarray}}
\newcommand{\feqn}{\end{eqnarray}}
\newcommand{\arr}{\begin{eqnarray*}}
\newcommand{\farr}{\end{eqnarray*}}

\newcommand{\A}{{\cal A}}
\newcommand{\F}{{\cal F}}

\newcommand{\p}{\partial}
\newcommand{\w}{\wedge}

\newcommand{\lp}{\left(}
\newcommand{\rp}{\right)}
\font\mybb=msbm10 at 12pt
\def\bb#1{\hbox{\mybb#1}}

\def\bI {\bb{I}}
\def\bR {\bb{R}}

\def\bC {\bb{C}}
\newcommand{\HH}{{\mathbb{H}}}

\newcommand{\da}{\bar\partial}

\def\al{\alpha}
\def\be{\beta}
\def\ga{\gamma}
\def\de{\delta}
\def\ep{\epsilon}

\def\ze{\zeta}

\def\la{\lambda}

\def\si{\sigma}

\def\Ga{\Gamma}

\def\La{\Lambda}

\title{Geometry of four-dimensional Killing spinors}

\author{Sergio L.~Cacciatori,$^{ad}$ Marco M.~Caldarelli,$^b$ Dietmar
Klemm,$^{cd}$ Diego S.~Mansi$^{cd}$ and Diederik Roest$^e$ \\
$^a$ Dipartimento di Scienze Fisiche e Matematiche, \\
\hspace*{0.15cm} Universit\`a dell'Insubria, \\
\hspace*{0.15cm} Via Valleggio 11, I-22100 Como. \\
$^b$ Departament de F\'{i}sica Fonamental, \\
\hspace*{0.15cm} Universitat de Barcelona, \\
\hspace*{0.15cm} Diagonal, 647, 08028 Barcelona, Spain. \\
$^c$ Dipartimento di Fisica dell'Universit\`a di Milano, \\
\hspace*{0.15cm} Via Celoria 16, I-20133 Milano. \\
$^d$ INFN, Sezione di Milano, Via Celoria 16, I-20133 Milano. \\
$^e$ Departament Estructura i Constituents de la Materia, \\
\hspace*{0.15cm} Facultat de F\'{i}sica, Universitat de Barcelona, \\
\hspace*{0.15cm} Diagonal, 647, 08028 Barcelona, Spain. \\
E-mail: \email{sergio.cacciatori@uninsubria.it, caldarelli@ub.edu,
dietmar.klemm@mi.infn.it, diego.mansi@mi.infn.it, droest@ecm.ub.es}}
\preprint{IFUM-890-FT \\
UB-ECM-PF-07-05}

\abstract{The supersymmetric solutions of ${\cal N}=2$, $D=4$ minimal ungauged and gauged supergravity are classified according to the fraction of preserved supersymmetry using spinorial geometry techniques. Subject to a reasonable assumption in the 1/2-supersymmetric time-like case of the gauged theory, we derive the complete form of all supersymmetric solutions. This includes a number of new 1/4- and 1/2-supersymmetric possibilities, like gravitational waves on bubbles of nothing in AdS$_4$.}


\begin{document}

\section{Introduction}
\label{intro}

Throughout the history of string and M-theory an important part in
many developments in the subject has been played by supersymmetric
solutions of supergravity, i.e.~by backgrounds which admit a number
of Killing spinors $\epsilon$ which are parallel with respect to the
supercovariant derivative\footnote{For the purpose of this
discussion we will ignore possible additional Killing spinor
equations coming from the variation of dilatinos and gauginos.}:
$D_\mu \epsilon = 0$. Due to their ubiquitous role it has long been
realised that it would be advantageous to have classifications of
all supersymmetric solutions of a given theory.

For purely gravitational backgrounds the supersymmetric possibilities follow from the Berger
classification of the possible Riemannian holonomies \cite{Berger} (see
\cite{Bryant, Figueroa-O'Farrill:1999tx} for
an extension to the Lorentzian case). However, in the presence of additional
force fields (carried by e.~g.~scalars, gauge potentials or a
cosmological constant) it has proven very difficult to obtain
knowledge of all supersymmetric possibilities.

The reason for the complication in the presence of additional fields lies in the
holonomy of the supercurvature $R_{\mu \nu} = D_{[\mu} D_{\nu ]}$. For purely
gravitational backgrounds the holonomy of the supercurvature is generically
given by $H=$ Spin$(d-1,1)$ in $d$
dimensions, and hence coincides with the Lorentz group. In such cases the
Lorentz gauge freedom allows one to choose constant Killing spinors. Another
simplification is that if there is one Killing spinor with a specific stability
subgroup, i.e.~it is invariant under some  Lorentz subgroup, all other
spinors with the same stability subgroup are Killing as well.

For more general solutions including fields other than gravity, the holonomy
is generically extended to a larger group $H \supset$ Spin$(d-1,1)$. For
example, in the present paper we consider gauged minimal four-dimensional
${\cal N}=2$
supergravity, which has $H=$ GL(4,$\bC$) \cite{Batrachenko:2004su}. In such cases one
cannot choose constant Killing spinors nor are all spinors with the same
stability subgroup automatically Killing. For these reasons the classification
of the backgrounds that allow for Killing spinors is more
convoluted, or richer, in such cases. For a long time the only classification
available was in ungauged minimal four-dimensional ${\cal N}=2$
supergravity \cite{Tod:1983pm, Kowalski-Glikman}, which has $H=$ SL(2,$\mathbb{H}$).

A new impulse was given to the subject with the introduction of $G$-structures
and the method of spinor bilinears to solve the Killing spinor equations
\cite{Gauntlett:2002sc}. In this approach, space-time forms are constructed as bilinears from a Killing spinor and
one analyses the constraints that these forms imply for the background.
Using this framework, a number of complete classifications \cite{Gauntlett:2002nw, Gutowski:2003rg, Meessen:2006tu} and many
partial results (see e.g.~\cite{Gauntlett:2002fz, Gauntlett:2003fk,
  Caldarelli:2003pb,
  Gauntlett:2003wb, Cariglia:2004kk, Cacciatori:2004rt, Cariglia:2004qi, Gutowski:2005id,
  Bellorin:2005zc, Huebscher:2006mr, Bellorin:2006yr} for an
incomplete list) have been obtained. By complete we mean that the most general
solutions for all possible fractions of supersymmetry have been obtained,
while for partial classifications this is only available for some
fractions. Note that the complete classifications mentioned above involve
theories with eight supercharges and $H=$ SL(2,$\mathbb{H}$), and allow for either
half- or maximally supersymmetric solutions.

An approach which exploits the linearity of the Killing spinors has
been proposed \cite{Gillard:2004xq} under the name of spinorial
geometry. Its basic ingredients are an explicit oscillator basis for
the spinors in terms of forms and the use of the gauge symmetry to
transform them to a preferred representative of their orbit.
In this way one can construct a linear system for the background
fields from any (set of) Killing spinor(s) \cite{Gran:2005wu}. This
method has proven fruitful in e.g.~the challenging case of IIB
supergravity \cite{Gran:2005wn, Gran:2005kg, Gran:2005ct}. In
addition, it has been adjusted to impose 'near-maximal'
supersymmetry and thus has been used to rule out certain large fractions of
supersymmetry \cite{Gran:2006ec, Grover:2006ps, Grover:2006wy,
  Gran:2006cn}. Finally, a complete classification for type I supergravity in ten
dimensions has been obtained \cite{Gran:2007fu}.

In the present paper we would like to address the classification of
supersymmetric solutions in four-dimensional minimal ${\cal N}=2$ supergravity. As
will also be reviewed in section 2, the
ungauged case has been classified completely \cite{Tod:1983pm,
  Kowalski-Glikman}. For the gauged case, the discussion of 1/4
supersymmetry splits up in a time-like and a light-like class (depending on the
causal nature of the Killing vector associated to the Killing spinor). The
time-like class is completely specified by a single complex function depending on
three spatial coordinates $b = b(z,w, \bar w)$, subject to a second-order
differential equation which can not be solved in general \cite{Caldarelli:2003pb}. The
light-like class can be given in all generality, and in addition its
restriction to 1/2-BPS solutions has been derived
\cite{Cacciatori:2004rt}. Furthermore, there are no backgrounds with 3/4
supersymmetry \cite{Grover:2006wy} and AdS$_4$ is the unique possibility with maximal
supersymmetry. Therefore the remaining open question concerns half-supersymmetric
backgrounds in the gauged theory\footnote{The addition of external matter was considered in \cite{Caldarelli:2003wh}.}.

In the following, we will first re-analyse the 1/4-supersymmetric
backgrounds using the method of spinorial geometry, and in fact find an additional possibility in
the light-like case: a half supersymmetric bubble of nothing in AdS$_4$ and its Petrov type II generalization, a new 1/4 BPS configuration that has the interpretation of gravitational waves propagating on the bubble of nothing. This completes the analysis of the null class in all its generality. Then we will derive the constraints for half-supersymmetric backgrounds for the timelike class. Subject to a single assumption on the time-dependence of the second Killing spinor these will be solved in general, up to a second order ordinary differential equation. The assumption will be justified by solving the full set of conditions in a number of examples which illustrate the possible spatial dependence of $b$. All these cases turn out to have time-dependence of the assumed form. The different examples are:
 \begin{itemize}
  \item the $b=b(z)$ family of solutions, comprising part of the Reissner-Nordstr\"om-Taub-NUT-AdS$_4$ backgrounds,
  \item waves on the previous backgrounds with $b=b(z,w)$,
  \item solutions with $b$ imaginary and their $PSL(2,\bR)$ transformed counterparts,
  \item solutions of the dimensionally reduced gravitational Chern-Simons model that can be embedded in the equations for a timelike Killing spinor \cite{Cacciatori:2004rt}.
 \end{itemize}
We determine when these backgrounds preserve 1/2 supersymmetry and provide the explicit Killing spinors. Moreover, in the subcases consisting of AdS$_4$ and AdS$_2\times\HH^2$, the action of the isometries of these backgrounds on the Killing spinors is given explicitly.

The outline of this paper is as follows. In section 2, we discuss the orbits
of Killing spinors and review the known classification results in the theory
at hand. In section 3, we go through the complete classification of the null
class. In section 4, we discuss the constraints for 1/4 and 1/2 supersymmetry in the
timelike class. We derive the time-dependence of the second Killing spinor and solve the equations for the case of linear time-dependence ($G_0 = 0$).
A number of examples of the 1/2 BPS timelike class are provided in section 5. Finally,
in section 6 we present our conclusions and outlook. In appendix A we review
our notation and conventions for spinors, while in appendix B the associated
bilinear forms are given. Appendix C deals with the special case $P' = 0$, to
be defined in section 4.4. Finally, in appendix D, we will give the details of the $G_{0}=0$ case.

\section{$G$-invariant Killing spinors in 4D}

\subsection{Orbits of Dirac spinors under the gauge group}
\label{orbits}

In order to obtain the possible orbits of Spin(3,1) in the space
of Dirac spinors $\Delta_c$, we first consider the most general
positive chirality spinor\footnote{Our conventions for spinors and
their description in terms of forms can be found in appendix A.} $a1
+ be_{12}$ ($a,b \in \bC$) and determine its stability subgroup.
This is done by solving the infinitesimal equation
\begin{equation}
\alpha^{cd}\Gamma_{cd}(a1 + be_{12}) = 0\,. \label{stab}
\end{equation}
First of all, notice that $a1 + be_{12}$ is in the same orbit as 1,
which can be seen from
\begin{displaymath}
e^{\gamma\Gamma_{13}}e^{\psi\Gamma_{12}}e^{\delta\Gamma_{13}}e^{h\Gamma_{02}}\,1
= e^{i(\delta+\gamma)}e^h\cos\psi\,1 + e^{i(\delta-\gamma)}e^h\sin\psi\,e_{12}\,.
\end{displaymath}
This means that we can set $a=1$, $b=0$ in (\ref{stab}), which implies
then $\alpha^{02} = \alpha^{13} = 0$, $\alpha^{01} = -\alpha^{12}$,
$\alpha^{03} = \alpha^{23}$. The stability subgroup of 1 is thus generated by
\begin{equation}
X = \Gamma_{01} - \Gamma_{12}\,, \qquad Y = \Gamma_{03} + \Gamma_{23}\,. \label{XY}
\end{equation}
One easily verifies that $X^2 = Y^2 = XY = 0$, and thus $\exp(\mu X
+ \nu Y) = 1 + \mu X + \nu Y$, so that $X,Y$ generate $\bR^2$.

Spinors of negative chirality are composed of odd forms, i.e.~$ae_1
+ be_2$. One can show in a similar way that they are in the same
orbit as $e_1$, and the stability subgroup is again $\bR^2$, with
the above generators $X,Y$.

For definiteness and without loss of generality we will always
assume that the first Killing spinor has a non-vanishing positive
chirality component, and use (part of) the Lorentz symmetry to bring
this to the form $1$. Hence we can write a general spinor as $1 +
ae_1 + be_2$. Now act with the stability subgroup of $1$ to bring
$ae_1 + be_2$ to a special form:
\begin{displaymath}
(1 + \mu X + \nu Y)(1 + ae_1 + be_2) = 1 + be_2 + [a + 2b(\nu- i\mu)]e_1\,.
\end{displaymath}
In the case $b=0$ this spinor is invariant, so the representative is
$1 + ae_1$, with isotropy group $\bR^2$. If $b \neq 0$, one can
bring the spinor to the form $1 + be_2$, with isotropy group $\bI$.
The representatives\footnote{Note the difference in form compared to the Killing spinors of the corresponding theories in
five and six dimensions: in six dimensions these can be chosen constant \cite{Gutowski:2003rg} while in five dimensions they are constant up to
an overall function \cite{Grover:2006ps}. In four dimensions such a choice is generically not possible.} together with the stability subgroups are
summarized in table \ref{tab:orbits}.

In the ungauged theory, we therefore can have the following
$G$-invariant Killing spinors. The $\bR^2$-invariant Killing spinors
are spanned by $1$ and $e_1$ and there can be up to four of these.
The $\bI$-invariant Killing spinors are spanned by all four basis
elements and there can be up to eight of these. In the first two case, the vector
$V_a$ bilinear in the spinor
$\epsilon$ is lightlike, whereas in the last case it is timelike,
see table~\ref{tab:orbits}. The existence of a globally defined
Killing spinor $\epsilon$, with isotropy group $G \in$ Spin(3,1),
gives rise to a $G$-structure. This means that we have an
$\bR^2$-structure in the null case and an identity structure in the
timelike case.

In U(1) gauged supergravity, the local Spin(3,1) invariance is
actually enhanced to Spin(3,1) $\times$ U(1). Thus, in order to
obtain the stability subgroup, one determines the Lorentz
transformations that leave a spinor invariant up to an arbitrary
phase factor, which can then be gauged away using the additional
U(1) symmetry. For the representative $1$, one gets in this way an
isotropy group generated by $X, Y$ and $\Gamma_{13}$ obeying
\begin{displaymath}
[\Gamma_{13}, X] = -2Y\,, \qquad [\Gamma_{13}, Y] = 2X\,, \qquad [X, Y] = 0\,,
\end{displaymath}
i.~e.~$G\cong$ U(1)$\ltimes \bR^2$. For $\epsilon = 1 + ae_1$ with $a \neq 0$,
the stability subgroup $\bR^2$ is not enhanced, whereas the $\bI$ of
the representative $1 + be_2$ is promoted to U(1) generated by
$\Gamma_{13} = i\Gamma_{\bar\bullet\bullet}$. The Lorentz
transformation matrix $a_{AB}$ corresponding to $\Lambda =
\exp(i\psi\Gamma_{\bar\bullet\bullet}) \in$ U(1), with
$\Lambda\Gamma_B\Lambda^{-1} = {a^A}_B\Gamma_A$, has nonvanishing
components
\begin{equation}
a_{+-} = a_{-+} = 1\,, \qquad a_{\bullet\bar\bullet} = e^{2i\psi}\,, \qquad
a_{\bar\bullet\bullet} = e^{-2i\psi}\,. \label{residualU1}
\end{equation}
Finally, notice that in U(1) gauged supergravity one can choose the
function $a$ in $1 + ae_1$ real and positive: Write $a =
R\exp(2i\delta)$, use
\begin{displaymath}
e^{\delta\Gamma_{13}}(1 + ae_1) = e^{i\delta}1 + e^{-i\delta}ae_1
= e^{i\delta}(1 + Re_1)\,,
\end{displaymath}
and gauge away the phase factor $\exp(i\delta)$ using the
electromagnetic U(1).

\begin{table}[ht]
\begin{center}
\begin{tabular}{||c||c|c||c||}
\hline $\epsilon$ & $G\subset$ Spin(3,1) & $G\subset$ Spin(3,1)
$\times$ U(1) &
$V_a = D(\epsilon,\Gamma_a\epsilon)$ \\
\hline\hline
$1$ & $\bR^2$ & U(1)$\ltimes \bR^2$$\,\,\,\,$ & $(1,0,-1,0)$ \\
\hline
$1 + ae_1$ & $\bR^2$ & $\bR^2 \,\,\,\, (a \in \bR)$ & $(1 + |a|^2, 0, -1-|a|^2, 0)$ \\
\hline
$1 + be_2$ & $\bI$ & U(1)$\,\,\,\,$ &  $(1+|b|^2, 0, -1 + |b|^2, 0)$ \\
\hline
\end{tabular}
\end{center}
\caption{The representatives $\epsilon$ of the orbits of Dirac
spinors and their stability subgroups $G$ under the gauge groups
Spin(3,1) and Spin(3,1) $\times$ U(1) in the ungauged and gauged
theories, respectively. The number of orbits is the same in both
theories, the only difference lies in the stability subgroups and
the fact that $a$ is real in the gauged theory. In the last column
we give the vectors constructed from the spinors.}\label{tab:orbits}
\end{table}

In the gauged theory the classification of $G$-invariant spinors is
therefore slightly more complicated.
There can be at most two U(1)$\ltimes \bR^2$-invariant Killing spinors, spanned
by $1$. The four $\bR^2$-invariant spinors are spanned by $1$ and
$e_1$. Then there are the U(1)-invariant spinors, spanned by $1$
and $e_2$. Finally, for generic enough Killing spinors, one does not
fall in any of the above classes and the common stability subgroup
is $\bI$. Note that in the gauged theory the presence of $G$-invariant Killing
spinors will in general not lead to a $G$-structure on the manifold
but to stronger conditions. The structure group is in fact reduced
to the intersection of $G$ with Spin(3,1), and hence is equal to
the stability subgroup in the ungauged theory.

We will now consider the possible supersymmetric solutions to the
equation $D_\mu \epsilon = 0$ in various sectors of ${\cal N}=2$, $D=4$ in
terms of the stability subgroup $G$ of the Killing spinors.

\subsection{The ungauged theory}

The supercovariant derivative of ungauged minimal ${\cal N}=2$ supergravity
in four dimensions reads
 \begin{align}
  D_{\mu} = \partial_{\mu} + \tfrac{1}{4} \omega_{\mu}^{ab}
  \Gamma_{ab} + \tfrac{i}{4} \F_{ab} \Gamma^{ab} \Gamma_{\mu} \,.
 \end{align}
As mentioned in the introduction, a first point to notice is that
there is no complex conjugation on the Killing spinor. Therefore,
the number of supersymmetries that are preserved is always even: if
$\epsilon$ is Killing, then so is $i \epsilon$.

First consider purely gravitational solutions with $\F = 0$. In this
case the supercovariant connection truncates to the Levi-Civita
connection and has Spin(3,1) holonomy. This implies the following. If $\epsilon$ is Killing,
then so are\footnote{These operations anti-commute and commute with
the $\Gamma$-matrices, respectively.} $\Gamma_3 * \epsilon$
and $\Gamma_{012} * \epsilon$ (where $*$ denotes complex
conjugation). Together, the operations $i$, $\Gamma_3 *$ and
$\Gamma_{012} *$ generate four linearly independent Killing spinors
from any null spinor $\epsilon = 1$ or $\epsilon = 1 + a e_1$ and
eight from any time-like spinor $\epsilon = 1 + b e_2$. This illustrates the
general statement in the introduction: if the gauge group equals the holonomy, as
in this case, then there is only one possible
number of Killing spinors for every stability subgroup. Therefore
there are only two classes of supersymmetric solutions, which are
listed in table~\ref{ung-grav}, and which consist of the
gravitational wave and Minkowski space-time, respectively.

\begin{table}[ht]
\begin{center}
$\begin{array}{||c||c|c||}
 \hline
 G= \; \backslash \; N= & 4 & 8 \\
 \hline \hline
 \mathbb{R}^2 & \surd & \times \\
 \mathbb{I} & \times & \surd \\
 \hline
\end{array}$
\end{center}
\caption{Gravitational solutions with
$G$-invariant Killing spinors in the ungauged theory.} \label{ung-grav}
\end{table}

Now let us also allow for fluxes $\F$. The supercovariant connection
no longer equals the Levi-Civita connection due to the flux term. In
particular, this implies that $\Gamma_{012} *$ no longer commutes
with $D_{\mu}$. However, this does still hold for the other
operation: $\Gamma_3 * \epsilon$ is Killing provided $\epsilon$ is.
The combined operations of $i$ and $\Gamma_3 *$ generate four
linearly independent spinors from any null or time-like spinor. Thus
the number of supersymmetries is always $N=4p$, as illustrated in
table~\ref{ung-flux}. Indeed the generalised holonomy of the
supercovariant connection in the ungauged case is SL(2,$\mathbb{H}$)
\cite{Batrachenko:2004su}, consistent with the supersymmetries
coming in quadruplets.

\begin{table}[ht]
\begin{center}
$\begin{array}{||c||c|c||}
 \hline
 G= \; \backslash \; N= & 4 & 8 \\
 \hline \hline
 \mathbb{R}^2 & \surd & \times \\
 \mathbb{I} & \surd & \surd \\
 \hline
\end{array}$
\end{center}
\caption{General solutions with $G$-invariant Killing
spinors in the ungauged theory.} \label{ung-flux}
\end{table}

The half-supersymmetric solution have been classified by Tod
\cite{Tod:1983pm} and consist of the plane wave and the
Israel-Wilson-Perjes metric, respectively. The maximally
supersymmetric solutions are AdS$_2$ $\times$ S$^2$ and its Penrose
limits, the Hpp wave and Minkowski space-time
\cite{Kowalski-Glikman}.

\subsection{The gauged theory}

The supercovariant derivative of gauged minimal ${\cal N}=2$ supergravity
in four dimensions reads
 \begin{align}
  D_{\mu} = \partial_{\mu} + \tfrac{1}{4} \omega_{\mu}^{ab}
  \Gamma_{ab} - i \ell^{-1} \A_{\mu} + \tfrac{1}{2} \ell^{-1}
  \Gamma_{\mu} + \tfrac{i}{4} \F_{ab} \Gamma^{ab} \Gamma_{\mu} \,.
  \label{supcovder}
 \end{align}
Due to the gauging the structure of $\Gamma$-matrices is richer, but
there still is no complex conjugation on the Killing spinor.
Therefore, the number of supersymmetries that are preserved is
always even: if $\epsilon$ is Killing, then so is $i \epsilon$.

Again, we first consider the purely gravitational solutions. In this case the supercovariant derivative
has SO(3,2) holonomy. The
operation $\Gamma_{012} *$ commutes with $D_{\mu}$ and therefore
generates additional Killing spinors. Together, the operations $i$
and $\Gamma_{012} *$ generate four linearly independent Killing
spinors from generic null or time-like spinors. The exception is the
null spinor $\epsilon = 1 + e_1$, in which case
$\epsilon$ and $\Gamma_{012} *$ are linearly dependent, and hence allows for two
instead of four Killing spinors. The possibilities allowed for by this analysis
of the supercovariant derivative can be found in table~\ref{g-grav}.

\begin{table}[ht]
\begin{center}
$\begin{array}{||c||c|c|c|c ||}
 \hline
 G= \; \backslash \; N= & 2 & 4 & 6 & 8 \\
 \hline \hline
 {\mathrm{U(1)}\ltimes \bR^2} & \times & \times & \times & \times \\
 \bR^2 & \surd & \circ & \times & \times \\
 {\mathrm{U(1)}} & \times & \circ & \times & \times \\
 \bI & \times & \circ & \circ & \surd \\
 \hline
\end{array}$
\end{center}
\caption{Gravitational solutions with $G$-invariant Killing
spinors in the gauged theory. Check marks indicate entries with actual solutions, while circles stand for allowed entries which are not realized.} \label{g-grav}
\end{table}

However, although all these entries are allowed for by the spinor orbit
structure and the crude analysis of the supercurvature above, not all of them
have an actual field theoretic realisation in supergravity. In other words,
there are no solutions to the Killing spinor equations for all of the above
sets of Killing spinors. The lightlike cases were
considered in \cite{Cacciatori:2004rt}: The 1/4-BPS case is the
Lobatchevski wave while imposing more
supersymmetries leads to the maximally supersymmetric AdS$_4$
solution (with $G$=1). The $N=4$ and $G=\mathbb{R}^2$ entry is thus
effectively empty. In particular, this implies that imposing
a single Killing spinor $1 + a e_1$ with $a \neq 1$ leads
to AdS$_4$. Also note that the $N=6$ and $G=1$ entry must be empty
since any time-like spinor plus $1 + e_1$ leads to
maximal supersymmetry, while all other Killing spinors come in groups of four.
The only remaining entries are $N=4$ and $G=$ U(1) or $G=\bI$.
Using the results of \cite{Caldarelli:2003pb,Cacciatori:2004rt}, it
is straightforward to show that in these purely gravitational timelike
cases the geometry is given by
\begin{displaymath}
ds^2 = -\frac{z^2+n^2}{\ell^2}(dt - 2n\cosh\theta d\phi)^2 +
\frac{\ell^2 dz^2}{z^2+n^2} + (z^2+n^2)(d\theta^2+\sinh^2\theta d\phi^2)\,,
\end{displaymath}
where $n=\pm\ell/2$. But this is simply AdS$_4$ written as a line bundle
over a three-dimensional base manifold, so both $N=4$ entries are empty
as well. We conclude that there are no 1/2-supersymmetric gravitational solutions
in the gauged theory, only the 1/4-supersymmetric Lobatchevski waves and maximally
supersymmetric AdS$_4$.

We now come to the general supersymmetric solutions in the gauged case. Due to
the gauging and flux terms, neither $\Gamma_{012} *$ nor $\Gamma_3 *$ commute with
$D_{\mu}$. Therefore we have the cases as listed in table~\ref{g-flux}. The
supercovariant connection in the gauged case has generalized
holonomy GL$(4,\mathbb{C})$ \cite{Batrachenko:2004su}, again consistent
with the supersymmetries coming in doublets.

\begin{table}[ht]
\begin{center}
$\begin{array}{||c||c|c|c|c||}
 \hline
 G= \; \backslash \; N= & 2 & 4 & 6 & 8 \\
 \hline \hline
 {\mathrm{U(1)}\ltimes \bR^2} & \circ & \times & \times & \times \\
 \bR^2 & \surd & \circ & \times & \times \\
 {\mathrm{U(1)}} & \surd & \surd & \times & \times \\
 \bI & \times & \surd & \circ & \surd \\
 \hline
\end{array}$
\end{center}
\caption{General solutions with $G$-invariant Killing
spinors in the gauged theory. Check marks indicate entries with actual solutions, while circles stand for allowed entries which are not realized.} \label{g-flux}
\end{table}

The 1/4-BPS solutions with $G=\bR^2$ and $G=U(1)$ were derived in \cite{Caldarelli:2003pb}, and we will show there is no solution with $G={\mathrm{U(1)}\ltimes \bR^2}$. In
addition, it was shown in \cite{Cacciatori:2004rt} that any additional
supersymmetries in the null case are always timelike, i.e.~end up
in the $N=4$ and $G=1$ entry. Again, the $N=4$ and $G=\mathbb{R}^2$
entry is empty. It would be interesting to see if there is a nice explanation
for this. In addition, the maximally supersymmetric case is always AdS$_4$.
Recently, it has been shown in \cite{Grover:2006wy} that the $N=6$ and
$G=1$ entry is empty as well, because imposing three complex Killing spinors
implies that the spacetime is AdS$_4$ and thus maximally supersymmetric.
The most general 1/2-BPS solution in the timelike case remains an open issue
and will be studied in this paper.

\subsection{Generalized holonomy}

In minimal gauged supergravity theories with eight supercharges, the
generalized holonomy group for vacua preserving $N$ supersymmetries,
where $N=0,2,4,6,8$, is GL$(\frac{8-N}2,\bC)$ $\ltimes\frac N2\bC^{\frac{8-N}2}$
\cite{Batrachenko:2004su}. To see this, assume that there exists a Killing
spinor $\epsilon_1$. By a local GL$(4,\bC)$ transformation, $\epsilon_1$ can be
brought to the form $\epsilon_1 = (1,0,0,0)^T$. This is annihilated by matrices of
the form
\begin{displaymath}
\cal A = \left(\begin{array}{cc} 0 & {\underline a}^T \\
         \underline 0 & A \end{array}\right)\,,
\end{displaymath}
that generate the affine group A$(3,\bC) \cong$ GL$(3,\bC) \ltimes \bC^3$.
Now impose a second Killing spinor $\epsilon_2 =
(\epsilon_2^0,{\underline\epsilon}_2)^T$. Acting with the stability subgroup
of $\epsilon_1$ yields
\begin{displaymath}
e^{\cal A}\epsilon_2 = \left(\begin{array}{c} \epsilon_2^0 + {\underline b}^T
{\underline\epsilon}_2 \\ e^A{\underline\epsilon}_2\end{array}\right)\,,
\qquad\mathrm{where}\quad {\underline b}^T = {\underline a}^T A^{-1}(e^A - 1)\,.
\end{displaymath}
We can choose $A \in$ gl$(3,\bC)$ such that $e^A{\underline\epsilon}_2 = (1,0,0)^T$,
and $\underline b$ such that $\epsilon_2^0 + {\underline b}^T{\underline\epsilon}_2
= 0$. This means that the stability subgroup of $\epsilon_1$ can be used to bring
$\epsilon_2$ to the form $\epsilon_2 = (0,1,0,0)$. The subgroup of A$(3,\bC)$
that stabilizes also $\epsilon_2$ consists of the matrices
\begin{displaymath}
\left(\begin{array}{cccc} 1 & 0 & b_2 & b_3 \\ 0 & 1 & B_{12} & B_{13} \\
0 & 0 & B_{22} & B_{23} \\ 0 & 0 & B_{32} & B_{33} \end{array}\right)
\quad \in {\mathrm{GL}}(2,\bC) \ltimes 2\bC^2\,.
\end{displaymath}
Finally, imposing a third Killing spinor yields GL$(1,\bC) \ltimes 3\bC$
as maximal generalized holonomy group, which is however not realized in
${\cal N}=2$, $D=4$ minimal gauged supergravity \cite{Cacciatori:2004rt,Grover:2006wy}.
It would be interesting to better understand why such preons actually do not exist.
In section \ref{halfBPS}, we explicitely compute the generalized holonomy
group for ${\cal N}=2$, $D=4$ minimal gauged supergravity in the case $N=2$ and show that
it is indeed contained in A$(3,\bC)$, supporting thus the classification
scheme of \cite{Batrachenko:2004su}.

\section{Null representative $1 + a e_1$}

In this section we will analyse the conditions coming from a single
null Killing spinor. As we saw in section \ref{orbits}, there are
two orbits of such spinors, one with representative $\epsilon = 1$
and stability subgroup $G =$ U(1)$\ltimes \bR^2$ and one with $\epsilon = 1 + a
e_1$ and $G = \mathbb{R}^2$. Owing to local U(1) gauge invariance,
it is always possible to choose the function $a$ real and positive,
so in the following we set $a = e^{\chi}, \chi \in \bR$. The Killing
spinor equations become
 \eqn
-\frac{i}\ell\A+\frac12\Omega+\frac{e^\chi}{\sqrt2}\left[\left(\frac1\ell+i\phi\right)
E^{\bar\bullet}-2i\F^{+\bar\bullet}E^{-}\right]&=&0\,, \nonumber \\
\dd\chi+\frac{i}\ell\A+\frac12\Omega+\frac{e^{-\chi}}{\sqrt2}
\left[\left(\frac1\ell-i\phi\right)E^{\bar\bullet}+2i\F^{+\bar\bullet}E^{-}
\right]&=&0\,, \nonumber \\
\omega^{-\bullet}+\frac{e^\chi}{\sqrt2}\left[2i\F^{-\bullet}E^{\bar\bullet}+
\left(\frac1\ell-i\phi\right)E^{-}\right]&=&0\,, \nonumber \\
\omega^{-\bullet}+\frac{e^{-\chi}}{\sqrt2}\left[-2i\F^{-\bullet}E^{\bar\bullet}+
\left(\frac1\ell+i\phi\right)E^{-}\right]&=&0\,,
\label{killinglight}
 \feqn
where $\phi\equiv\F^{+-} + \F^{\bar\bullet\bullet}$ and $\Omega\equiv\omega^{+-} +
\omega^{\bar\bullet\bullet}$.

The conditions for the special U(1)$\ltimes \bR^2$-orbit with $\epsilon = 1$ can be
obtained as the singular limit $\chi \rightarrow - \infty$ of the
above equations. Note however that, in this limit, the second line
implies the constraint $\ell^{-1} - i \phi = 0$, while the fourth
line leads to $\ell^{-1} + i \phi = 0$. Clearly, for $\ell^{-1} \neq
0$ this does not allow for a solution. Hence, in the gauged theory,
{\it there are no backgrounds with} U(1)$\ltimes \bR^2$-{\it invariant Killing
spinors}.

The only null possibility is therefore given by the
$\bR^2$-invariant Killing spinor $\epsilon = 1 + e^{\chi} e_1$. We
will now analyse the above conditions for the generic case with
$\chi$ finite. In fact, we will furthermore assume it is positive.
This does not constitute any loss of generality since one can flip
the sign of $\chi$ by changing chirality (a spinor $1 + e^{\chi}
e_1$ with $\chi$ negative is gauge equivalent to a spinor $e_1 +
e^{\tilde \chi} 1$ with $\tilde \chi = - \chi$ positive), and hence
the resulting background will not depend on this sign.

From the last two equations one obtains the constraints
 \eq \F^{-\bullet}=\F^{-\bar\bullet}=0\,,\qquad
  \phi=-\frac{i}\ell\tanh\chi \label{constrF} \feq on the field
strength, as well as \eq
\omega^{-\bullet}=\omega^{-\bar\bullet}=-\frac1{\sqrt2\ell\cosh\chi}E^- \label{omega-light}
\feq for the spin connection. (\ref{constrF}) implies $\F^{+-}=0$
and $\F^{\bar\bullet\bullet}=-\frac{i}\ell\tanh\chi$. The first two
equations of (\ref{killinglight}) yield then \eqn
\omega^{+-}&=&2e^{\chi}H_{3}E^{-}-\frac1\ell\frac{e^{2\chi}}{\cosh\chi}E^{1}\,,
\nonumber \\
\omega^{\bar\bullet\bullet} &=& 2i\sinh\chi H_{1}E^{-}+\frac i{\ell}
\frac{\cosh2\chi}{\cosh\chi}E^{3}\,, \nonumber \\
\A&=&-\ell\cosh\chi H_{1}E^{-}-\sinh\chi E^{3}\,, \nonumber \\
\dd\chi&=&-2\cosh\chi H_{3}E^{-}+\frac2\ell\sinh\chi E^{1}\,, \label{eqchi}
\feqn
where $E^1 = (E^{\bullet} + E^{\bar\bullet})/\sqrt 2$, $iE^3 = (E^{\bullet} -
E^{\bar\bullet})/\sqrt 2$, and we defined
\begin{displaymath}
\frac{\F^{+\bullet}+\F^{+\bar\bullet}}{\sqrt2}=H_{1}\,,\qquad
\frac{\F^{+\bullet}-\F^{+\bar\bullet}}{\sqrt2}=iH_{3}\,.
\end{displaymath}
In order to proceed, we distinguish two subcases, namely $\dd\chi = 0$ and
$\dd\chi\neq 0$.

\subsection{Constant Killing spinor, $\dd a = 0$}

If $a$ and hence $\chi$ are constant, eqn.~(\ref{eqchi}) implies $\chi = H_3 = 0$.
Next we impose vanishing torsion. The torsion two-form reads
\eqn
T^{-}&=&\dd E^{-}+\frac2\ell E^{1}\wedge E^{-}\,, \nonumber \\
T^{+}&=&\dd E^{+}-E^{1}\wedge\left(\omega^{+1}+\frac{E^{+}}\ell\right)+
\omega^{+3}\wedge E^{3}\,, \nonumber \\
T^{1}&=&\dd E^{1}+E^{-}\wedge\left(\omega^{+1}+\frac{E^{+}}\ell\right)\,,
\nonumber \\
T^{3}&=&\dd E^{3}+\frac1\ell E^{1}\wedge E^{3}-\omega^{+3}\wedge E^{-}\,.
\feqn
From $T^-=0$ one gets $E^{-}\wedge\dd E^{-}=0$, so by Fr\"obenius' theorem
there exist two functions $\eta$ and $u$ such that locally
\begin{displaymath}
E^{-}=\eta\dd u\,.
\end{displaymath}
Plugging this into $T^-=0$ yields
\begin{displaymath}
\eta\left(\dd\log\eta+\frac2\ell E^{1}\right)\wedge\dd u=0\,,
\end{displaymath}
so that there exists a function $\xi$ such that
\begin{displaymath}
E^{1}=-\frac{\ell}{2\eta}\dd\eta+\xi\dd u\,.
\end{displaymath}
The gauge field and its field strength can now be written as
\begin{displaymath}
\A=-\ell\eta H_{1}\dd u\,,\qquad\F=\frac{\ell}2H_{1}\dd\eta\wedge\dd u\,,
\end{displaymath}
and the Bianchi identity $\F=\dd\A$ implies
\begin{displaymath}
\left(\dd H_{1}+\frac32H_{1}\dd \log\eta\right)\wedge\dd u=0\,.
\end{displaymath}
This means that $H_{1}\eta^{3/2}$ can depend only on $u$,
\begin{displaymath}
H_{1}\eta^{3/2}=-\frac{\varphi'(u)}\ell\,,
\end{displaymath}
where the prefactor and the derivative were chosen in order to conform with
the notation of \cite{Caldarelli:2003pb}.
Let us define a new coordinate $x=-\eta^{-1/2}$, so that
$E^1=\frac\ell{x}\dd x+\xi\dd u$, $E^-=x^{-2}\dd u$ and
\begin{equation}
\A=-x\varphi'(u)\dd u\,. \label{gaugepot}
\end{equation}
One can now use part of the residual gauge freedom, given by the stability
subgroup $\bR^2$ of the null spinor $1+ae_1$, in order to simplify $E^{1}$.
To this end, consider an $\bR^2$ transformation with group element
\begin{displaymath}
\Lambda = 1+\mu X+\nu Y\,,
\end{displaymath}
where $X$ and $Y$ are given in (\ref{XY}). Defining $\alpha = \mu+i\nu$,
this can also be written as
\begin{equation}
\Lambda = 1+\alpha\Gamma_{+\bullet}+\bar\alpha\Gamma_{+\bar\bullet}\,.
\end{equation}
Given the ordering $A,B=+,-,\bullet,\bar\bullet$, the Lorentz transformation
matrix $a_{AB}$ corresponding to $\Lambda\in\bR^2\subseteq$ Spin(3,1) reads
\eq
a_{AB}=\left(
\begin{array}{cccc}
0&1&0&0\\
1&-4|\alpha|^2&2\bar\alpha&2\alpha\\
0&-2\bar\alpha&0&1\\
0&-2\alpha&1&0
\end{array}\right)\,. \label{R2transf}
\feq
The transformed vielbein ${}^{\alpha}E^A=a^{A}{}_{B}E^B$ is thus given by
\eqn
{}^{\alpha}E^{\bullet}&=&E^{\bullet}-2\alpha E^{-}\,,\qquad
{}^{\alpha}E^{1}=E^{1}-\sqrt2\left(\alpha+\bar\alpha\right)E^{-}\,, \nonumber\\
{}^{\alpha}E^{\bar\bullet}&=&E^{\bar\bullet}-2\bar\alpha E^{-}\,, \qquad
{}^{\alpha}E^{3}=E^{3}+\sqrt2i\left(\alpha-\bar\alpha\right)E^{-}\,,\nonumber\\
{}^{\alpha}E^{-}&=&E^{-}\,,\qquad \qquad \qquad
{}^{\alpha}E^{+}=E^{+}+2\bar\alpha E^{\bullet}+2\alpha E^{\bar\bullet}
-4|\alpha|^{2}E^{-}\,.
\feqn
Choosing $\alpha+\bar\alpha=\xi x^2/\sqrt 2$, we can eliminate $E^1_u$,
so one can set $\xi=0$ without loss of generality. Note that this still leaves
a residual gauge freedom associated to the imaginary part of $\alpha$,
which will be used below.

From $\dd T^3=0$ we get $\dd(\omega^{+3}/x)\wedge\dd u=0$, and thus there
exist two functions $\beta,\tilde\beta$ such that
\begin{displaymath}
\omega^{+3}=-x\dd \beta+\tilde\beta \dd u\,.
\end{displaymath}
Plugging this into $T^3=0$ yields $\dd(xE^{3}+\beta\dd u)=0$, which is solved by
\eq
E^{3}=-\frac\ell{x}\dd y+\beta\dd u\,,
\feq
where $y$ denotes some function that we shall use as a coordinate.
Using the remaining gauge freedom (\ref{R2transf}) with
${\mathrm{Im}}\alpha = -\beta x^2/2\sqrt 2$ allows to set also $\beta = 0$.
The equation $T^{1}=0$ tells us that $\omega^{+1}+E^+/\ell=\gamma\dd u$ for
some function $\gamma$. Using this together with $T^+=0$, one shows that
\begin{displaymath}
\dd\left(E^{-}\wedge E^{+}\right)=-\frac2x\dd x\wedge\left(E^{-}\wedge E^{+}
\right)\,,
\end{displaymath}
which means that the surface described by $E^{-}$ and $E^{+}$ is integrable,
so that
\eq
E^{+}=\ell^{2}\frac{{\cal G}}2\dd u+h\dd V\,,
\feq
for some functions ${\cal G}, h, V$. The metric becomes then
\eqn
\dd s^{2}&=&2E^{-}E^{+}+\left(E^{1}\right)^{2}+\left(E^{3}\right)^{2}\nonumber\\
&=&\frac{\ell^{2}}{x^{2}}\left({\cal G}\dd u^{2}+\frac{2h}{\ell^{2}}\dd u\dd V
+\dd x^{2}+\dd y^{2}\right)\,.
\feqn
Finally, the equation $T^{+}=0$ implies
\eqn
\partial_{x}h&=&\partial_{y}h=0\,, \qquad
\partial_{V}{\cal G}=\frac2{\ell^{2}}\partial_{u}h\,, \label{partialVG} \\
\gamma&=&\frac{x\ell}2\partial_{x}{\cal G}\,, \qquad
\tilde\beta=-\frac{x\ell}2\partial_{y}{\cal G}\,. \nonumber
\feqn
$h$ can be eliminated by introducing a new coordinate $v(u,V)$ with
$\partial_V v = h/\ell^2$ and shifting ${\cal G}\to{\cal G} + 2\partial_u v$,
which leads to
\eq
\dd s^{2}=\frac{\ell^{2}}{x^{2}}\left({\cal G}\dd u^{2}+2\dd u\dd v+
\dd x^{2}+\dd y^{2}\right)\,. \label{metriclight}
\feq
Note that, due to (\ref{partialVG}), ${\cal G}$ is independent of $v$, therefore
$\partial_v$ is a Killing vector. One easily verifies that it coincides
with the Killing vector constructed from the Killing spinor as $-\frac{\ell^2}{2\sqrt 2}D(\epsilon,\Gamma_{\mu}\epsilon)$.

All that remains is to impose the Maxwell and Einstein equations. One finds
that the former are automatically satisfied by the gauge potential (\ref{gaugepot}).
The same holds for the Einstein equations, except for the $uu$-component,
which gives the Siklos equation with sources
\eq
\Delta{\cal G}-\frac2x\partial_{x}{\cal G}=-\frac{4x^{2}}{\ell^{2}}\varphi'(u)^{2}\,.
\label{siklos}
\feq
This family of solutions enjoys a large group of diffeomorphisms which leave the solution invariant in form but change the function $\cal G$. This is the Siklos-Virasoro invariance, discussed in \cite{Siklos:1985,Cacciatori:2004rt}.
In conclusion, the geometry of solutions admitting the constant null spinor
$1+e_1$ is given by the Lobachevski waves with metric (\ref{metriclight}) and
gauge field (\ref{gaugepot}), where ${\cal G}$ satisfies (\ref{siklos}) and
$\varphi(u)$ is arbitrary. This coincides exactly with the results
of \cite{Caldarelli:2003pb}, where it was shown moreover that there is a second
covariantly constant spinor iff the wave profiles ${\cal G}$ and $\varphi$ have
the form
\eq
{\cal G}_\al (x,y,u) = -\frac{x^4}{\ell^2} + 2\alpha x^3 - \alpha^2\ell^2(x^2+y^2)\,,
\qquad \varphi(u) = u\,, \label{half-susy-constant}
\feq
up to Siklos-Virasoro transformation, with $\alpha\in\bR$ constant. In this case, the solution does also belong to the timelike class \cite{Caldarelli:2003pb}. While the $\al \neq 0$ solution only has the obvious Killing vectors $\partial_v$ and $\partial_y$, the special $\al=0$ case is maximally symmetric with a five-dimensional isometry group.

\subsection{Killing spinor with $\dd a \neq 0$}

If $\dd a$ and  hence also $\dd\chi$ do not vanish, one can use the $\bR^2$ stability subgroup of the spinor
$1+e^{\chi}e_1$ to eliminate the fluxes $\F^{+\bullet}$ and $\F^{+\bar\bullet}$.
To see this, observe that under an $\bR^2$ transformation (\ref{R2transf}),
\begin{displaymath}
{}^{\alpha}\F^{+\bullet} = \F^{+\bullet}-\frac{2i\alpha}{\ell}\tanh\chi\,,
\qquad {}^{\alpha}\F^{\bar\bullet\bullet} = \F^{\bar\bullet\bullet}\,,
\end{displaymath}
so by choosing $\alpha = -\frac{i\ell}2\F^{+\bullet}\coth\chi$ one can achieve
${}^{\alpha}\F^{+\bullet} = 0$. Note that this would not be possible if
$\chi = 0$. With this gauge fixing, one has
\begin{equation}
\dd\chi = \frac2\ell\sinh\chi\,E^1\,, \qquad \A = -\sinh\chi\,E^3\,, \qquad
\F=-\frac1\ell\tanh\chi\,E^1\wedge E^3\,. \label{dchi}
\end{equation}
Next we impose vanishing torsion. Using (\ref{dchi}), one easily shows that
$T^-=0$ leads to
\begin{displaymath}
\dd\left[\left(e^{2\chi}-1\right)E^{-}\right] = 0\,,
\end{displaymath}
and therefore one can introduce a function $u$ with
\begin{equation}
\left(e^{2\chi}-1\right)E^{-}=\dd u\,. \label{E-}
\end{equation}
Before we come to the other torsion components, let us consider the Bianchi
identity and the Maxwell equations. The gauge field strength reads
\begin{displaymath}
\F=\frac{\dd\chi}{\sinh2\chi}\wedge\A\,.
\end{displaymath}
Requiring it to be equal to $\dd\A$ implies that $\A/\sqrt{\tanh\chi}$ is
closed, so that locally
\begin{equation}
\A=\sqrt{\tanh\chi}\dd\Psi\,. \label{gaugepotchi}
\end{equation}
Note that the functions $\chi$, $u$ and $\Psi$ must be independent, because
otherwise $E^1$, $E^-$ and $E^3$ would not be linearly independent. We can thus
use these three functions as coordinates.

Using
\begin{displaymath}
{}^{*}\F=-\frac1\ell\tanh\chi E^{-}\wedge E^{+}\,,
\end{displaymath}
the Maxwell equations $\dd{}^{*}\F=0$ imply
\begin{displaymath}
\dd\left(E^-\wedge E^+\right)+2\frac{\dd\chi}{\sinh2\chi}\wedge\left(E^-\wedge E^+
\right)=0\,.
\end{displaymath}
By Fr\"obenius' theorem and (\ref{E-}), $E^+$ can thus be written as
\begin{displaymath}
E^+ = \frac{\tilde{\cal K}}2\dd u+h\dd V\,,
\end{displaymath}
where $\tilde{\cal K}$, $h$ and $V$ are some functions, and we can use $V$ as
the remaining coordinate. Substituing $E^+$ into the Maxwell equations one
obtains a constraint on the function $h$,
\begin{displaymath}
\dd\left(\frac{h}{e^{2\chi}+1}\right)\wedge\dd u\wedge\dd V=0\,,
\end{displaymath}
and hence
\begin{displaymath}
h=h_{0}(u,V)\left(e^{2\chi}+1\right)\,.
\end{displaymath}
In what follows, we define ${\cal K}=\tilde{\cal{K}}/(e^{2\chi}+1)$ and use
$\omega^{+1}=(\omega^{+\bullet}+\omega^{+\bar\bullet})/\sqrt2$, $\omega^{+3}=
(\omega^{+\bullet}-\omega^{+\bar\bullet})/\sqrt2i$. We now come to the remaining
torsion components. From $T^3=0$ and $T^1=0$ one obtains respectively
\begin{displaymath}
\omega^{+3}=AE^{-}\,, \qquad \omega^{+1}=-\frac{E^+}{\ell\cosh\chi}+BE^{-}\,,
\end{displaymath}
where $A$ and $B$ are some functions to be determined. Finally, $T^+=0$ yields
\eqn
\partial_{V}{\cal K}&=&2\partial_u h_0\,, \label{partialVG2} \quad
A=-\frac12\left(e^{4\chi}-1\right)\frac{\sinh\chi}{\sqrt{\tanh\chi}}
\partial_{\Psi}{\cal K}\,, \quad
B=\frac1\ell\left(e^{4\chi}-1\right)\sinh\chi\partial_{\chi}{\cal K}\,. \nonumber
\feqn
The line element is given by
\eqn
\dd s^{2}&=&2E^{-}E^{+}+\left(E^{1}\right)^{2}+\left(E^{3}\right)^{2}\nonumber\\
&=&\coth\chi\left({\cal K}\dd u^{2}+2h_0\dd u\dd V\right) + \frac{\ell^2\dd\chi^2}
{4\sinh^2\chi} + \frac{\dd\Psi^2}{\sinh\chi\cosh\chi}\,.
\feqn
As before, one can eliminate $h_0$ by introducing a new coordinate $v(u,V)$ with
$\partial_V v = h_0$ and shifting ${\cal K} \to {\cal K}+ 2\partial_u v$,
whereupon the metric becomes
\begin{equation}
\dd s^2 = \coth\chi\left({\cal K}\dd u^{2}+2\dd u\dd v\right) + \frac{\ell^2\dd\chi^2}
{4\sinh^2\chi} + \frac{\dd\Psi^2}{\sinh\chi\cosh\chi}\,. \label{metricchi}
\end{equation}
Notice that, owing to (\ref{partialVG2}), ${\cal K}$ is independent of $v$, therefore
$\partial_v$ is a Killing vector. It coincides with the Killing vector
$-\sqrt 2 D(\epsilon,\Gamma_{\mu}\epsilon)$ constructed from the Killing
spinor. All that remains now is to impose Einstein's equations. One finds that
they are all satisfied except for the $uu$ component, which yields again a Siklos-type
equation for ${\cal K}$,
\eq
\partial_{\Psi}^2{\cal K}+4\tanh\chi\partial_{\chi}^2{\cal K}-\frac 2{\cosh^{2}\chi}
\partial_{\chi}{\cal K}=0\,. \label{sikloschi}
\feq
In conclusion, the bosonic fields for a configuration admitting a null Killing
spinor with $\dd\chi\neq 0$ are given by (\ref{gaugepotchi}) and (\ref{metricchi}),
with ${\cal K}$ satisfying (\ref{sikloschi})\footnote{This solution escaped a
  majority of the present authors in
\cite{Caldarelli:2003pb}. The reason for this is that equ.~(4.32) of
\cite{Caldarelli:2003pb} is not correct; it must be $R_{+-ij}=0$, which yields
no information on the constant $\kappa$. Thus, in addition to the solutions
with $\kappa=0$ found in \cite{Caldarelli:2003pb} (the Lobachevski waves),
there are also the $\kappa=1$ solutions, which are exactly the ones found here
with $\dd\chi\neq 0$.}. As we will discuss in section \ref{im-b}, the ${\cal K} = 0$ solution is of Petrov type D and represents a bubble of nothing in anti-De Sitter space-time. When ${\cal K} \neq 0$, the metric becomes of Petrov type II and the Weyl scalar signalling the presence of gravitational radiation acquires a non-vanishing value. Hence the general solution represents a gravitational wave on a bubble of nothing. To our knowledge these solutions have not featured in the literature before.

\subsection{Half-supersymmetric backgrounds}

In the previous subsections we have addressed the conditions for preserving
one null Killing spinor of the form $\ep_1 = 1$ or $\ep_1 = 1 + e^{\chi} e_1$. It is natural
to enquire about the possibility of these backgrounds admitting an additional
Killing spinor with the same $\bR^2$ stability subgroup, i.e.~of the form $\ep_2 = c_0 1 + c_1 e_1$.
Using the fact that $\ep_1$ is Killing, the second Killing spinor equation
$D_\mu \ep_2 = 0$ can then be rewritten as
 \eq
  (c_0 - c_1) D_\mu 1 + \partial_\mu c_0 1 + \partial_\mu c_1 e_1 = 0 \,,
 \feq
in the U(1)$\ltimes \bR^2$ case and
 \eq
  (c_0 - c_1 e^{-\chi}) D_\mu 1 + \partial_\mu c_0 1 + (\partial_\mu c_1 - c_1
  \partial_\mu \chi) e_1 = 0 \,,
 \feq
in the $\bR^2$ case. Furthermore, we can assume that $(c_0 - c_1) \neq 0$ and
$(c_0 - c_1 e^{-\chi}) \neq 0$ in the two cases, respectively, since otherwise
the second Killing spinor would be linearly dependent on the first and there
would not be any additional constraints. Hence the $e_2$ and $e_{12}$
components of $D_{\mu} 1$ have to vanish separately. In particular, this
implies that $\omega^{- \bullet} = 0$ (as can be seen from the third line of \eqref{killinglight} in the
singular limit $\chi \rightarrow - \infty$). However, this is clearly
incompatible with \eqref{omega-light}. We conclude that, in the gauged theory,
{\it there are no backgrounds with four $\bR^2$-invariant Killing spinors}. In other
words, there are no half-supersymmetric backgrounds with an
$\bR^2$-structure. This is unlike the ungauged case, where the
half-supersymmetric gravitational waves provide such solutions.

Therefore, the only possibility to augment the supersymmetry of the null
solutions above is to add a Killing spinor which breaks the $\mathbb{R}^2$
invariance, i.e.~with a non-vanishing $e_2$ and/or $e_{12}$ component. From a linear combination of the first and second Killing spinor
one can then always construct a time-like Killing spinor, and hence this
brings us to the next section. For the convenience of the reader, we will already
summarise how to restrict the 1/4-supersymmetric null solutions to allow
for a time-like Killing spinor as well.

For the case with constant null Killing spinors, $\dd\chi = 0$, the restriction
was already discussed in \cite{Caldarelli:2003pb} and is given in \eqref{half-susy-constant}. For the other case,
with $\dd\chi\neq 0$, it is straightforward to show that the solution \eqref{gaugepotchi},
\eqref{metricchi} admits a second Killing spinor iff
$\partial_{\chi}{\cal G} = \partial_{\Psi}{\cal G} = 0$, so that ${\cal G}$
depends only on $u$. By a simple diffeomorphism one can then set ${\cal G}=0$.
The general solution to the Killing spinor equations reads in this case
\eq
\epsilon = \lambda_1 (1 + e^{\chi}e_1) + \frac{\lambda_2}{\sqrt{e^{4\chi}-1}}
           (e_2 + e^{\chi}e_1\wedge e_2)\,,
\feq
where $\lambda_{1,2} \in \bC$ are constants. The invariants constructed from
$\epsilon$, as defined in appendix B, are
\begin{eqnarray}
V &=& \sqrt 2\coth\chi(|\lambda_2|^2\dd v
- |\lambda_1|^2\dd u) - \frac{2i}{\sinh 2\chi}(\lambda_2{\bar\lambda}_1-{\bar\lambda}_2
\lambda_1)\,\dd\Psi\,, \nonumber \\
B &=& -\sqrt 2(|\lambda_1|^2\dd u
+ |\lambda_2|^2\dd v) + \frac{\ell e^{\chi}}{\sqrt{e^{4\chi}-1}\sinh\chi}
({\bar\lambda}_1\lambda_2 + \lambda_1{\bar\lambda}_2)\,\dd\chi\,, \nonumber \\
f &=& i(\lambda_1{\bar\lambda}_2-{\bar\lambda}_1\lambda_2)\sqrt{\tanh\chi}\,,
\qquad g = ({\bar\lambda}_1\lambda_2 + \lambda_1{\bar\lambda}_2)
\sqrt{\coth\chi}\,. \nonumber
\end{eqnarray}
The norm of the Killing vector $V$ is given by
\begin{displaymath}
V^2 = -\frac 2{\sinh 2\chi}({\bar\la}_1\la_2+\la_1{\bar\la}_2)^2 - 4|\la_1\la_2|^2
\tanh\chi\,.
\end{displaymath}
Since $\chi>0$, this is negative unless $\la_1=0$ or $\la_2=0$, so indeed the solution
\eqref{gaugepotchi}, \eqref{metricchi} with ${\cal G} = 0$ must belong also to the timelike class.
It turns out that it is identical to the bubble of nothing of section \ref{im-b} with imaginary $b$ and $L<0$.
The coordinate transformation
\begin{eqnarray}
u &=& \sqrt 2 A^2 (t-Ly) - \frac z{2\sqrt 2 A^2}\,, \qquad
v = - \sqrt 2 A^2 (t-Ly) -\frac z{2\sqrt 2 A^2}\,, \nonumber \\
\Psi &=& -2A^2 t\,, \qquad \chi = {\mathrm{artanh}}\frac{X^2}{A^4}
\end{eqnarray}
with $A^8 = -1/4L$
brings the metric \eqref{metricchi} (with ${\cal G}=0$) to \eqref{metr-imb},
and the field strength of \eqref{gaugepotchi} to \eqref{gaugefield-imb}.
Note that, in the new coordinates, the above invariants become $V = \partial_t$
as a vector, and $B = \dd z$, in agreement with section \ref{geom}.

\section{Timelike representative $1 + b e_2$}

We will now turn to the timelike case and first recover the general 1/4-BPS solutions \cite{Caldarelli:2003pb}. Afterwards we will study the conditions for 1/2 supersymmetry. This will complete the classification since we already know that no 3/4-supersymmetric solutions can arise and AdS$_4$ is the unique maximally supersymmetric possibility.

\subsection{Conditions from the Killing spinor equations}

Acting with the supercovariant derivative (\ref{supcovder}) on the
representative $1 + b e_2$ yields the linear system
\begin{eqnarray}
\partial_+ b + \frac b2\,\omega^{\bar\bullet\bullet}_+ - \frac b2\,
\omega^{+-}_+ - \frac i\ell b\,\A_+ &=& 0\,, \nonumber \\
\frac 12 \omega^{\bar\bullet\bullet}_+ + \frac 12 \omega^{+-}_+ - \frac i\ell
\A_+ + \frac b{\ell\sqrt 2} + \frac{ib}{\sqrt 2} \F^{\bar\bullet\bullet} +
\frac{ib}{\sqrt 2} \F^{+-} &=& 0\,, \nonumber \\
\omega^{\bullet -}_+ + i\sqrt 2 b\,\F^{\bullet -}\,=
\,\omega^{\bullet +}_+ &=& 0\,, \label{+}
\end{eqnarray}
\begin{eqnarray}
\partial_- b + \frac b2\,\omega^{\bar\bullet\bullet}_- - \frac b2\,
\omega^{+-}_- - \frac i\ell b\,\A_- + \frac 1{\ell\sqrt 2} + \frac i{\sqrt 2}
\F^{\bar\bullet\bullet} - \frac i{\sqrt 2} \F^{+-} &=& 0\,, \nonumber \\
\frac 12 \omega^{\bar\bullet\bullet}_- + \frac 12 \omega^{+-}_- -\frac i\ell
\A_- &=& 0\,, \nonumber \\
b\,\omega^{\bullet +}_- + i\sqrt 2\,\F^{\bullet +}\,=
\,\omega^{\bullet -}_- &=& 0\,, \label{-}
\end{eqnarray}
\begin{eqnarray}
\partial_{\bullet} b + \frac b2\,\omega^{\bar\bullet\bullet}_{\bullet} -
\frac b2\,\omega^{+-}_{\bullet} - \frac i\ell b\,\A_{\bullet} -
i\sqrt 2\,\F^{\bar\bullet -} &=& 0\,, \nonumber \\
\frac 12 \omega^{\bar\bullet\bullet}_{\bullet} + \frac 12 \omega^{+-}_{\bullet}
- \frac i\ell \A_{\bullet} - i\sqrt 2 b\,\F^{\bar\bullet +} &=& 0\,, \nonumber \\
\omega^{\bullet -}_{\bullet} + \frac b{\ell\sqrt 2} - \frac{ib}{\sqrt 2}
\F^{\bar\bullet\bullet} - \frac{ib}{\sqrt 2} \F^{+-} &=& 0\,, \nonumber \\
b\,\omega^{\bullet +}_{\bullet} + \frac 1{\ell\sqrt 2} - \frac i{\sqrt 2}
\F^{\bar\bullet\bullet} + \frac i{\sqrt 2} \F^{+-} &=& 0\,, \label{bullet}
\end{eqnarray}
\begin{eqnarray}
\partial_{\bar\bullet} b + \frac b2\,\omega^{\bar\bullet\bullet}_{\bar\bullet}
- \frac b2\,\omega^{+-}_{\bar\bullet} - \frac i\ell b\,
\A_{\bar\bullet} &=& 0\,, \nonumber \\
\frac 12 \omega^{\bar\bullet\bullet}_{\bar\bullet} + \frac 12
\omega^{+-}_{\bar\bullet} - \frac i\ell \A_{\bar\bullet} &=& 0\,, \nonumber \\
\omega^{\bullet -}_{\bar\bullet}\,=
\,b\,\omega^{\bullet +}_{\bar\bullet} &=& 0\,. \label{barbullet}
\end{eqnarray}
From eqns.~(\ref{+}) - (\ref{barbullet}) one obtains the gauge
potential and the fluxes in terms of the spin connection and the
function $b$,
\begin{eqnarray}
\A_+ &=& \frac{i\ell}2\left(\frac{\partial_+ \bar b}{\bar b}-
\frac{\partial_+ b}b - \omega^{\bar\bullet\bullet}_{+}\right)\,,
\quad \A_- = \frac{i\ell}2 \omega^{\bullet\bar\bullet}_-\,, \quad
\A_{\bullet} = \frac{i\ell}2(\omega^{\bullet\bar\bullet}_{\bullet} +
\omega^{+-}_{\bullet})\,, \nonumber \\
\F^{+-} &=& \frac i{\sqrt 2}(b\,\omega^{\bullet +}_{\bullet} - b^{-1}
\omega^{\bullet -}_{\bullet})\,, \qquad \F^{\bullet +} = \frac i{\bar b\sqrt 2}
\omega^{+-}_{\bar\bullet}\,, \nonumber \\
\F^{\bullet\bar\bullet} &=& \frac i{\sqrt 2}(b\,\omega^{\bullet +}_{\bullet}
+ b^{-1}\omega^{\bullet -}_{\bullet}) + \frac i\ell\,, \qquad
\F^{\bullet -} = \frac i{b\sqrt 2}\omega^{\bullet -}_{+}\,. \label{AF}
\end{eqnarray}
Furthermore, the system (\ref{+}) - (\ref{barbullet}) determines almost
all components of the spin connection (with the exception of
$\omega^{\bullet\bar\bullet}$) in terms of the function $b$ and its
spacetime derivatives,
\begin{eqnarray}
\omega^{+-}_+ &=& \frac{\partial_+ b}b + \frac{\partial_+ \bar b}{\bar b}\,,
\qquad \omega^{+-}_- = 0\,, \qquad \omega^{+-}_{\bullet} =
\frac{\partial_{\bullet}\bar b}{\bar b}\,, \nonumber \\
\omega^{+\bullet}_+ &=& \omega^{+\bullet}_{\bar\bullet} = 0\,, \qquad
\omega^{+\bullet}_- = -\frac{\partial_{\bar\bullet} b}{b^2\bar b}\,, \qquad
\omega^{+\bullet}_{\bullet} = \frac{\partial_- b}b +
\frac{\sqrt 2}{b\ell}\,, \nonumber \\
\omega^{-\bullet}_+ &=& - b\,\partial_{\bar\bullet}\bar b\,, \qquad
\omega^{-\bullet}_- = \omega^{-\bullet}_{\bar\bullet} = 0\,, \qquad
\omega^{-\bullet}_{\bullet} = \frac{\partial_+ \bar b}{\bar b} +
\frac{b\sqrt 2}\ell\,. \label{spinconn}
\end{eqnarray}
In what follows, we assume $b \neq 0$.
One easily shows that $b = 0$ leads to $\ell^{-1} = 0$,
so this case appears only in ungauged supergravity.

\subsection{Geometry of spacetime}
\label{geom}

In order to obtain the spacetime geometry, we consider the spinor bilinears
\begin{equation}
V_{\mu} = D(\epsilon, \Gamma_{\mu}\epsilon)\,, \qquad
B_{\mu} = D(\epsilon, \Gamma_5 \Gamma_{\mu}\epsilon)\,, \nonumber
\end{equation}
whose nonvanishing components are
\begin{equation}
V_+ = \sqrt 2\,\bar b b\,, \qquad V_- = -\sqrt 2\,, \qquad
B_+ = \sqrt 2\,\bar b b\,, \qquad B_- = \sqrt 2\,. \nonumber
\end{equation}
As $V^2 = -4\bar b b = - B^2$, $V$ is timelike and $B$ is spacelike.
Using eqns.~(\ref{+}) - (\ref{barbullet}), it is straightforward to
show that $V$ is Killing and $B$ is closed, i.~e.~,
\begin{eqnarray}
\partial_A V_B + \partial_B V_A - {\omega^C}_{B|A} V_C - {\omega^C}_{A|B}
V_C &=& 0\,, \nonumber \\
\partial_A B_B - \partial_B B_A - {\omega^C}_{B|A} B_C + {\omega^C}_{A|B}
B_C &=& 0\,. \nonumber
\end{eqnarray}
There exists thus a function $z$ such that $B = dz$ locally. Let us
choose coordinates $(t,z,x^i)$ such that $V = \partial_t$ and $i=1,2$.
The metric will then be independent of $t$. Note also that the system
(\ref{+}) - (\ref{barbullet}) yields
\begin{displaymath}
\partial_t b = \sqrt 2\,(|b|^2\partial_- - \partial_+)b = 0\,,
\end{displaymath}
so $b$ is time-independent as well. In terms of the vierbein $E^A_{\mu}$
the metric is given 
\begin{equation}
ds^2 = 2 E^+ E^- + 2 E^{\bullet} E^{\bar\bullet}\,, \label{metricE}
\end{equation}
where
\begin{displaymath}
E^+_{\mu} = \frac{B_{\mu} + V_{\mu}}{2\sqrt 2 |b|^2}\,, \qquad
E^-_{\mu} = \frac{B_{\mu} - V_{\mu}}{2\sqrt 2}\,.
\end{displaymath}
From $V^2 = -4|b|^2$ and $V = \partial_t$ as a vector we get
$V_t = -4|b|^2$, so that $V = - 4|b|^2 (dt + \sigma)$ as a one-form,
with $\sigma_t = 0$. Furthermore, $V^{\bullet} = 0$ yields
$E^{\bullet}_t = 0$, and thus
\begin{displaymath}
E^{\bullet} = E^{\bullet}_z dz + E^{\bullet}_i dx^i\,.
\end{displaymath}
The component $E^{\bullet}_z$ can be eliminated by a diffeomorphism
\begin{displaymath}
x^i = x^i(x'^j, z)\,,
\end{displaymath}
with
\begin{displaymath}
E^I_i \frac{\partial x^i}{\partial z} = - E^I_z\,, \qquad I = \bullet,
                                        \bar\bullet\,.
\end{displaymath}
As the matrix $E^I_i$ is invertible\footnote{One has
$\det(E^I_i) = -\det(E^A_{\mu})$, and the latter is always nonzero.},
one can always solve for $\partial x^i/\partial z$. Note that the metric
is invariant under
\begin{displaymath}
t \to t + \chi(x^i, z)\,, \qquad \sigma \to \sigma - d\chi\,,
\end{displaymath}
where $\chi(x^i, z)$ denotes an arbitrary function. This second gauge
freedom can be used to eliminate $\sigma_z$. Hence, without loss of
generality , we can take $\sigma = \sigma_i dx^i$, and the metric
(\ref{metricE}) becomes
\begin{equation}
ds^2 = - 4|b|^2 (dt + \sigma_i dx^i)^2 + \frac{dz^2}{4|b|^2}
       + 2 E^{\bullet}_i dx^i E^{\bar\bullet}_j dx^j\,.
\end{equation}
Next one has to impose vanishing torsion,
\begin{displaymath}
\partial_{\mu} E^A_{\nu} - \partial_{\nu} E^A_{\mu} + \omega^A_{\mu B}
E^B_{\nu} - \omega^A_{\nu B} E^B_{\mu} = 0\,.
\end{displaymath}
One finds that some of these equations are already identically
satisfied, while the remaining ones yield (using the expressions
(\ref{spinconn}) for the spin connection) the constraints
\begin{eqnarray}
\partial_z \sigma_i &=& -\frac 1{4|b|^2}(E^{\bar\bullet}_i E^j_{\bar\bullet}
                        - E^{\bullet}_i E^j_{\bullet})\partial_j
                        \ln(b/\bar b)\,, \label{dzsigma} \\
\partial_i \sigma_j - \partial_j \sigma_i &=& (E^{\bullet}_i
E^{\bar\bullet}_j - E^{\bullet}_j E^{\bar\bullet}_i)\left(\partial_z
\ln(b/\bar b) + \frac 1{b\ell} - \frac 1{\bar b\ell}\right)\,,
\label{disigma} \\
\omega^{\bullet\bar\bullet}_t &=& -2|b|^2 \partial_z\ln(b/\bar b)
+ \frac{2b}\ell - \frac{2\bar b}\ell\,, \label{omegat} \\
\partial_i E^{\bullet}_j - \partial_j E^{\bullet}_i &=& (E^{\bullet}_i
E^{\bar\bullet}_j - E^{\bullet}_j E^{\bar\bullet}_i)
\omega^{\bullet\bar\bullet}_{\bar\bullet} \,, \label{diE}
\end{eqnarray}
as well as
\begin{equation}
\left[\partial_z + \omega^{\bullet\bar\bullet}_z + \frac 12\partial_z
\ln(\bar b b) + \frac 1{2\ell}\left(\frac 1b + \frac 1{\bar b}\right)
\right] E^{\bullet}_i = 0\,. \label{dzE}
\end{equation}
In (\ref{dzsigma}), $E^i_I$ denotes the inverse of $E^J_j$. In order to
obtain the above equations, one has to make use of the inverse tetrad
\begin{displaymath}
E_+ = -\frac 1{2\sqrt 2}\partial_t + \sqrt 2 |b|^2\partial_z\,, \qquad
E_- = \frac 1{2\sqrt 2 |b|^2}\partial_t + \sqrt 2\,\partial_z\,, \qquad
E_{\bullet} = E^i_{\bullet}(\partial_i - \sigma_i\partial_t)\,.
\end{displaymath}
(\ref{dzE}) can be solved to give
\begin{equation}
E^{\bullet}_i = \frac 1{|b|} {\hat E}^{\bullet}_i\exp\left[-\int dz\,
\omega^{\bullet\bar\bullet}_z - \frac 1{2\ell}\int dz \left(\frac 1b +
\frac 1{\bar b}\right)\right]\,,
\end{equation}
where ${\hat E}^{\bullet}_i$ is an integration constant that depends
only on the coordinates $x^j$. At this point it is convenient to use
the residual U$(1)$ gauge freedom of a combined local Lorentz and
gauge transformation to eliminate $\omega^{\bullet\bar\bullet}_z$.
This is accomplished by the transformation (\ref{residualU1}),
with
\begin{displaymath}
\psi = \frac i2 \int dz\,\omega^{\bullet\bar\bullet}_z\,.
\end{displaymath}
Note that $\psi$ is real, as it must be. Defining
\begin{equation}
\Phi := -\frac 1{2\ell}\int dz \left(\frac 1b + \frac 1{\bar b}\right)\,,
        \label{Phi}
\end{equation}
we have thus
\begin{equation}
E^{\bullet}_i = \frac 1{|b|}{\hat E}^{\bullet}_i\exp\Phi\,. \label{relEhatE}
\end{equation}
Using (\ref{relEhatE}) in (\ref{diE}), one gets for the only remaining
unknown component $\omega^{\bullet\bar\bullet}_{\bullet}$ of the
spin connection
\begin{displaymath}
\omega^{\bullet\bar\bullet}_{\bullet} = \left[
{\hat \omega}^{\bullet\bar\bullet}_{\bullet} - {\hat E}^i_{\bullet}\partial_i
\right] |b|\exp(-\Phi)\,,
\end{displaymath}
where ${\hat \omega}^{\bullet\bar\bullet}_{\bullet}$ denotes the spin
connection following from the zweibein ${\hat E}^I_i$.

In what follows, we shall choose the conformal gauge for the two-metric
$h_{ij} = {\hat E}_{Ii}{\hat E}^I_j$, i.~e.~,
\begin{equation}
h_{ij} = e^{2\xi}[(dx^1)^2 + (dx^2)^2]\,.
\end{equation}
with $\xi$ depending only on the coordinates $x^i$. Furthermore, we
choose an orientation such that
\begin{displaymath}
{\hat E}^{\bullet}_i {\hat E}^{\bar\bullet}_j -
{\hat E}^{\bullet}_j {\hat E}^{\bar\bullet}_i = -i e^{2\xi}\epsilon_{ij}\,,
\end{displaymath}
where $\epsilon_{12} = 1$. To be concrete, we shall take
\begin{displaymath}
({\hat E}^I_i) = \frac 1{\sqrt 2}e^{\xi}\left(\begin{array}{lr} 1 & i \\
                            1 & -i \end{array}\right)\,.
\end{displaymath}
The eqns.~(\ref{dzsigma}) and (\ref{disigma})
then simplify to
\begin{eqnarray}
\partial_z \sigma_i &=& -\frac i{4|b|^2}\epsilon_{ij}\partial_j
                        \ln(b/\bar b)\,, \label{dzsigmanew} \\
\partial_i \sigma_j - \partial_j \sigma_i &=& -\frac i{|b|^2} e^{2(\Phi + \xi)}
\epsilon_{ij}\left(\partial_z \ln(b/\bar b) + \frac 1{b\ell} -
\frac 1{\bar b\ell}\right)\,. \label{disigmanew}
\end{eqnarray}
Moreover, one has
\begin{equation}
\omega^{\bullet\bar\bullet}_{\bullet} = -\partial_{\bullet}
\ln\left(|b|e^{-\Phi - \xi}\right)\,.
\end{equation}
In \cite{Caldarelli:2003pb} it has been shown that in the case where
the Killing vector constructed from the Killing spinor is timelike,
the Einstein equations follow from the Killing spinor equations,
so all that remains to do at this point is to impose the Bianchi identity
and the Maxwell equations. Using the spin connection (\ref{spinconn}) and
(\ref{omegat}) in (\ref{AF}), the gauge potential and the field
strength become
\begin{eqnarray}
\A &=& i(dt + \sigma)(b - \bar b) + \frac{\ell}2 \epsilon_{ij}
\partial_j (\Phi + \xi)\,dx^i - \frac{i\ell}4 d\ln(b/\bar b)\,, \nonumber \\
\F &=& i(dt + \sigma)\wedge d\,(\bar b - b) + \frac 1{4|b|^2} dz\wedge dx^i
\epsilon_{ij}\partial_j (b + \bar b) \nonumber \\
&& + \frac 1{2|b|^2}\left[\partial_z
(b + \bar b) + \frac 1\ell\right] e^{2(\Phi + \xi)}\epsilon_{ij} dx^i
\wedge dx^j\,. \label{fieldstr}
\end{eqnarray}
The Bianchi identity $\F = d\A$ yields
\begin{equation}
\Delta (\Phi + \xi) = \frac 2\ell e^{2(\Phi + \xi)}\left[\partial_z \frac 1b +
\partial_z\frac 1{\bar b} - \frac 1{b^2\ell} - \frac 1{{\bar b}^2\ell} +
\frac 1{\bar b b \ell}\right]\,, \label{bianchi}
\end{equation}
with $\Delta = \partial_i\partial_i$ denoting the flat space Laplacian
in two dimensions. As for the Maxwell equations,
\begin{displaymath}
\partial_{\mu}(\sqrt{-g}\F^{\mu\nu}) = 0\,,
\end{displaymath}
the only nontrivial information comes from the $t$-component,
which gives
\begin{equation}
4e^{2(\Phi + \xi)}\left[b^2 \partial^2_z \frac 1b - {\bar b}^2 \partial^2_z
\frac 1{\bar b} - \frac{3b}\ell\partial_z \frac 1b + \frac{3\bar b}\ell
\partial_z \frac 1{\bar b} + \frac 1{b\ell^2} - \frac 1{\bar b\ell^2}\right]
+ b^2 \Delta \frac 1b - {\bar b}^2 \Delta \frac 1{\bar b} = 0\,,
\label{maxwell}
\end{equation}
where we used eqns.~(\ref{dzsigmanew}) and (\ref{disigmanew}).

Let us now show that the equations (\ref{bianchi}) and (\ref{maxwell}) are
actually the same as the ones in \cite{Cacciatori:2004rt}. If we set
\begin{equation}
F = -\frac 1{\ell\bar b}\,, \qquad e^{\phi} = 2e^{\Phi+\xi}\,,
\end{equation}
(\ref{bianchi}) yields exactly equation (2.3) of \cite{Cacciatori:2004rt}.
On the other hand, deriving (\ref{bianchi}) with respect to $z$ and using
(\ref{Phi}), one obtains
\begin{equation}
\Delta A + e^{2\phi}\left[3A\partial_z A - 3B\partial_z B + A^3 - 3AB^2 +
\partial^2_z A\right] = 0\,, \label{DeltaA}
\end{equation}
where $A$ and $B$ denote the real and imaginary part of $F$ respectively.
This can be used in (\ref{maxwell}) to get
\begin{displaymath}
\Delta B + e^{2\phi}\left[\partial_z^2 B + 3B\partial_z A + 3A\partial_z B
- B^3 + 3A^2 B\right] = 0\,,
\end{displaymath}
which, together with (\ref{DeltaA}), yields
\begin{equation}
\Delta F + e^{2\phi}\left[F^3 + 3F\partial_z F + \partial^2_z F\right] = 0\,,
\label{DeltaF}
\end{equation}
i.~e.~, equation (2.2) of \cite{Cacciatori:2004rt}. For a complete
identification of the present results with the ones in \cite{Cacciatori:2004rt},
one also has to set $\sigma = \omega$.

In conclusion, the metric of the general 1/4-supersymmetric solution
is given by
\begin{equation}
ds^2 = -4|b|^2(dt+\sigma)^2 + \frac 1{4|b|^2}\left(dz^2 + 4 e^{2( \Phi + \xi) }dw\,
       d\bar w\right)\,,
\end{equation}
where $b$ and $\phi$ are determined by the system (\ref{bianchi}), (\ref{maxwell})
and $w=x^1+ix^2\equiv x+iy$.
The one-form $\sigma$ follows then from (\ref{dzsigmanew})
and (\ref{disigmanew}), and the gauge field strength is given by (\ref{fieldstr}).
Note that (\ref{maxwell}) represents also the integrability condition for
(\ref{dzsigmanew}), (\ref{disigmanew}). As noted in \cite{Cacciatori:2004rt}, this system of equations is invariant under PSL$(2,\bR)$ transformations\footnote{It might be of interest to investigate the possible relation between this 'hidden symmetry' and the Ehlers group for solutions of four-dimensional vacuum gravity with a Killing vector.}. If we define a new coordinate $z'$ through the M\"obius transformation
\eq
z'=\frac{\al z+\be}{\ga z+\de}\,,
\label{psl1}\feq
with $\al$, $\be$, $\ga$ and $\de$ arbitrary real constants satisfying $\al\de-\be\ga=1$,
then the functions $\tilde b(z',x^i)$ and $\tilde\Phi(z',x^i)$ defined by
\eq
\frac{1}{\tilde b}=\frac1{(\ga z'-\al)^2b}-\frac{2l\ga}{\ga z'-\al}\,,
\qquad
e^{\tilde\Phi}=\lp\ga z'-\al\rp^2e^\Phi\,,
\label{psl2}\feq
solve the system in the new coordinate system $(z',x^i)$, with the function $\xi(x^i)$ left invariant and $z$ seen as a function of $z'$. This symmetry allows to generate new BPS solutions from the known ones. Note however that it is only a symmetry of the equations for 1/4 supersymmetry, and if we apply it to solutions with additional Killing spinors, it will in general not preserve them, as we shall show explicitely in some examples.

\subsection{Half-supersymmetric backgrounds}
\label{halfBPS}

We now would like to investigate the possibility of adding a
second Killing spinor. Since the first Killing spinor $\epsilon_1$ has
stability subgroup $1$, one cannot use Lorentz transformations to bring the
second spinor to a preferred form. Therefore we use the most general
form
 \begin{align}
  \epsilon_2 = c_0 1 + c_1 e_1 + c_2 e_2 + c_{12} e_1 \wedge e_2 \,.
 \end{align}
The corresponding linear system simplifies significantly after
inserting the results from $\epsilon_1$. These determine all the
fluxes and the spin connection in terms of the functions $b$, $\xi$ and their
derivatives. First it is convenient to introduce the new basis\footnote{Note that
$\epsilon_1 = (1,0,0,0)$ in this basis.}
\begin{displaymath}
\alpha = \left(\begin{array}{c} \alpha_0 \\ \alpha_1 \\ \alpha_2 \\ \alpha_{12}
         \end{array}\right) = \left(\begin{array}{c} c_0 \\ b^{-1}c_2 - c_0 \\
         \bar b c_1 \\ c_{12}\end{array}\right)\,,
\end{displaymath}
in which the Killing spinor equations for $\epsilon_2$ read
\begin{equation}
(\partial_A + M_A)\alpha = 0\,, \label{linsys}
\end{equation}
with the connection $M_A$ given by
\begin{displaymath}
M_+ = \left(\begin{array}{cccc} 0 & -\partial_+\ln\bar b & 0 & 0 \\
      0 & \partial_+\ln\bar b & -\partial_{\bullet}\ln b &
      \partial_{\bullet}\ln b \\
      0 & 0 & \frac{\bar b - b}{\sqrt 2\ell} + \frac 12\partial_+
      \ln\frac b{\bar b} & -\frac{\sqrt 2}{\ell}\bar b - \partial_+\ln b \\
      0 & -|b|^2\partial_{\bar\bullet}\ln\bar b & 0 & \frac{\bar b - b}
      {\sqrt 2\ell} - \frac 12\partial_+\ln(\bar b b) \end{array}\right)\,,
\end{displaymath}
\begin{displaymath}
M_- = \left(\begin{array}{cccc} 0 & 0 & |b|^{-2}\partial_{\bullet}
      \ln\bar b & -|b|^{-2}\partial_{\bullet}\ln\bar b \\
      0 & \partial_-\ln b & -|b|^{-2}\partial_{\bullet}\ln\bar b &
      |b|^{-2}\partial_{\bullet}\ln\bar b \\
      0 & \partial_{\bar\bullet}\ln b & \frac{b - \bar b}{\sqrt 2\ell |b|^2}
      -\frac 12\partial_-\ln(\bar b b) & 0 \\
      0 & 0 & -\frac{\sqrt 2}{\ell\bar b} - \partial_-\ln\bar b &
      \frac{b - \bar b}{\sqrt 2\ell |b|^2} + \frac 12\partial_-\ln\frac{\bar b}b
      \end{array}\right)\,,
\end{displaymath}
\begin{displaymath}
M_{\bullet} = \left(\begin{array}{cccc} 0 & -\partial_{\bullet}\ln\bar b & 0 & 0 \\
              0 & \partial_{\bullet}\ln(\bar b b) & 0 & 0 \\
              0 & -\frac{\sqrt 2}{\ell}\bar b - \partial_+\ln b & -
              \partial_{\bullet}\ln\left(|b|e^{-\Phi - \xi}\right) & 0 \\
              0 & \frac{\sqrt 2}{\ell}b + \partial_+\ln\bar b & 0 &
              -\partial_{\bullet}\ln\left(|b|e^{-\Phi - \xi}\right) \end{array}
              \right)\,,
\end{displaymath}
\begin{displaymath}
M_{\bar\bullet} = \left(\begin{array}{cccc} 0 & 0 & -\partial_-\ln\bar b &
                  \partial_-\ln\bar b + \frac{\sqrt 2}{\ell\bar b} \\
                  0 & 0 & \partial_-\ln(\bar b b) + \frac{\sqrt 2}{\ell b} &
                  -\partial_-\ln(\bar b b) - \frac{\sqrt 2}{\ell\bar b} \\
                  0 & 0 & \partial_{\bar\bullet}\ln\left(\sqrt{\frac b{\bar b}}
                  e^{-\Phi - \xi}\right) & -\partial_{\bar\bullet}\ln b \\
                  0 & 0 & -\partial_{\bar\bullet}\ln\bar b & \partial_{\bar\bullet}
                  \ln\left(\sqrt{\frac{\bar b}b}e^{-\Phi - \xi}\right) \end{array}
                  \right)\,.
\end{displaymath}

Let us first of all consider the simpler possibility of a second Killing spinor of the form $\ep_2 = c_0 1 + c_2 e_2$.
As discussed in section \ref{orbits}, both $\ep_1$ and $\ep_2$ are invariant under the same U(1) symmetry, and hence this case constitutes
the $G=$ U(1) case with four supersymmetries. As can easily be seen from the above Killing spinor equations with $\al_1 \neq 0$ and
$\al_2 = \al_{12} = 0$, this restricts the derivatives of the coefficient
$b$ to be
 \begin{align}
  \partial_- b = - \frac{\sqrt{2}}{\ell} \,, \quad \partial_+ b = -
  \frac{\sqrt{2} b \bar b}{\ell} \,, \quad \partial_\bullet b =
  \partial_{\bar \bullet} b = 0 \,.
 \end{align}
Hence this corresponds to $\partial_z b = - 1 / \ell$. As will be
discussed in section 5.1, this restriction uniquely leads to the
half-supersymmetric anti-Nariai space-time. Hence {\it $AdS_2 \times
H^2$ is the only possibility for backgrounds with four}
U(1)-{\it invariant Killing spinors}.

In the more general case with $\al_2$ and $\al_{12}$ non-vanishing, i.e.~with trivial stability subgroup, the Killing spinor equations do not so readily provide information
about $b$ and one has to resort to their integrability conditions. The first integrability conditions for the linear system (\ref{linsys}) are
 \begin{equation}
  N_{\mu\nu}\alpha \equiv (\partial_{\mu} M_{\nu} - \partial_{\nu} M_{\mu} +
  [M_{\mu}, M_{\nu}])\alpha = 0\,, \label{1st-intcond}
 \end{equation}
where the matrices $M_{\mu} = E^A_{\mu}M_A$ are given by
\begin{displaymath}
M_t = \sqrt 2(|b|^2 M_- - M_+)\,, \qquad M_z = \frac 1{2\sqrt 2 |b|^2}(M_+ + |b|^2 M_-)\,,
\end{displaymath}
\begin{displaymath}
M_w = \sigma_w M_t + \frac 1{\sqrt 2 |b|}e^{\Phi+\xi} M_{\bullet}\,, \qquad
M_{\bar w} = \sigma_{\bar w} M_t + \frac 1{\sqrt 2 |b|}e^{\Phi+\xi}M_{\bar\bullet}\,,
\end{displaymath}
and we introduced the complex coordinates $w=x+iy$, $\bar w=x-iy$.
For half-supersymmetric solutions, the six matrices $N_{\mu\nu}$ must have rank
two. (As at least one Killing spinor exists, namely $\epsilon_1 = (1,0,0,0)$, we
already know that the $N_{\mu\nu}$ can have at most rank three. Rank one is not
possible, because 3/4 BPS solutions cannot exist \cite{Grover:2006wy}.
Rank zero corresponds to the maximally supersymmetric case, which implies that the
spacetime geometry is AdS$_4$ \cite{Caldarelli:2003pb}.)
Let us define
\begin{displaymath}
\tilde N_{\mu\nu} \equiv S N_{\mu\nu} T\,,
\end{displaymath}
with
\begin{displaymath}
S = \left(\begin{array}{cccc} 1 & 0 & 0 & 0 \\ 1 & 1 & 0 & 0 \\ 0 & 0 & 1 & 0 \\
0 & 0 & 0 & 1 \end{array}\right)\,, \qquad
T = \left(\begin{array}{cccc} 1 & 0 & 0 & 0 \\ 0 & 1 & 0 & 0 \\ 0 & 0 & 1 & 0 \\
0 & 0 & 1 & 1 \end{array}\right)\,.
\end{displaymath}
The similarity transformation $S$ corresponds to adding the first line
to the second one and $T$ adds the last column to the third one. This does not
alter the rank of $N_{\mu\nu}$. One finds
\begin{displaymath}
\tilde N_{wt} =
\left(\begin{array}{c@{\quad}c@{\quad}c@{\quad}c}
& 2b\partial\partial_z\bar b + \frac 2{\ell}\partial\bar b & & -\frac{2|b|}{\bar b}
e^{-\Phi-\xi}\left(\partial^2\bar b + \frac 1{\bar b}\partial\bar b\partial\bar b\right. \\
\raisebox{1.5ex}[-1.5ex]{0} &  + 2b\left(\partial_z\bar b + \frac 1{\ell}\right)
\partial\ln\bar b & \raisebox{1.5ex}[-1.5ex]{0} &  \left.- 2\partial(\Phi+\xi)\partial
\bar b\right) \vspace*{0.2cm} \\
& 2\bar b\partial\partial_z b + \frac 2{\ell}\partial b & & -\frac{2|b|}b e^{-\Phi-\xi}
\left(\partial^2 b + \frac 1b\partial b\partial b\right. \\
\raisebox{1.5ex}[-1.5ex]{0} & + 2\bar b\left(\partial_z b + \frac 1{\ell}\right)
\partial\ln b & \raisebox{1.5ex}[-1.5ex]{0} & \left.- 2\partial(\Phi+\xi)
\partial b\right) \vspace*{0.2cm} \\
& 2|b|^3 e^{-\Phi-\xi}\da\partial\ln b & & 2\bar b\partial\partial_z b \\
\raisebox{1.5ex}[-1.5ex]{0} & -2|b|^3 e^{\Phi+\xi}b^{-2}\left(2\partial_z b +
\frac 1{\ell}\right)\left(\partial_z b + \frac 1{\ell}\right) & \raisebox{1.5ex}[-1.5ex]
{$\frac 2{\ell}\partial b$} & + 2\bar b\left(\partial_z b + \frac 1{\ell}\right)\partial
\ln b \vspace*{0.2cm} \\
& 2|b|^3 e^{-\Phi-\xi}\da\partial\ln\bar b & & 2b\partial\partial_z\bar b - \frac 2{\ell}
\partial\bar b \\
\raisebox{1.5ex}[-1.5ex]{0} & -2|b|^3 e^{\Phi+\xi}{\bar b}^{-2}\left(2\partial_z\bar b
+ \frac 1{\ell}\right)\left(\partial_z\bar b + \frac 1{\ell}\right) & \raisebox{1.5ex}
[-1.5ex]{$-\frac 2{\ell}\partial\bar b$} & + 2b\left(\partial_z\bar b + \frac 1{\ell}
\right)\partial\ln\bar b
\end{array}\right), \nonumber 
\end{displaymath}
\newpage
$\tilde N_{\bar wt} =$ 
\begin{displaymath}
\left(\begin{array}{c@{\quad}c@{\quad}c@{\quad}c}
& 2b\da\partial_z\bar b & & -2|b|e^{-\Phi-\xi}\da\partial\ln\bar b \\
\raisebox{1.5ex}[-1.5ex]{0} & + 2b\left(\partial_z\bar b +
\frac 1{\ell}\right)\da\ln\bar b & \raisebox{1.5ex}[-1.5ex]{$-\frac 2{\ell |b|}e^{\Phi+\xi}
\left(2\partial_z\bar b + \frac 1{\ell}\right)$} & + \frac{2b}{|b|\bar b}e^{\Phi+\xi}
\left(2\partial_z\bar b + \frac 1{\ell}\right)\left(\partial_z\bar b + \frac{b-
\bar b}{\ell b}\right) \vspace*{0.2cm} \\
& 2\bar b\da\partial_z b & & -2|b|e^{-\Phi-\xi}\da\partial\ln b \\
\raisebox{1.5ex}[-1.5ex]{0} & +2\bar b\left(\partial_z b + \frac 1{\ell}\right)
\da\ln b & \raisebox{1.5ex}[-1.5ex]{$\frac 2{\ell |b|}e^{\Phi+\xi}
\left(2\partial_z b + \frac 1{\ell}\right)$} & +\frac{2\bar b}{|b|b}e^{\Phi+\xi}
\left(2\partial_z b + \frac 1{\ell}\right)\left(\partial_z b + \frac 1{\ell}
\right) \vspace*{0.2cm} \\
& 2|b|\bar b e^{-\Phi-\xi}\left(\da^2 b + \frac 1b \da b\da b\right. & &
2\bar b\da\partial_z b + \frac 2{\ell}\da b \\
\raisebox{1.5ex}[-1.5ex]{0} & \left.-2\da(\Phi+\xi)\da b\right) & \raisebox{1.5ex}
[-1.5ex]{$\frac 6{\ell}\da b$} & +2\bar b\left(\partial_z b + \frac 1{\ell}\right)
\da\ln b \vspace*{0.2cm} \\
& 2|b|b e^{-\Phi-\xi}\left(\da^2\bar b + \frac 1{\bar b}\da\bar b\da\bar b\right. & &
2b\da\partial_z\bar b - \frac 4{\ell}\da\bar b \\
\raisebox{1.5ex}[-1.5ex]{0} & \left.-2\da(\Phi+\xi)\da\bar b\right) & \raisebox{1.5ex}
[-1.5ex]{$-\frac 6{\ell}\da\bar b$} & +2b\left(\partial_z\bar b + \frac 1{\ell}
\right)\da\ln\bar b
\end{array}\right), \nonumber
\end{displaymath}
where $\partial = \partial_w$, $\da = \partial_{\bar w}$. The other four integrability
conditions give no additional information, because the lines of the corresponding
matrices are proportional to the lines of $\tilde N_{wt}$ and
$\tilde N_{\bar w t}$\footnote{In order to show this, one has to make use of
eqns.~(\ref{bianchi}) and (\ref{DeltaF}).}.

As the upper right $3\times 3$ determinant of $\tilde N_{wt}$ must vanish, we obtain
$\partial b = 0$ or
\begin{equation}
\partial_z\left(e^{-2(\Phi+\xi)}\bar b\partial\bar b\right)\partial\left(e^{-2(\Phi
+\xi)}b\partial b\right) - \partial_z\left(e^{-2(\Phi+\xi)}b\partial b\right)
\partial\left(e^{-2(\Phi+\xi)}\bar b\partial\bar b\right) = 0\,. \label{poissonbr}
\end{equation}
Let us assume that the expression in (\ref{poissonbr}) does not vanish. One has then
$\partial b = 0$ as well as $\partial\bar b = 0$\footnote{This follows from the vanishing
of the $3\times 3$ determinant that is obtained from $\tilde N_{wt}$ by deleting the
first column and the third line.}. But then also (\ref{poissonbr}) holds, which leads
to a contradiction. Thus (\ref{poissonbr}) must be satisfied in any case.

Note that the vanishing of the first column of ${\tilde N}_{\mu\nu}$ implies that also
the first column of $T^{-1}N_{\mu\nu}T$ is zero, and thus $T^{-1}N_{\mu\nu}T \in$ a$(3,\bC)$,
hence the generalized holonomy in the case of one preserved complex supercharge is
contained in the affine group A$(3,\bC)$. This supports the classification scheme
of \cite{Batrachenko:2004su}. Of course, depending on the particular solution, the
generalized holonomy may also be a subgroup of A$(3,\bC)$.

\subsection{Time-dependence of second Killing spinor}

In this section we will utilize the above Killing spinor equations
to derive the time-dependence of the second Killing spinor. In
addition, we will show that the Killing spinor equations can be completely
solved when the second Killing spinor is time-independent.

Let us first simplify the Killing spinor equations \eqref{linsys}. In the following we
set $b=re^{i\varphi}$ and define $\psi = \Phi+\xi$, $\psi_1=r^2\alpha_1$,
$\psi_2=re^{-\psi}\alpha_2$, $\psi_{12}=re^{-\psi}\alpha_{12}$ and
$\psi_{\pm}=\psi_2\pm\psi_{12}$.
First of all, use the integrability conditions \eqref{1st-intcond}, that can be
rewritten as ${\tilde N}_{\mu\nu}T^{-1}\alpha=0$. Defining $P=e^{-2\psi}b\partial b$,
the second component for $\mu=w$, $\nu=t$ gives
\begin{equation}
\psi_1 P' + \psi_-\partial P = 0\,, \label{intcondP1-}
\end{equation}
with $'=\partial_z$. Let us assume $P'\neq 0$ (the case $P'=0$ is considered
in appendix \ref{app-P'=0} and will lead to the same conclusions). If we define $g(t,z,w,\bar w)=-\psi_-/P'$, we get
\begin{displaymath}
\psi_- = -gP'\,, \qquad \psi_1 = g\partial P\,.
\end{displaymath}
The third component of the $(w,t)$ integrability condition is of the form
\begin{displaymath}
\psi_1 f_1 + \psi_2\partial b + \psi_- f_- = 0\,,
\end{displaymath}
for some functions $f_1$, $f_-$ that depend on $z,w,\bar w$ but not on $t$.
Using the above form of $\psi_1$ and $\psi_-$, this becomes
\begin{equation}
f_1 g \partial P + \psi_2\partial b - f_- g P' = 0\,. \label{f1f-}
\end{equation}
Now, if $g=0$, the latter equation implies $\psi_2\partial b=0$, and hence
(since $\partial b\neq 0$ due to $P'\neq 0$) $\psi_2=0$. Furthermore, $\psi_1=\psi_-=0$
in this case, so there exists no other Killing spinor. Thus, $g\neq 0$ and we can
write $g = \exp G$. Dividing \eqref{f1f-} by $g$ and deriving with respect to $t$
yields $\partial_t(\psi_2/g)=0$ and hence
\begin{displaymath}
\psi_2 = e^G\psi_2^0(z,w,\bar w)\,.
\end{displaymath}
It is then plain that $\partial_t\psi_i = \psi_i\partial_t G$, $i=1,2,12$. The
Killing spinor equations are of the form
$\partial_{\mu}\psi_i = {\cal M}_{\mu ij}\psi_j$, for
some time-independent matrices ${\cal M}_{\mu}$. Taking the derivative of this with
respect to $t$, one gets $\partial_{\mu}\partial_t G=0$, whence
\begin{displaymath}
G = G_0 t + \tilde G(z,w,\bar w)\,,
\end{displaymath}
with $G_0\in\bC$ constant. We have thus $\partial_t\psi_i = G_0\psi_i$ and
hence also $\partial_t\al_i = G_0\al_i$. Furthermore, the time-dependence of $\al_0$ can be easily deduced from the Killing spinor equations: if $G_0$ does not vanish it is of the same exponential form as the other components of the second Killing spinor, i.e.~$\partial_t\al_0 = G_0\al_0$, while if $G_0$ vanishes there can be a linear part in $t$, i.e. $\partial_t \al_0 = c$ for some constant $c$. Hence, in terms of the basis elements, the time-dependence of the second Killing spinor takes the form\footnote{We will loosely refer to Killing spinors with $G_0 = 0$ as time-independent, despite the possible linear time-dependence, to distinguish from the $G_0 \neq 0$ exponential time-dependence.}
 \begin{align}
   & G_0 = 0 : \; \; \; \epsilon_2 = c_0 1 + c_1 e_1 + c_2 e_2 + c_{12} e_1 \wedge e_2 + c t (1 + b e_2) \,, \notag \\
   & G_0 \neq 0 : \; \; \; \epsilon_2 = e^{G_0 t} (c_0 1 + c_1 e_1 + c_2 e_2 + c_{12} e_1 \wedge e_2) \,,
 \label{time-dep}
 \end{align}
where $c_0,c_1,c_2,c_{12}$ are time-independent functions of the spatial coordinates, and $c$ is a constant. This was derived assuming $P'$ does not vanish, but as we show in appendix C is in fact a completely general result. Hence, adding a second Killing spinor to $\epsilon_1 = 1 + b e_2$, the Killing spinor equations imply that $\epsilon_2$ always has the above time-dependence.

Plugging this time-dependence into the subsystem of the Killing spinor equations not containing $\alpha_0$ one obtains in terms of $\psi_i$
\eqn
\psi_1'-\left(\frac{G_0}{4r^2}+\frac{b'}b\right)\psi_1-\frac{\partial b}b\psi_-&=&0\,,
\label{KSE1.1} \\
\psi_2'-\left(\frac{G_0}{4r^2}+\frac{\bar b'}{\bar b}+\frac1{\ell\bar b}\right)\psi_2-
\left(\frac{b'}b+\frac1{\ell b}\right)\psi_{12}&=&0\,, \label{KSE1.2} \\
\psi_{12}'-e^{-2\psi}\frac{\bar\partial\bar b}{\bar b}\psi_1-\left(\frac{G_0}{4r^2}+
\frac{b'}b+\frac{\bar b'}{\bar b}+\frac1{\ell\bar b}\right)\psi_{12}&=&0\,, \label{KSE1.3}
\feqn
\eqn
\psi_1'-\left(-\frac{G_0}{4r^2}+\frac{\bar b'}{\bar b}\right)\psi_1-\frac{\partial\bar b}
{\bar b}\psi_-&=&0\,, \label{KSE2.1} \\
\psi_2'+e^{-2\psi}\frac{\bar\partial b}{b}\psi_1-\left(-\frac{G_0}{4r^2}+\frac{b'}b+
\frac{\bar b'}{\bar b}+\frac1{\ell b}\right)\psi_2&=&0\,, \label{KSE2.2} \\
\psi_{12}'-\left(\frac{\bar b'}{\bar b}+\frac1{\ell\bar b}\right)\psi_2-\left(-\frac{G_0}
{4r^2}+\frac{b'}b+\frac1{\ell b}\right)\psi_{12}&=&0\,, \label{KSE2.3}
\feqn
\eqn
\partial\psi_1-\sigma_w G_0\psi_1&=&0\,, \label{KSE3.1} \\
\partial\psi_2-\left(\frac{b'}b+\frac1{\ell b}\right)\psi_1-\left(\sigma_w G_0+
\frac{\partial b}b+\frac{\partial\bar b}{\bar b}-2\partial\psi\right)\psi_2&=&0\,,
\label{KSE3.2} \\
\partial\psi_{12}+\left(\frac{\bar b'}{\bar b}+\frac1{\ell\bar b}\right)\psi_1-
\left(\sigma_w G_0+\frac{\partial b}b+\frac{\partial\bar b}{\bar b}-2\partial\psi\right)
\psi_{12}&=&0\,, \label{KSE3.3}
\feqn
\eqn
\bar\partial\psi_1-\left(\sigma_{\bar w} G_0+\frac{\bar\partial b}b+
\frac{\bar\partial\bar b}{\bar b}\right)\psi_1+e^{2\psi}\left[\left(\frac{b'}b+
\frac{\bar b'}{\bar b}\right)\psi_- + \frac1\ell\left(\frac{\psi_2}b-\frac{\psi_{12}}
{\bar b}\right)\right]&=&0\,, \label{KSE4.1} \\
\bar\partial\psi_2-\left(\sigma_{\bar w} G_0+\frac{\bar\partial\bar b}{\bar b}\right)
\psi_2-\frac{\bar\partial b}b\psi_{12}&=&0\,, \label{KSE4.2} \\
\bar\partial\psi_{12}-\frac{\bar\partial\bar b}{\bar b}\psi_2-\left(\sigma_{\bar w} G_0 +
\frac{\bar\partial b}b\right)\psi_{12}&=&0\,. \label{KSE4.3}
\feqn

For $G_0=0$, these equations simplify significantly, and allow for a complete
solution. As is shown in appendix \ref{app-G0=0}, under the additional assumption
$\psi_-\neq 0$, $\psi_1\neq 0$, the metric and the field strength for
half-supersymmetric solutions with $G_0=0$ are given in terms of a single real
function $H$ depending only on the combination $Z-w-\bar w$
and satisfying the second order differential equation
\begin{equation}
2\left(1+e^{-2H}\right)\ddot H + {\dot H}^2\left[1-\frac{3\alpha^2}{e^{2H}+1-\alpha^2}
\right]
=\frac{\gamma^2}{\ell^2}\,, \label{equ_H}
\end{equation}
where $\alpha\in\bR$ denotes an arbitrary constant and $\gamma=0,1$. The new coordinate
$Z$ is defined by $Z=z$ for $\gamma=0$ and $Z=\ell\ln\left(1+\frac z{\ell}\right)$
for $\gamma=1$. Furthermore, in the remainder of this section and in
appendix \ref{app-G0=0}, a dot denotes a derivative with respect to $Z-w-\bar w$.
Given a solution of \eqref{equ_H}, one defines the functions $\chi, \rho$ by
\eq
\chi = \frac{i\alpha}{\sqrt{e^{2H}+1-\alpha^2}}\,, \qquad
\frac 1{\ell^2\rho^2} = \left(\frac{\gamma}{\ell}+\dot H\right)^2 - {\dot H}^2\chi^2\,.\label{def-of-chi-rho}
\feq
Note that $\chi$ is imaginary and $\rho$ is real. $b$ and $\psi$ are then given by
\begin{displaymath}
b = e^{\gamma Z/\ell}\rho\,e^{i\varphi}\,, \qquad e^{2\psi} = e^{2(H+\gamma Z/\ell)}\,,
\end{displaymath}
where
\begin{displaymath}
\tan\varphi = \frac{i\dot H\chi}{\frac{\gamma}{\ell}+\dot H}\,,
\end{displaymath}
so that the metric reads
\eq
ds^2 = -4\rho^2 e^{2\gamma Z/\ell}(dt + \sigma)^2 + \frac 1{\rho^2}\left(\frac{dZ^2}4 +
       e^{2H}dw d\bar w\right)\,, \label{metric-G0=0}
\feq
where the shift vector satisfies
\begin{displaymath}
\partial_Z\sigma_w = \frac 14 e^{-\gamma Z/\ell}\left(\frac{\chi}{\rho^2}\right)^{\cdot}\,,
\qquad \partial\sigma_{\bar w} - \bar\partial\sigma_w = -\frac 12 e^{-\gamma Z/\ell}
\left(e^{2H}\frac{\chi}{\rho^2}\right)^{\cdot}\,.
\end{displaymath}
Finally, the gauge field strength is given by \eqref{fieldstr}.

Equation \eqref{equ_H} is actually the Euler-Lagrange equation for the following standard action for the scalar $H$
\eq
S=\int\dd\left(Z-w-\bar w\right)\left[\frac12M(H)\dot H^{2}-V(H)\right]\,,
\feq
where
\eq
M(H)=\frac{\left(e^{2H}+1\right)^{2}}{\left(e^{2H}+1-\alpha^{2}\right)^{3/2}}\,,\qquad V(H)=-\frac{\gamma^{2}}{2\ell^{2}}\frac{e^{2H}+1-2\alpha^{2}}{\left(e^{2H}+1-\alpha^{2}\right)^{1/2}}\,.
\feq
Thus it is possible to use the energy conservation law of that model in order to evaluate the ``velocity'' $\dot H$ in terms of $H$. Since $\dd H=\dot H\dd\left(Z-w-\bar w\right)$ one has
\eq
\frac{\dd}{\dd H}\left(\frac12M(H)\dot H^{2}+V(H)\right)=0\,,
\feq
so that there must exist a constant $E$ such that
\eq
\dot H=\sqrt{\frac2{M(H)}\left[E-V(H)\right]}=\sqrt2\frac{\left(e^{2H}+1-\alpha^{2}\right)^{3/4}}{e^{2H}+1}\left[E+\frac{\gamma^{2}}{2\ell^{2}}\frac{e^{2H}+1-2\alpha^{2}}{\sqrt{e^{2H}+1-\alpha^{2}}}\right]^{1/2}\,.\label{dotHeq}
\feq
The key-point is to consider now, as a new coordinate, the function $H$ in place of $w+\bar w$\footnote{This is possible by simply requiring that $\dot H\neq0$.} and to write down the full solution, say metric plus gauge field, in terms of $H$. Using $w=x+iy$, the general solution is given by
\eqn
\dd s^{2}&=&-4\rho^{2}e^{2\gamma Z/\ell}\left[\dd t+e^{-\gamma Z/\ell}\hat\sigma_{y}\dd y\right]^{2}+\frac{e^{2H}}{4\rho^{2}}\dd y^{2}+\frac1{4\rho^{2}}\left[\dd Z^{2}+e^{2H}\left(\dd Z-\frac{\dd H}{\dot H}\right)^{2}\right]\,,\nonumber\\
{\cal A}&=&\ell\dot H\left[-2i\rho^{2}\chi e^{\gamma Z/\ell}\dd t+\left(1-e^{2H}\chi^{2}\right)\dd y\right]-i\frac\ell4\dd\log\frac{b}{\bar b}\,,
\feqn
where $\dot H$ is given in equation \eqref{dotHeq}, the functions $\chi$ and $\rho$ are defined in \eqref{def-of-chi-rho} and the shift vector reads
\eqn
\hat\sigma_{y}&=&-\frac{i}2e^{2H}\frac{\chi}{\rho^{2}}\nonumber\,.
\feqn

If $\gamma=1$, a simple example of this set of solutions can be obtained by setting $\alpha=0$, so
 \begin{align}
   \dot H = 1 / \ell \,, \qquad b = \frac{1}{2}\left(1 + \frac{z}{\ell}\right) \,.
 \end{align}
As will be shown in section 5.1.2, this corresponds to the maximally supersymmetric AdS$_4$ solution. More general $\gamma=1$ solutions will be two-parameter deformations thereof, the parameters being $\alpha$ and the \emph{energy} $E$ of the associated scalar system.

Setting $\gamma=0$ the potential $V(H)$ vanishes and the parameter $E$ can be fixed by a simple rescaling of the coordinates. Thus we are left with a one-parameter family of solutions. Since the metric does no more depend explicitly on $Z$, it is useful to replace the coordinate $Z$ instead of $x$ by $H$. Defining a new coordinate $r$ such that $r^{4}\equiv16\left(e^{2H}+1-\alpha^{2}\right)$ and a new parameter $Q=\frac4\ell\alpha$, the complete solution reads
\eqn
\dd s^{2}&=&-\left(\frac{r^{2}}{\ell^{2}}+\frac{Q^{2}}{r^{2}}\right)\left[\dd t-\frac{2\ell^{3}Q}{r^{4}+\ell^{2}Q^{2}}\dd y\right]^{2}+\nonumber\\
&&+\left(\frac{r^{2}}{\ell^{2}}+\frac{Q^{2}}{r^{2}}\right)^{-1}\left[h(r)^{2}\left(\dd r+\frac2{h(r)}\dd x\right)^{2}+\frac14\left(r^{4}+\ell^{2}Q^{2}-16\right)\left(\dd x^{2}+\dd y^{2}\right)\right]\,,\nonumber\\
{\cal A}&=&-\frac{Q}r\dd t+\frac{2\ell}r\dd y-i\frac\ell4\dd\log\frac{b}{\bar b}\,,
\feqn
where
\eq
h(r)=\frac{r^{4}+\ell^{2}Q^{2}}{r^{4}+\ell^{2}Q^{2}-16}\,.
\feq
The parameter $Q$ can thus be interpreted as an electric charge. The Petrov type of the solution is $D$ or simpler. If one sets $Q=4/\ell$ the Petrov type is reduced to $N$, so that there is a gravitational wave.

In order to complete the classification of $G_0 = 0$ solutions, we need
to study separately the cases where either $\psi_1$ or $\psi_-$
vanishes (it can easily be seen from \eqref{KSE1.2} and \eqref{KSE1.3} that there is no solution if both vanish). As one can see by looking at equations (\ref{KSE1.1}) and (\ref{KSE2.1}), the condition $\psi_1=0$ leads to
$b=b(z)$, which is studied in detail in section \ref{b(z)}. 
The other possibility, $\psi_{-}=0$, is more involved, but as we show in appendix~\ref{app-G0=0} it boils down to three different cases, that can be completely solved: the AdS$_2\times\HH^2$ anti-Nariai spacetime studied in section \ref{ssan}, the imaginary $b$ case solved in section~\ref{im-b}, and finally the half BPS solution coming from the gravitational Chern-Simons model, that we analyse in section~\ref{gcs}.

We would like to remark that the assumption $G_0=0$ on the overall time-dependence of the second Killing spinor seems a reasonable choice since all known 1/2-supersymmetric solutions to be studied in the next section are contained in this class, or can be brought to this class by a general coordinate transformation. Hence we expect the $G_0 = 0$ class to form an important subclass of all 1/2-supersymmetric solutions.

\section{Timelike half-supersymmetric examples}

The problem of finding all half BPS configurations in the timelike class involves the solution of the integrability conditions we obtained above. To obtain explicit examples of half BPS solutions, we shall restrict to some simple subclasses with particular $b$. This will determine the fraction of preserved supersymmetry for the solutions which are already known to be 1/4 supersymmetric, and will also lead to new solutions.

\subsection{Static Killing spinors and $b=b(z)$}
\label{b(z)}

The timelike vector field $V$, constructed as a bilinear of the Killing spinor, is static if the associated one-form $V=\dd t+\si$ satisfies the Fr\"obenius condition $V\w\dd V=0$.  Obviously, there can be static BPS solutions with $V$ not being static itself, due to the choice of coordinates; we shall loosely refer the Killing spinors whose vector bilinear is static as {\em static Killing spinors}. The staticity condition, in turn, implies $\dd\sigma=0$ and puts strong constraints on the function $b$. Indeed, equation (\ref{dzsigmanew}) implies that
the phase $\varphi$ of $b$ depends only on $z$. Then, (\ref{disigmanew}) gives the
modulus $r$ of $b$ in terms of its phase,
\eq
r=\frac{\sin\varphi(z)}{l\varphi'(z)}\,.
\label{stat}\feq
As a consequence, $r$ and therefore the complete complex function $b$, depend
on the single variable $z$.
The full solution is therefore determined by the single real function $\varphi$,
which has to satisfy the equations for supersymmetry (together with the
conformal factor $\psi$).

However, since the equations can be exactly solved for arbitrary $b(z)$, we will stick to this more general case and eventually comment on the static subcase.

If $b$ depends only on $z$, the equations of motion simplify to
\eq
\mathrm{Im}\left(b^2\p_z^2\frac1b-\frac{3b}l\p_z\frac1b+\frac1{bl^2}\right)=0\,,
\label{risolvimi}\feq
\eq
e^{-2\xi}\Delta\xi=\frac2le^{2\Phi}\left[\p_z\left(\frac1b+\frac1{\bar b}\right)
-\frac1l\lp\frac1b+\frac1{\bar b}\right)^2+\frac3{lb\bar b}\right]\,.
\label{separiamola}\feq
Here we have used the fact that $\Phi$, defined in \eqref{Phi}, depends only on the
coordinate $z$. In principle there is also an integration constant $K(w,\bar w)$ with
arbitrary dependence on the transverse coordinates, but since $\Phi$ appears only in the
combination $\Phi+\xi$ in all the equations, we can always absorb the $(w,\bar w)$
dependence into the conformal factor $\xi$.
Now the left hand side of equation (\ref{separiamola}) depends only on the coordinates
$w$ and $\bar w$, while the right hand side depends only on $z$. This equation can be
therefore satisfied only if both sides are equal to some constant $\kappa$. The system
of equations is then
\eq
\Delta\xi+\frac\kappa2e^{2\xi}=0\,,
\label{eqxi}\feq
\eq
e^{2\Phi(z)}
\left[\p_z\left(\frac1b+\frac1{\bar b}\right)
-\frac1l\lp\frac1b+\frac1{\bar b}\right)^2+\frac3{lb\bar b}\right]=-\frac{l}4\kappa\,.
\label{risolviancheme}\feq
Note that the first one is the Liouville equation, whose solution describes the
transverse two-dimensional manifold, which has therefore constant curvature $\kappa$.

Equations (\ref{risolvimi}) and (\ref{risolviancheme}) can easily be
solved \cite{Cacciatori:2004rt}. Their solution is given by\footnote{
With this definition, the constants $\al$, $\be$ and $\ga$ coincide with
$a$, $b$ and $c$ of \cite{Cacciatori:2004rt} respectively.
}
\eq
 \bar b = - \frac{\al z^2+ \be z+ \ga}{\ell (2 \al z+ \be)}\,,
\label{solz}\feq
with $\al, \be, \ga \in\bC$. Then $\xi$ solves the Liouville equation for a constant
curvature two-manifold with scalar curvature\footnote{
This scalar curvature differs from the one given in \cite{Cacciatori:2004rt}
for the case in which all coefficients are real, $k=4 \al \ga - \be^2=\kappa/4$. The factor of $4$ comes from the different definition of the conformal factor of the transverse metric, our $\xi$ is related to the old $\gamma$ by $\xi=\ga-\ln2$.}
\eq
\kappa = 8 (\al \bar \ga + \bar \al \ga)-4 \be \bar \be\,.
\feq
This solution generically belongs to the supersymmetric
Reissner-Nordstr\"om-Taub-NUT-AdS$_4$ family
of spacetimes. The values $\al = 0$ and $\be^2 = 4 \al \ga$ are special cases and will be treated separately in the following. Note that the coefficients $\al$, $\be$ and $\ga$ are not three independent parameters, as they can be rescaled
without changing the function $b$: the solutions depend only on their
ratios. For example, if $\al \neq0$, one can use $\be/ \al$ and $\ga/\al$ as independent
complex parameters of the family of solutions.

The solutions with static Killing spinor form a subset of this family. For \eqref{solz} the staticity condition (\ref{stat}) yields the condition
$\al\bar\be - \bar\al\be = 0$. 
Recalling the expression for the NUT charge of these solutions,
\eq
 n=\frac i4\lp\frac{\bar \be}{\bar \al}-\frac{\be}{\al} \rp\,,
\feq
this charge must vanish for non-vanishing $\al$, as one could have guessed.
On the other hand, for $\al=0$ the solution is anti-Nariai, as we shall see below.
We conclude that the most general supersymmetric configuration with static Killing
vector constructed as a Killing spinor bilinear is either of the form
(\ref{solz}) -- i.~e.~in the fourth row of table 1 of \cite{Alonso-Alberca:2000cs} -- with
vanishing NUT charge, or it is anti-Nariai spacetime.

The supersymmetric static solutions discussed so far are generically 1/4-BPS. We want to see what further condition ensures the presence of an
additional Killing spinor. Inserting the staticity ansatz $b=b(z)$  into the
integrability equations and requiring these matrices to be of rank smaller or equal to two,
one finds the following condition (in particular this is obtained from the
vanishing of the minor of the last row of $\tilde N_{wt}$ and the first two
rows of $\tilde N_{\bar w t}$)
\eq
\lp b'+\frac1l\rp
\lp 2b'+\frac1l\rp
\left[\p_z\left(\frac1b+\frac1{\bar b}\right)
-\frac1l\lp\frac1b+\frac1{\bar b}\right)^2+\frac3{lb\bar b}\right]=0\,,
\label{1/2-susy-static}
\feq
As an aside, note that we have only
used the ansatz $b=b(z)$ so far and not the staticity condition
\eqref{stat}, i.e.~the precise relation between $r$ and $\varphi$. The static
solutions are therefore in general still a subset of the solutions under consideration.

Condition (\ref{1/2-susy-static}) calls for the following three different cases, corresponding to the vanishing of its three factors.

\subsubsection{AdS$_2\times\HH^2$ space-time ($\al = 0$)}\label{ssan}

Requiring the first factor of \eqref{1/2-susy-static} to vanish leads to
$b=-\frac{z}{\ell}+ i c$ with constant $c$, corresponding to $\al = 0$ in
\eqref{solz}. We can absorb the imaginary part of $c$ by a shift of the
coordinate $z$ and henceforth will assume $c \in \bR$.

In this case $\kappa=-4$ and we have a hyperbolic transverse space. As a solution
of (\ref{eqxi}) we can take
\eq
e^{2\xi}=\frac1{2x^2}\,.
\feq
Moreover, $e^\Phi=l|b|$ and $\si=0$, therefore giving the metric
\eq
ds^2=-4\left(\frac{z^2}{\ell^2}+c^2\right)\,dt^2+
\frac{dz^2}{4\left(\frac{z^2}{\ell^2}+c^2\right)}+\frac{\ell^2}{2x^2}
\left(dx^2+dy^2\right)\,.
\feq
This is the anti-Nariai AdS$_2\times\HH^2$ solution, with the AdS$_2$ factor
written in Poincar\'e coordinates for $c=0$ and in global coordinates for
$c\neq0$. The
coordinate transformations between Poincar\'e coordinates $(t_P, z_P)$ (with
$c=0$) to global ones $(t_{gl}, z_{gl})$ (with
$c \neq 0$) is given by
 \begin{align}
  & z_P = \frac{1}{2 c } (z_{gl} - \sqrt{z_{gl}^2 + \ell^2 c^2} \cos(4 c t_{gl}
  / \ell)) \,, \notag \\
  & t_P = - \frac{\ell}{2} \frac{\sqrt{z_{gl}^2 + \ell^2 c^2} \sin(4 c t_{gl}
  / \ell)}{z_{gl} - \sqrt{z_{gl}^2 + \ell^2 c^2} \cos(4 c t_{gl}/
  \ell)} \,. \label{Poincare-global}
 \end{align}
The electromagnetic field strength \eqref{fieldstr} in this case is given by
 \eq
  {\cal F} = - \frac{1}{\ell x^2} dx \wedge dy \,,
 \feq
i.e.~only lives on the hyperbolic part and is independent of the
 coordinates of the AdS part of space-time.

This solution preserves precisely 1/2 of the supersymmetries, as was already
shown in \cite{Cacciatori:1999rp}.
To obtain the form of the Killing spinors admitted by this metric we first
observe that the integrability conditions
impose $\al_2=\al_{12}=0$. Then the Killing spinor equations are easily solved, but one
should treat separately the cases $c=0$ and $c\neq0$:
\begin{itemize}
\item If $c = 0$, then
\eq
\al_0=\la_1+\la_2\lp\frac{2t}\ell-\frac1{2b}\rp\,,\qquad
\al_1=\frac{\la_2}b\,,
\feq
where $\la_{1,2}\in\bC$ are integration constants.
This yields the following Killing spinors, spanning a two-dimensional complex space,
\eq
\ep=\left[\la_1+\la_2\lp\frac{2t}\ell-\frac1{2b}\right)\right]1
+b\left[\la_1+\la_2\lp\frac{2t}\ell+\frac1{2b}\rp\right]e_2\,.
\feq
Note that $\la_1=1$, $\la_2=0$ corresponds to the original Killing
spinor. Also note that the constant $G_0$, corresponding to the
time-dependence of the second Killing spinor with $\la_2 \neq 0$, is zero. The form of
the scalar invariant corresponding to the general spinor $\ep$ is
\eq
\tilde b=b\left[
|\la_1|^2+|\la_2|^2\left(\frac{4t^2}{\ell^2}-\frac1{4b^2}\right)
+\frac{2t}\ell\lp\bar\la_1\la_2+\la_1\bar\la_2\rp
\right]
+\frac{1}2\lp\bar\la_1\la_2-\la_1\bar\la_2\rp\,. \label{b-poinc}
\feq
Here the first term is real, while the second is imaginary. Note that the
latter is in fact constant. Then the Killing vector $\tilde V$ built from $\ep$ will
have a norm $\tilde V^2=-4|\tilde b|^2$, and will be timelike unless $\tilde b$ vanishes.
This is however not possible, because both the real and imaginary parts of $\tilde b$ should
vanish, but since $\la_{1,2}$ do not depend on the coordinates, the real part cannot
vanish. Therefore, every Killing spinor of this solution belongs to the timelike class.
\item
If $c\neq0$ we have
\eq
\al_0  =  \frac{1}{2\sqrt{c}} [ \la_1 - i \la_2 + (\la_1 + i \la_2)
\frac b{|b|}e^{-4ic t/\ell}\,, \qquad
\al_1  =  - \frac{i \sqrt{c}}{|b|} (\la_1 + i \la_2) e^{-4ic t/\ell}\,,
\feq
and the most general Killing spinor is parametrized by $\tilde\la_{1,2}\in\bC$ as follows
\eq
\ep= \frac{1}{2 \sqrt{c}} (\la_1 - i \la_2) (1 + b e_2) +
 \frac{b}{2|b|} (\la_1 + i \la_2) e^{-4ic t/\ell} (1 + b^* e_2)\,.
\feq
Note that the combination $\la_1 - i \la_2$ corresponds to the first Killing
spinor $1 + b e_2$, while the orthogonal combination $\la_1 + i \la_2$ gives rise
to the second Killing spinor proportional to $1 + b^* e_2$. Any combination
with $\la_2 \neq 0$ has $G_0 = - 4 i c / \ell$.

In this case, the real part of the invariant $\tilde b$ is given by
\begin{align}
 \text{Re} (\tilde b) = & \frac{|\la_1|^2}{2 \ell c} (-z + \sqrt{z^2 + \ell^2 c^2} \cos(
4 c t / \ell)) +  \frac{|\la_2|^2}{2 \ell c} (-z - \sqrt{z^2 + \ell^2 c^2} \cos(
4 c t / \ell)) +  \notag \\
 & + \frac{1}{2 \ell c} ( \la_1 \la_2^* + \la_2 \la_1^*)
\sqrt{z^2 + \ell^2 c^2} \sin (4 c t / \ell)) \,, \label{b-global}
\end{align}
while the imaginary part is identical to that of \eqref{b-poinc}.
\end{itemize}
It can easily be checked that the coordinate transformation \eqref{Poincare-global}
indeed relates the complex scalar $\tilde b$, which is composed of spinor bilinears, in
\eqref{b-poinc} and \eqref{b-global} to each other.

Let's now check how the isometries of AdS$_2$ act on the Killing spinors. It is useful
to do this by embedding AdS$_2$ with metric
\eq
ds^2=-4\left(\frac{z^2}{\ell^2}+c^2\right)\,dt^2+
\frac{dz^2}{4\left(\frac{z^2}{\ell^2}+c^2\right)}
\label{ads2}\feq
into the three-dimensional flat space $X^a=(U,T,X)$ with metric
\eq
ds^2=-dU^2-dT^2+dX^2\,.
\feq
Then, AdS$_2$ is obtained as the hyperboloid defined by
\eq
-U^2-T^2+X^2=\frac{\ell^2}4\,,
\feq
and its isometry group SO(2,1) will act as the three-dimensional Lorentz group on the
embedding coordinates $X^a$ (here $a$ is a three-dimensional Lorentz index).

If $c=0$, the AdS$_2$ metric (\ref{ads2}) is in the Poincar\'e form, and can be seen to
be the induced metric on the hyperboloid by parameterizing it with the coordinates
$(t,z)$ given by
\eq
  z = U + X \,,\qquad t=\frac{\ell T}{2(U + X)}\,.
\feq
Then, if one defines the 3d Lorentz vector
\eq
\La^a=
\left(
\frac1\ell\lp|\la_1|^2-|\la_2|^2\rp,
\frac1\ell\lp\la_1^*\la_2+\la_1\la_2^*\rp,
-\frac1\ell\lp|\la_1|^2+|\la_2|^2\rp
\right)\,,
\feq
one explicitly checks that the invariant $\tilde b$ can be put in the form
\eq
\tilde b=X_a\La^a+\frac{i\ell}2\sqrt{\La_a\La^a}.
\label{manifest}\feq
Now, the real and imaginary part of $\tilde b$ are independently manifestly invariant
under the AdS$_2$ isometries, as they should be (since they transform respectively as
pseudoscalar and scalar under diffeomorphism\footnote{Note that $\La$ doesn't depend
on the sum of the phases of $\la_{1,2}$; this is diffeomorphism invariant but
transforms under U(1) gauge transformations.}).

If $c\neq0$ we have AdS$_2$ in global coordinates, and the embedding is modified to
\eq
U = - \frac\ell{2c}\sqrt{\frac{z^2}{\ell^2}+c^2}\,\cos\frac{4c t}\ell\,,\quad
T = - \frac\ell{2c}\sqrt{\frac{z^2}{\ell^2}+c^2}\,\sin\frac{4c t}\ell\,,\quad
X = \frac z{2c}\,.
\feq
The invariant (\ref{b-global}) takes again the manifestly invariant form
(\ref{manifest}), as expected, and the isometries of AdS$_2$ are realized linearly
on the Killing spinors through their action on $\La^a$.

This result may be useful to study in detail quotients of AdS$_2$ and to see
whether this operation breaks some supersymmetry.

%%%%%%%%%%%%%%%%%%%%%%%%%%%%%%%%%%
\subsubsection{AdS$_4$ space-time ($\be^2 = 4 \al \ga$)}

The following subcase corresponds to the vanishing of the second factor of the
integrability condition \eqref{1/2-susy-static}. The function $b$ is then
given by $b=-\frac{z}{2l}+ i c$, which can be obtained as the special case $\be^2
= 4 \al \ga$ from \eqref{solz}.
This corresponds to AdS$_4$, the only maximally supersymmetric solution of the
theory. Indeed the integrability condition
matrices vanish in this case.

Let's see in detail the form of the metric arising from different values of
$c$.  As in the previous case we can take the
constant $c$ to be real. If $c=0$, the metric is static, $\si=0$, $\xi=0$ and
$e^{2\Phi}=|b|^4$, and we obtain anti-de~Sitter in Poincar\'e coordinates,
\eq
ds^2=-\frac{z^2}{\ell^2}\left(dt^2-dx^2-dy^2\right)+\frac{\ell^2}{z^2}dz^2\,.
\feq
On the other hand, for $c\neq0$, the metric appears in non-static coordinates,
\eq
\si=-\frac{\ell dy}{4c x^2}\,,\qquad
e^{2\xi}=\frac{\ell^2}{4c^2x^2}\,,\qquad
e^{2\Phi}=|b|^4\,,
\feq
which give
\begin{align}
ds^2=-\left(
\frac{z^2}{l^2}+4c^2
\right)
\left[
\lp dt-\frac{\ell dy}{4c x^2}\rp^2
-\frac{\ell^2}{16c^2x^2}\lp dx^2+dy^2\rp\right]
+\left(\frac{z^2}{l^2}+4c^2
\right)^{-1}dz^2 \,.
\end{align}
The field strength \eqref{fieldstr} vanishes in this case.

We shall now obtain the form of the Killing spinors for AdS$_4$, and will do this in the
simpler $c=0$ case. The solution of the Killing spinor equations yields
\begin{alignat}{2}
& \al_0=\la_1-\lp\frac t\ell+\frac\ell z\rp\la_2+\frac{\bar w}\ell\la_3\,,
\quad && \al_2=-\frac{wz}{2\ell^2}\la_2+\frac12\lp1+\frac{zt}{\ell^2}\rp\la_3-\frac{z}{2
  \ell} \la_4
\,, \notag \\
& \al_1=\frac{2\ell}z\la_2 \,,
\qquad &&
\al_{12}=\frac{wz}{2\ell^2}\la_2+\frac12\lp1-\frac{zt}{\ell^2}\rp\la_3+ \frac{z}{2
  \ell} \la_4\,, \label{AdS4-Ks}
\end{alignat}
where the coefficients $\la_{1,\ldots,4}$ span a four dimensional complex space,
as expected in the case of maximal supersymmetry. In the form basis of the spinors
$\ep=c_0 1+c_1e_1+c_2e_2+c_{12}e_1\w e_2$, we obtain
\begin{alignat}{2}
& c_0=\la_1-\lp\frac t\ell+\frac\ell z\rp\la_2+\frac{\bar w}\ell\la_3\,,
\quad
&& c_2=-\frac z{2\ell}\la_1+\frac z{2\ell}\lp\frac t\ell-\frac\ell z\rp\la_2-
\frac{z\bar w}{2\ell^2}\la_3\,, \notag \\
& c_1=\frac w\ell\la_2-\lp\frac t\ell+\frac\ell z\rp\la_3+\la_4 \,,
\quad
&& c_{12}=\frac{wz}{2\ell^2}\la_2-\frac{z}{2\ell}\lp\frac t\ell-\frac\ell z\rp\la_3
+\frac{z}{2\ell} \la_4\,.
\end{alignat}
The new Killing spinors corresponding to $\la_2$ and $\la_4$ both
have\footnote{Note that this does not hold for $\la_3$, whose time-dependence
  is not of the form derived in section 4.4. There is no contradiction
  however, since all solutions in this class have $P=0$ and hence are treated
  separately in appendix C. It is interesting to find that nevertheless the
  time-dependence of many Killing spinors in this class have the canonical
  $G_0$ time-dependence.} $G_0 = 0$.
To study the action of the AdS$_4$ isometries it is useful to embed the hyperboloid in a
five-dimensional flat space $(U,V,T,X,Y)$ with metric
\eq
ds^2=-dU^2+dV^2-dT^2+dX^2+dY^2.
\label{met32}\feq
Then, AdS$_4$ is the hypersurface $-U^2+V^2-T^2+X^2+Y^2=-\ell^2 / 4$ and its isometries
are realized as the SO(3,2) isometries of the embedding space. The relation with the
Poincar\'e coordinates is
\eq
\frac t\ell=\frac T{U-V}\,,\qquad
\frac x\ell=\frac X{U-V}\,,\qquad
\frac y\ell=\frac Y{U-V}\,,\qquad
z=2(U-V)\,.
\feq
If we define the vectors
\eq
\ell \La^a=\lp\begin{array}{c}
\displaystyle|\la_1|^2-|\la_2|^2+|\la_3|^2 -|\la_4|^2 \\
\\
\displaystyle|\la_1|^2+|\la_2|^2-|\la_3|^2 -|\la_4|^2 \\
\\
\displaystyle\la_3\bar\la_4+\bar\la_3\la_4-\bar\la_1\la_2-\la_1\bar\la_2\\
\\
\displaystyle\la_2\bar\la_4+\bar\la_2\la_4-\bar\la_1\la_3-\la_1\bar\la_3\\
\\
\displaystyle i \lp\la_2\bar\la_4-\bar\la_2\la_4+\bar\la_1\la_3-\la_1\bar\la_3\rp
\end{array}\rp\,,\qquad
X^a=\left(
\begin{array}{c}
U\\V\\T\\X\\Y
\end{array}
\right)\,,
\feq
where the index $a=1,\ldots,5$ is an SO(3,2) index raised and lowered using the metric
(\ref{met32}), then
\eq
\La_a\La^a=- \frac1{\ell^2} \left[
\la_3\bar\la_4-\bar\la_3\la_4+\bar\la_1\la_2-\la_1\bar\la_2
\right]^2\geq0\,,
\feq
and the invariant $\tilde b$ for the Killing spinors reads
\eq
\tilde b=c_0^*c_2+c_1c_{12}^*=X_a\La^a+\frac{i\ell}{2} \sqrt{\La_a\La^a}\,.
\feq
This form of $\tilde b$ is manifestly invariant under the AdS$_4$ isometries, and shows
that under $\La^a$ transforms in the fundamental representation of SO(3,2) under these
transformations. Note that it has precisely the same form (\ref{manifest}) as in the anti-Nariai case. Again, the explicit knowledge of the AdS$_4$ isometry group action on the Killing spinors is important to study the supersymmetry of its quotients.

\subsubsection{The Reissner-Nordstr\"om-Taub-NUT-AdS$_4$ family}

The last subcase corresponds to the vanishing of the third factor of the
integrability condition \eqref{1/2-susy-static}. Note that this is precisely the
expression in square brackets of equation (\ref{risolviancheme}) and the condition
reads simply $\kappa=0$. Then $\xi$ is an harmonic function and the transverse space
is flat. In particular, the solution (\ref{solz}) admits a second Killing spinor if
\eq
|\be|^2=2(\al \bar \ga +\bar \al \ga) \,.
\feq
Since $\al \neq 0$ we can define $\zeta=\mathrm{Im}(\be/\al) $ and
$\delta=\mathrm{Im}(\ga/\al)$.
Moreover, all equations are invariant under rigid translations in the $z$
directions, since the coordinate $z$ never appears explicitly in them. One can
use this freedom to eliminate the real part of $\be/\al$ by performing the
redefinition $z\mapsto z - \tfrac{1}{2} \mathrm{Re}(\be/\al)$. Hence this complete
family of 1/2 BPS solutions is determined by two real parameters $\zeta$ and $\delta$,
\eq
 b = -\frac1\ell \frac{z^2- i \zeta z+ \tfrac14 \zeta^2 - i\delta}{2z- i \zeta
 } \,.
\feq
Then $\si= - 2 \zeta (r/\ell)^2 d \vartheta$ and the resulting metric is
\begin{eqnarray}
ds^2=-\frac{\lp z^2+\frac{\zeta^2}4\rp^2+\lp\zeta z+\delta\rp^2}{\ell^2\lp z^2+
\frac{\zeta^2}4\rp}\left(dt- \frac{2 \zeta}{\ell^2}  r^2 d \vartheta\right)^2
\qquad\qquad\qquad\quad\nonumber\\
+\frac{\ell^2\lp z^2+\frac{\zeta^2}4\rp dz^2}{\lp z^2+\frac{\zeta^2}4\rp^2+\lp\zeta z+
\delta\rp^2} + \frac{4}{\ell^2} \lp z^2+\frac{\zeta^2}4\rp\left(dr^2+r^2\,d\vartheta^2\right)\,,
\label{mezzasusanna}\end{eqnarray}
where we used polar coordinates $(r,\vartheta)$ in the $(w,\bar w)$ plane.
The charges of the solution are
\eq
M=-\frac{\delta\zeta}{\ell^2},\qquad
n=\frac\zeta2,\qquad
P=-\frac{\zeta^2}{2\ell},\qquad
Q=-\frac\delta \ell.
\label{halfsusycharges}\feq
Essentially, the imaginary part of $\ga$ gives the electric charge and the
imaginary part of $\be$ determines the NUT charge. Note that the quantization
condition $P=-(k \ell ^2+4n^2)/2\ell$ is also satisfied. In terms of the charges, the solution is
given by
\eq
 b= - \frac{1}{\ell} \frac{\left(z-in\right)^2+2n^2+i \ell Q}{2\left(z-in\right)}\,.
\feq

The subfamily of static half BPS configurations is obtained by imposing the staticity
condition $\zeta=0$ or equivalently vanishing NUT charge. It is parameterized by the
single parameter left, $\delta\in\bR$ and the solutions are restricted to have the
following charges
\eq
M=0\,, \qquad
n=0\,, \qquad
P=0\,, \qquad
Q=-\frac\delta \ell \,.
\nonumber\feq
In terms of the charges, the solution is given by
\eq
 b= - \frac{1}{\ell} \frac{z^2+i \ell Q}{2z}\,.
\feq
The metric and electromagnetic field strength for this solution read
\eq
ds^2=-\lp\frac{Q^2}{z^2}+\frac{z^2}{\ell^2}\rp\,dt^2
+\frac{dz^2}{\frac{Q^2}{z^2}+\frac{z^2}{\ell^2}}
+4\ell^2 z^2\,dw d\bar w\,,
\label{mezzasusannastatica}\feq
and
\eq
\F = -\frac{Q}{z^2}\,\dd t\w\dd z\,.
\label{mssf}\feq
This is simply the backreacted AdS$_4$ filled with the electric field generated by an electric charge $Q$ placed in its center $\zeta=0$. The solution has a singularity there.
Note that this solution was already shown to be 1/2 supersymmetric in
\cite{Caldarelli:1998hg}. It was also shown there that the Killing spinors are
preserved if one compactifies the transverse two-dimensional plane to a two-torus.

We will now discuss the Killing spinors for these metrics. The integrability conditions
impose $\al_2=0$ and
\eq
\lp b'+\frac1\ell-\frac b{\ell\bar b}\rp\al_3=\lp b'+\frac 1\ell\rp\al_4\,.
\feq
With these constraints, the Killing spinor equations simplify, and can be solved to give
\eq
\al_0=\la_1+2i\zeta\bar w\la_2\,,\qquad
\al_1=0\,,
\feq
\eq
\al_2=\frac{z^2+i\zeta z+\frac{\zeta^2}4+i\delta}{\sqrt{4z^2+\zeta^2}}\,,\qquad
\al_{12}=\al_2-\frac{\la_2}2\sqrt{4z^2+\zeta^2}\,,
\feq
where $\la_{1,2}\in\bC$ parameterize the two dimensional space of Killing spinors.
Then the most general Killing spinor for these metrics is
\begin{align}
\ep=\lp\la_1+2i\zeta\bar w\la_2\rp 1
-\ell\la_2\sqrt{\frac{2z+i\zeta}{2z-i\zeta}}\,e_1\qquad\qquad\qquad\qquad\nonumber\\
+b\lp\la_1+2i\zeta\bar w\la_2\rp e_2
-\frac{z^2-i\zeta z+\frac{\zeta^2}4-i\delta}{\sqrt{4z^2+\zeta^2}}\la_2\,e_1\w e_2\,.
\end{align}
Again the second Killing spinor has $G_0 = 0$ time-dependence.
Finally, the corresponding orbit of the Killing spinor is determined by the invariant
\eq
\tilde b=b|\la_1|^2+\lp\ell\frac{z^2+i\zeta z+\frac{\zeta^2}4+i\delta}{2z-i\zeta}
+4\zeta^2bw\bar w\rp|\la_2|^2+2i\zeta b\lp\bar w\bar\la_1\la_2-w\la_1\bar\la_2\rp\,.
\feq
It is easy to show now that $\tilde b$ is non vanishing for any choice of $\la_{1,2}$:
indeed if $\tilde b=0$, we have $\p\bar\p\tilde b=4\zeta^2b|\la_2|^2=0$
and either $\la_2=0$, which implies in turn $\la_1=0$, or $\zeta=0$. In the latter case,
it is very easy to see that $\tilde b=0$ iff $\ep=0$. Therefore, all Killing spinors of
this family of metrics belong to the timelike class, and the solution is
purely timelike.\\

\noindent{\bf Summary of the $b=b(z)$ case:}
\begin{enumerate}
\item
The only supersymmetric solutions with static Killing spinor (i.e. whose timelike Killing vector constructed as a Killing spinor bilinear is static) are AdS$_4$, the anti-Nariai spacetime and the Reissner-Nordstr\"om-AdS$_4$ solutions of the fourth row of table~1 of \cite{Alonso-Alberca:2000cs}, i.~e.~solutions of the form (\ref{solz})
with vanishing NUT charge.
\item
The only 1/2 BPS solutions with static Killing spinor are the anti-Nariai spacetime
and the solution (\ref{mezzasusannastatica}) with field strength (\ref{mssf}).
\item
The most general half BPS solution with $b=b(z)$ are the anti-Nariai spacetime and the
solution (\ref{mezzasusanna}) with charges (\ref{halfsusycharges}) describing an electric charge in the
center of AdS$_4$.
\end{enumerate}

The natural way to continue this approach is to study half BPS solutions with $b$
harmonic, and this will be the subject of the next paragraph.

\subsection{Harmonic $b$ solutions}

The previous class of solutions can be generalized by requiring $\Delta b=0$
instead of $b=b(z)$ \cite{Cacciatori:2004rt}. This implies that $\Delta 1/b = 0$ and hence
\eqref{maxwell} still simplifies in exactly the same way as in the $b=b(z)$
case. Indeed, the solution is
\eq
  \bar b = - \frac{\al z^2+ \be z+\ga}{\ell (2 \al z+ \be)}\,,
\label{ffff}\feq
where now $\al$, $\be$ and $\ga$ are no more constants but arbitrary functions of
$(w,\bar w)$. It is then easy to show that the $\Delta b=0$ condition
requires these functions to be harmonic and all (anti-)holomorphic, that is
 $\al$, $\be$ and $\ga$ all depending either only on $w$ or only on $\bar w$, and
this is the most general solution with $\Delta b = 0$. The $b=b(z)$
configurations are particular cases of this larger class, and are obtained for
$\al$, $\be$ and $\ga$ constant.
Note that also the $\p b=0$ and $\p\bar b=0$ subclasses fall into this family.

Let's take for definiteness $\al$, $\be$, $\ga$  all anti-holomorphic, then
$b=b(z,w)$. The requirement that the integrability conditions allow for an
extra Killing spinor, i.e.~that they are of rank $\leq 2$, in this case leads
to several conditions. One of these is obtained from the minor of the last
three lines of $\tilde N_{wt}$ and reads
\eq
\lp2\p_z\bar b+\frac1l\rp\lp\p_z\bar b+\frac1l\rp
\lp\p^2b+\frac1b\p b\p b-2\p(\Phi + \xi)\p b\right)\p b=0. \label{1/2-susy-harmonic}
\feq
This gives three different cases to be analysed, corresponding to the
vanishing of the first three factors of this equation (vanishing of the fourth factor
implies $b=b(z)$ and hence brings one back to the previous section).

\subsubsection{Deformations of AdS$_2\times\HH^2$}

The vanishing of the first factor in \eqref{1/2-susy-harmonic} implies
$b=-\frac z\ell + i c(w)$, where $c(w)$ is an arbitrary holomorphic function. These are the $\al(w) = 0$ supersymmetric Kundt solutions of Petrov type II, describing gravitational and electro-magnetic waves propagating on anti-Nariai space-time \cite{Cacciatori:2004rt}.  

The
remaining integrability conditions however imply $\al_1 = \al_2 = \al_{12} = 0$,
in which case there is no second Killing spinor, or $\partial c = 0$. Therefore there
are no new half BPS solutions with non constant $c$. In this class $c$ constant is the
half supersymmetric anti-Nariai spacetime and the other preserve only 1/4 of the
supersymmetries.

\subsubsection{Deformations of AdS$_4$}

The vanishing of the second factor in \eqref{1/2-susy-harmonic} implies
$b=-\frac z{2\ell}+ i c(w)$. In this case we are considering the $\be^2 = 4 \al \ga$ supersymmetric Kundt solutions, describing gravitational and electro-magnetic waves propagating on AdS$_4$ spacetime \cite{Cacciatori:2004rt}. 

Again the
remaining integrability equations have to solutions: $\al_1 = \al_2 = \al_{12} =
0$ or $\partial c = 0$. Hence, as in the previous case, we find that there
are no harmonic deformations of AdS$_4$ preserving half supersymmetry.

\subsubsection{Deformations of Reissner-Nordstr\"om-Taub-NUT-AdS$_4$}

Not considering the previous two special cases, the general solution represents expanding gravitational and electro-magnetic waves propagating on a Reissner-Nordstr\"om-Taub-NUT-AdS$_4$ spacetime \cite{Cacciatori:2004rt}. When Im$(\beta) = 0$, the solution can be put in Robinson-Trautman form and is of Petrov type II.

The vanishing of the third factor in \eqref{1/2-susy-harmonic} is given
by
 \eq
  \p^2b+\frac1b\p b\p b-2\p(\Phi + \xi)\p b=0 \,.
 \feq
With $b$ given in \eqref{ffff} this case can be solved for the derivative of $\Phi + \xi$
and implies
\eq
\bar\p(\Phi+\xi)=\frac1{2\p b}\lp\p^2b+\frac1b\p b\p b\right),
\feq
and therefore $\Delta(\Phi+\xi)=0$. Then (\ref{separiamola}) fixes the transverse
manifold to be flat and
\eq
\kappa(w)= 8 (\al \bar \ga + \bar \al \ga) -4 \be \bar \be=0.
\feq
But $\al$,$\be$ and $\ga$ being holomorphic, this last equation can be satisfied if
and only if they are constant, and we are back to the previous
case, i.~e.~there are no new 1/2 BPS solutions. \\

\noindent{\bf Summary of the harmonic case:}

There are no new half BPS solutions in the harmonic $b$ case. The only half
BPS solutions are those with $b=b(z)$, and as soon as one deforms these
solutions by adding some harmonic $(w,\bar w)$-dependence, one breaks
 supersymmetry further to 1/4.

\subsection{Imaginary $b$ solutions} \label{im-b}

Another subcase we want to study is $\bar b = -b$, i.~e.~$b$ purely
imaginary. For notational convenience we introduce\footnote{In the following we
will assume that $X$ is positive without loss of generality.}
\begin{displaymath}
b = i X \,,
\end{displaymath}
where $X$ is real. From \eqref{Phi} one gets $\Phi=0$.
All quantities in the Bianchi identity \eqref{bianchi}, apart from
$b$ and hence $X$, are then $z$-independent. The only consistent possibility is to take
$\partial_z X=0$. The remaining equations \eqref{bianchi} and \eqref{maxwell} read
 \begin{align}
  \Delta \xi = \frac{6}{\ell^2 X^2} e^{2 \xi} \,, \qquad \Delta \frac{1}{X} -
  \frac{4}{\ell^2 X^3} e^{2 \xi} = 0 \,. \label{field-eqs-b-im}
 \end{align}
Examples of 1/4 supersymmetric solutions of this class, i.e.~with imaginary $b$, that
were discussed in \cite{Cacciatori:2004rt}
are $X = (x / \ell)^\alpha$ with $\alpha = -2$ and $\alpha = \tfrac{1}{3}$. These
correspond to a particular Petrov type I solution and an electrovac AdS
travelling wave of Petrov type N, respectively. It was shown that the
latter actually preserves a second, null Killing spinor. In this section we
will derive the general condition for 1/2 supersymmetry in the case of
imaginary $b$ and will find that there is a one-parameter family of such solutions.

The condition for 1/2 supersymmetry is very simple in this case. Assuming
that $\partial X$ is not equal to zero, which would clearly be incompatible
with \eqref{field-eqs-b-im}, there is only one differential
constraint which needs to be satisfied for the existence of a second Killing
spinor, i.~e.~for the matrices of integrability conditions to have rank 2, namely
 \begin{align}
  \partial^2 X^{-1} - 2 \partial \xi \partial X^{-1} = 0 \,.
 \end{align}
The above three differential equations can be integrated to
 \begin{align}
  e^{2 \xi} = - i \bar K (\bar w) \partial X^{-1} \,, \quad
  \partial X^{-1} = \frac i{\ell^2} K(w) \left(\frac{1}{4 X^4}  + L\right) \,,
 \label{int-sol}
 \end{align}
where $K(w)$ is an arbitrary holomorphic function and $L$ is a real
constant. The function $K(w)$ corresponds to the freedom to choose holomorphic
coordinates on the two-dimensional space, and hence it can be gauged away. A
convenient gauge choice will be $K(w) = i \ell$. Note that, for this choice, the
imaginary part of the right hand side of the last equation vanishes, and therefore
that $\partial_y X = 0$.

For $L = 0$, \eqref{int-sol} can be integrated to give
 \begin{align}
  X^3 = \frac{3 x}{2 \ell}  \,,
 \end{align}
which is (up to a rescaling of the coordinate $x$) the example given above
with $\alpha = \tfrac{1}{3}$. This was already found to be 1/2 supersymmetric
in \cite{Cacciatori:2004rt}. Here we find that this solution is a special case
of the most general possibility.

For other values of the constant $L$ it is convenient to use $X$ as a new
coordinate instead of solving for $X(x)$.
From \eqref{dzsigmanew} and \eqref{disigmanew} it follows that $\sigma$ can be
chosen to be
 \begin{align}
  \sigma = \frac{\dd y}{4 X^4}\,.
 \end{align}
Then the metric reads
 \begin{align}
  ds^2 = - 4 X^2 \left(dt + \frac{dy}{4 X^4}\right)^2 + \frac{1}{4X^2}\, dz^2 + \frac{\ell^2dX^2}{X^2(1+4LX^4)} + \frac{1+4LX^4}{4X^6}\, dy^2 \,. \label{metr-imb}
 \end{align}
Finally, from (\ref{fieldstr}) we obtain the gauge field strength
\begin{align}
\F = 2 \dd t\wedge\dd X\,. \label{gaugefield-imb}
\end{align}
Note that the geometry \eqref{metr-imb} is generically of Petrov type D, and becomes
of Petrov type N for $L=0$.

Now let us turn our attention to the form of the second Killing spinor. First of all,
the integrability conditions imply that it takes the form
\begin{displaymath}
\alpha^T = ( \beta_1, \beta_2, i X^3 e^{\xi} \beta_2, i X^3 e^{\xi} \beta_2 )\,,
\end{displaymath}
where $\beta_1$ and $\beta_2$ are arbitrary space-time dependent functions.
The Killing spinor equations \eqref{linsys} yield
\begin{displaymath}
  \beta_1 = \lambda_1 - \tfrac{1}{2} \lambda_2 b^{-2} \,, \quad
  \beta_2 = \lambda_2 b^{-2} \,,
\end{displaymath}
where $\lambda_1$ and $\lambda_2$ are integration constants. This implies that
the new Killing spinor takes the form $\epsilon = \lambda_1 \epsilon_1 + \lambda_2
\epsilon_2$, where
 \begin{align}
 \epsilon_1 = 1 + i X e_2 \,, \quad
  \epsilon_2
  = \tfrac{1}{2} X^{-2} (1 - i X e_2) + \sqrt{\tfrac{1}{4} X^{-4} + L}\,
  (e_1 - i X e_1 \wedge e_2)\,.
 \end{align}
Note that $G_0 = 0$ as well in this class.

One interesting aspect of the second Killing spinor $\epsilon_2$ is the norm of
its associated Killing vector $V_\mu=D(\ep_2,\Ga_\mu\ep_2)$.
We find $V_\mu V^\mu = - 4 X^2 L^2$, hence the second Killing spinor is indeed null
for the case $L = 0$, as was noticed before, while it is timelike for $L \neq 0$. In
the latter case, to understand whether the solution belongs also to the null class of
supersymmetric solutions, we have therefore to study the most general linear
combination of the two Killing spinors. The Killing vector $\tilde V$ constructed
from $\epsilon = \lambda_1 \epsilon_1 + \lambda_2\epsilon_2$ has norm
\begin{displaymath}
\tilde V^2 = \frac 1{X^2}\lp{\bar\la}_1\la_2-\la_1{\bar\la}_2\rp^2
-4X^2\lp L|\la_1|^2+|\la_2|^2\rp^2\,,
\end{displaymath}
which can vanish only if $L\le 0$. We have therefore three cases:
\begin{enumerate}
\item $L>0$, pure timelike class, Petrov type D.
\item $L=0$, belongs to both null and timelike classes, Petrov type N. This is the
homogeneous half BPS pp-wave in AdS. (In the terminology of \cite{Cacciatori:2004rt}
it has a wave profile ${\cal G}_{\alpha}$ with $\al=0$).
\item $L<0$, belongs to both null and timelike classes, Petrov type D.
\end{enumerate}

Actually the solutions \eqref{metr-imb} with $L>0$ can be cast into a simpler form. This is done by trading the coordinate $y$ for a new variable
$\psi=Ly-t$. For convenience, let us also introduce the Schwarzschild coordinate $r$
and rescale $z$,
\eq
r=-\frac{\ell}{\sqrt LX}\,,\qquad
\zeta=\frac12\sqrt{L}\,z\,.
\feq
In the new coordinates, the metric and the gauge field strength read
\eq
ds^2=-\lp\frac{r^2}{\ell^2}+\frac{q_e^2}{r^2}\rp dt^2
+\frac{dr^2}{\frac{r^2}{\ell^2}+\frac{q_e^2}{r^2}}
+\frac{r^2}{\ell^2}\lp d\psi^2+d\zeta^2\rp\,, \qquad \F = \frac{q_e}{r^2}dt\w dr\,,
\feq
where we have defined $q_e=2\ell/\sqrt L$. This is precisely the half BPS solution
obtained in \cite{Caldarelli:1998hg}, the massless limit of an electrically charged
toroidal black hole, which forms a naked singularity.
It is also interesting to note that the charge $q_e$ diverges in the $L\rightarrow0$ limit. This limit is naively singular in these coordinates, but it can be taken if we perform a Penrose limit \cite{Penrose:1976,Gueven:2000ru}. The existence of this limit explains why we obtained a one-parameter family of geometries (\ref{metr-imb}) connecting the massless limit of toroidal black holes and a pp-wave. Indeed, define the new coordinates $(X^+,X^-,R,Z)$ and the rescaled charge $Q_e$ by
\eq
\psi+t=2\ep^2X^+\,,\qquad\psi-t=2X^-
\,,\qquad r=\frac1{\ep R}\,,\qquad\ze=\ep Z\,,\qquad
q_e=\frac{Q_e}\ep\,.
\feq
Then, the singular limit $\ep\rightarrow0$ yields is a regular solution of the theory and corresponds to the half supersymmetric solution (\ref{metr-imb}) with $L=0$,
\eq
ds^2=\frac{\ell^2}{R^2}\lp4\,dX^+dX^--\frac{Q_e^2R^4}{\ell^6}\,dX^{-2}+dR^2+dZ^2\right)\,,\quad
{\cal F}=\frac{Q_e}{\ell^2}\,\dd X^-\w\dd R\,.
\feq
In the procedure, we have blown up the metric in the neighborhood of a geodesic with $\psi+t$ constant near the boundary $r\rightarrow\infty$ of AdS.

We now turn to the $L<0$ case, which is both timelike and lightlike. Let
us define $L=-\mu^2$. We can perform a coordinate transformation inspired from the
previous one,
\eq
\psi=Ly-t\,,\qquad
r=-\frac\ell{\mu X}\,,\qquad
\zeta=\frac\mu2 z\,,
\feq
under which the metric and the field strength become
\eq
ds^2=\lp\frac{r^2}{\ell^2}-\frac{q_e^2}{r^2}\rp dt^2
+\frac{dr^2}{\frac{r^2}{\ell^2}-\frac{q_e^2}{r^2}}
+\frac{r^2}{\ell^2}\lp -d\psi^2+d\zeta^2\rp\,, \qquad \F = \frac{q_e}{r^2}dt\w dr\,,
\feq
where we have defined $q_e=2\ell/\mu$. We see that this is the precisely the metric for
$L>0$ after the double analytic continuation
\eq
t\mapsto it\,,\qquad
\psi\mapsto i\psi\,,\qquad
q_e\mapsto -iq_e\,.
\feq
This solution represents therefore a bubble of nothing in AdS
\cite{Witten:1981gj,Birmingham:2002st,Balasubramanian:2002am,Astefanesei:2005eq}.
Note that the metric is singular for $r=\sqrt{\ell q_e}$. One should
compactify $t$, in such a way to eliminate the conical singularity on the
$(t,r)$ hypersurface. Then, if we compactify also $\zeta$, this $S^1$ will
have a minimal radius for $r=\sqrt{\ell q_e}$ (the boundary of the bubble of nothing) and
then grow with $r$. Note that for $r\rightarrow\infty$ one locally recovers AdS
spacetime, and that the $L=0$ solutions can again be understood as a Penrose limit of this metric.

\subsection{Action of the PSL$(2,\bR)$ group on the imaginary $b$ solutions} \label{im-b-PSL}

We can now generate new supersymmetric solutions by acting with the PSL$(2,\bR)$ symmetry group (\ref{psl1})-(\ref{psl2}) on the known ones. It is easy to check that the AdS$_4$ and AdS$_2\times\HH^2$ solutions are invariant under this group (although it acts non trivially on the Killing spinors). Its action on the $b=b(z)$ subfamily of the RNTN-AdS$_4$ solutions was studied in \cite{Cacciatori:2004rt}, where it was shown that it acts non trivially on the charges, by mixing them. Here we want to apply it to the imaginary $b$ solutions of the previous paragraph.

The new solution solution of the supersymmetry equations (\ref{bianchi})-(\ref{maxwell}) generated by the transformation (\ref{psl1})-(\ref{psl2}) is
\eq
\tilde b=-\frac{\ga^2Xz^2}{2\ga^2\ell Xz+i}\,,\qquad
e^{2(\tilde\Phi+\xi)}=\frac{\ga^4z^4}{4X^4}\lp1+4LX^4\rp\,,
\feq
where, without loss of generality, we eliminated $\al$ by means of a translation of $z$\footnote{After this translation the limit $\ga\rightarrow0$ is not anymore well-defined. To perform it, one has to substitute preliminarily $z$ with $z-\al/\ga$ everywhere.}, and dropped the prime of the new coordinate $z'$.
The shift function is then determined by solving equations \eqref{dzsigmanew} and \eqref{disigmanew},
\eq
\si_x=0\,,\qquad
\si_{y}=\frac{1+4LX^4}{4\ga^2X^4z^2}+\frac{\ga^2\ell^2}{X^2}\,.
\feq
Then, defining the new coordinates $(T,\si,p,q)$ through
\eq
T=\frac t{2\ell^2\ga^2}\,,\qquad
\si=\frac y2\,,\qquad
p=-\frac{\ell}X\,,\qquad
q=2\ell^2\ga^2z\,,
\feq
the metric reads
\begin{align}
ds^2=-\frac{Q(q)}{q^2+p^2}\left[
dT+\lp\frac{P(p)}{q^2}+\frac{p^2}{\ell^2}\rp d\si\right]^2
+\frac{q^2+p^2}{Q(q)}\,dq^2
+\frac{q^2+p^2}{P(p)}\,dp^2
\nonumber\\
+\frac1{\ell^4}(q^2+p^2)P(p)\,d\si^2,
\label{pi1}\end{align}
with
\eq
Q(q)=\frac{q^4}{\ell^2}\,,\qquad P(p)=\frac1{\ell^2}\lp p^4+4L\ell^2\rp\,,
\feq
and the gauge field (\ref{fieldstr}) is
\eq
{\cal F}=\dd\lp\frac{pq^2}{\ell(q^2+p^2)}\rp\w\dd T
+\dd\lp\frac{4\ell Lp}{q^2+p^2}\rp\w\dd\si\,.
\label{pi2}\feq
The form of the metric suggests some connection with the Plebanski-Demianski family of solutions, and indeed these geometries are of Petrov type D for $L\neq0$, and of Petrov type N for $L=0$, but we were not able to find the precise relation. Note also that the parameter $\ga$ has been reabsorbed in the new variables, and we are left with a one-parameter ($L$) family of solutions.

The left hand side of the necessary condition (\ref{poissonbr}) for the existence of a second Killing spinor reads, for this solution,
\eq
-\frac{9iX^4\lp1+4LX^4\rp}{\ell^2\lp1+4\ga^4\ell^2X^2z^2\rp^4}\ga^2
\feq
which clearly vanishes only for $\ga=0$, i.e. if the PSL$(2,\bR)$ transformation is trivial. Therefore, the new solutions (\ref{pi1})-(\ref{pi2}) preserves only 1/4 of the supersymmetries, and we explicitly see that the PSL$(2,\bR)$ transformations can break any additional supersymmetry. Also note that if we perform the PSL$(2,\bR)$ transformation adapting the original metric to a different Killing spinor, we could in principle end up with other supersymmetric solutions. 

Surprisingly, we find that the $L=0$ solution can be cast in the Lobatchevski wave form, even though it only has a time-like Killing spinor. This can be seen by trading the coordinates $(q,p)$ for $(x,z)$ defined by
\eq
x=\frac{\ell^3}2\lp\frac1{q^2}-\frac1{p^2}\rp\,,\qquad
z=\frac{\ell^3}{qp}\,,
\feq
in the metric (\ref{pi1}) with $L=0$, which becomes
\eq
ds^2=\frac{\ell^2}{z^2}\lp
-2\,dTd\si+\frac{z^2}{2 \ell  \sqrt{x^2+z^2}}\frac{x-\sqrt{x^2+z^2}}{x+\sqrt{x^2+z^2}}\,dT^2
+dz^2+dx^2
\rp\,.
\feq
The field strength can be easily obtained from equation (\ref{pi2}) but the result is not particularly enlightening and therefore we do not report it. This metric represents a 1/4 BPS Lobatchevski wave, whose Killing spinor falls in the timelike class. This does not contradict the results obtained in the null case, since the null Lobatchevski had a field strength (\ref{gaugepot}) of the form ${\cal F}=\phi'(T)\dd T\w\dd z$, while this solution has a much more complicated gauge field. It is however interesting to note that the solutions of the null case do not exhaust all possible supersymmetric Lobatchevski waves.

\subsection{Gravitational Chern-Simons system and $G_0 = \psi_-=0$ solutions}\label{gcs}

A number of the previously studied subcases can be combined into the interesting Ansatz
\eq
b=-\frac1\ell\frac{\al z^2+\be z+\ga}{2\al z+\be-i\eta(w,\bar w)},
\label{bcs}\feq
where $\al$, $\be$ and $\ga$ are three real constants. For $\al = \be= 0$ this reduces to $b$ imaginary, while $\eta = 0$ leads to the real subcase of $b= b(z)$. With this assumption, the equations for a timelike Killing spinor reduce to
\eq
\Delta\xi+\frac12e^{2\xi}\lp k-3\eta\rp = 0 \,,\qquad
\Delta\eta+e^{2\xi}\lp k\eta-\eta^3\rp=0,
\label{eqgcs}\feq
where we have defined $k=4\al\ga-\be^2$ and $\Delta = 4 \partial \bar \partial$.
Interestingly, as shown in \cite{Cacciatori:2004rt}, this system of equations follows from the dimensionally reduced Chern-Simons action \cite{Deser:1981wh,Guralnik:2003we},
\eq
S=\int d^2x\sqrt{^{(2)}g}\lp{}^{(2)}R\eta+\eta^3\rp\,, \label{CS}
\feq
if we use the conformal gauge $^{(2)}g_{ij}dx^idx^j=e^{2\xi}\lp dx^2+dy^2\rp$ and $\eta$ is the curl of a vector potential, $\sqrt{^{(2)}g} \, \ep_{ij}\eta=\p_iA_j-\p_jA_i$. To obtain equations (\ref{eqgcs}) we vary the action with respect to $A_i$ and $\xi$. When varying the dimensionally reduced Chern-Simons action with respect to $g_{ij}$ there is however an additional equation to \eqref{eqgcs}.

Using the results of Grumiller and Kummer \cite{Grumiller:2003ad}, one obtains the most general solution to the dimensionally reduced Chern-Simons system \cite{Cacciatori:2004rt}
\eq
e^{2\xi}=\frac{L}{\ell^4} -\frac{k}2\eta^{2}+\frac14\eta^{4}, \label{solution}
\feq
where $L$ is an integration constant and $\dd\eta=e^{2\xi}\dd x$. Trading the coordinate $x$ for $\eta$, we get the following  configuration of the fields
\eqn
\dd s^{2}&=&-\frac4{\ell^{2}}\frac{P_{2}^{2}}{P_{2}'^{2}+\eta^{2}}\left[\dd t+\sigma\right]^{2}+\frac{\ell^{2}}4\frac{P_{2}'^{2}+\eta^{2}}{P_{2}^{2}}\left[\dd z^{2}+P_{2}^{2}\left(e^{-2\xi}\dd\eta^{2}+e^{2\xi}\dd y^{2}\right)\right]\,,\nonumber\\
{\cal A}&=&\frac2\ell\frac{P_{2}\eta}{P_{2}'^{2}+\eta^{2}}\left[\dd t+\sigma\right]+\frac\ell4{\cal V}\dd y-i\frac\ell4\dd\log\frac{b}{\bar b}\,,
\feqn
where $P_{2}(z)=\al z^{2}+\be z+\ga$, $k$ is defined as above and the shift function reads
\eq
\sigma=\frac{\ell^{2}}{2}\left(\al\eta^{2}+\frac{e^{2\xi}}{P_{2}}\right)\dd y\,.
\feq
These solutions preserve 1/4 of the original supersymmetry. In fact, the $k=0$ solutions coincide with the imaginary $b$ ones and their PSL$(2,\bR)$ transforms of sections \ref{im-b} and \ref{im-b-PSL}. For $k$ non-vanishing these are different solutions.

As can be seen from the Poisson bracket \eqref{poissonbr}, the only possibility to have 1/2 supersymmetry is $\al =0$ and hence $k \leq 0$. In fact, starting from any solution with $k \leq 0$, one can always obtain $\al=0$ by an appropriate PSL$(2,\bR)$ transformation. The non-trivial part of the PSL$(2,\bR)$ symmetry is $z \mapsto - 1 / (z + \delta)$, whose action on the parameters $\al$, $\be$ and $\ga$ of the Ansatz \eqref{bcs} is given by
 \eq
  \al \mapsto \al \delta^2 - \be \delta + \ga \,, \quad
  \be \mapsto 2 \al \delta - \be \,, \quad
  \ga \mapsto \al \,,
 \feq
which keeps $k$ fixed. Indeed, for $k \leq 0$, there is always a PSL$(2,\bR)$ transformation that sets $\al = 0$, while this is impossible for $k > 0$.

The requirement $\al = 0$ leads to the half-supersymmetric imaginary $b$ solution of section \ref{im-b} for $k=0$.
In the case of $k$ negative, when $\al = 0$ one can  scale $\be$ to $1$ in (\ref{bcs}) without loss of generality, and $\ga$ can be put to zero by a translation in $z$. Hence the function $b$ is given by
\eq
b=-\frac1\ell\frac{z}{1-i \eta }\,.
\feq
The metric is given in \eqref{metric} and is generically of Petrov type $D$. The second Killing spinor can be found in \eqref{Ks}. As shown in appendix~\ref{app-G0=0}, the $G_0=\psi_-=0$ solutions are either the imaginary $b$ ones, anti-Nariai spacetime or the above 1/2 supersymmetric solution with $k=-1$.
 
We would like to mention that \eqref{solution} is the most general solution to the dimensionally reduced Chern-Simons system, but not to the equations \eqref{eqgcs}. The reason for this is the additional constraint one obtains when varying \eqref{CS} with respect to $g_{ij}$.  An example of this is provided by the Petrov type I solution with $b= i (x / \ell)^{2}$ in section \ref{im-b} and its PSL$(2,\bR)$ transform given in eq.~(2.44) of \cite{Cacciatori:2004rt}. 

\section{Final remarks}
\label{finalrem}

In this paper, we applied spinorial geometry techniques to classify all
supersymmetric solutions of minimal ${\cal N}=2$ gauged supergravity in four dimensions.

In the presence of null Killing spinors, the problem can be completely solved,
and all 1/4- and 1/2-supersymmetric solutions have been written down
explicitly. We showed that there are no 1/4-BPS backgrounds with U(1)$\ltimes
\bR^2$-invariant Killing spinors and those with $\bR^2$-invariant Killing
spinors have been derived in sections 3.1 and 3.2. The backgrounds in the
latter section were previously unknown and are Petrov type II configurations describing gravitational waves propagating on a bubble of nothing in AdS$_4$. In addition, it turned out that there are no 1/2-BPS
backgrounds with $\bR^2$-invariant Killing
spinors and hence any additional Killing spinor is timelike. In section 3.3
we gave the backgrounds with one null and one timelike Killing spinor.

For a timelike Killing spinor we derived the
conditions for the corresponding backgrounds in section 4.1 and 4.2. We worked out the first
integrability conditions necessary for the existence of a second Killing
spinor in section 4.3. We explicitly solved these equations in a number of
subcases in section 5, and thereby found several new solutions, like
the bubbles of nothing in AdS$_4$, already obtained in the null formalism, and their PSL$(2,\bR)$-transformed configurations.
Furthermore, our results showed that the generalized holonomy in the case of one
preserved complex supercharge is contained in A$(3,\bC)$, supporting thus the
classification scheme of \cite{Batrachenko:2004su}.

In addition, the time-dependence of a second time-like Killing spinor was
shown to be an overall exponential factor with coefficient $G_0$ in section 4.4. In the case $G_0 = 0$ these equations have been solved in full
generality, up to a second order ordinary differential equation. 
We expect this class to comprise a large number of interesting
1/2-BPS solutions. Indeed, all the examples of section 5 either have vanishing
$G_0$ or can be transformed to that case by a coordinate transformation.

There are several interesting points that remain to be understood. First of all, it would
be desirable to get a deeper insight into the underlying geometric structure in
the case of U(1) invariant spinors. In five dimensions, spacetime is a fibration over
a four-dimensional Hyperk\"ahler or K\"ahler base for ungauged and gauged supergravity
respectively \cite{Gauntlett:2002nw,Gauntlett:2003fk}, whereas in four-dimensional
ungauged supergravity one has a fibration over a three-dimensional flat
space \cite{Tod:1983pm}. This suggests that the base for $D=4$ gauged supergravity
might be an odd-dimensional analogue of a K\"ahler manifold, i.~e.~, a Sasaki manifold.
From the equations \eqref{bianchi} and \eqref{maxwell} this is not obvious.

Secondly, in \cite{Cacciatori:2004rt}, a surprising relationship between the equations
\eqref{bianchi}, \eqref{maxwell} governing 1/4 BPS solutions and the gravitational
Chern-Simons theory \cite{Deser:1981wh} was found. Why such a relationship should exist
is not clear at all, and deserves further investigations.

The third point concerns preons, which were conjectured in \cite{Bandos:2001pu} to
be elementary constituents of other BPS states. In type II and eleven-dimensional
supergravity, it was shown that imposing 31 supersymmetries implies that the
solution is locally maximally supersymmetric \cite{Gran:2006ec, Bandos:2006xz,
  Gran:2006cn}. Similar
results in four- and five-dimensional gauged supergravity were obtained
in \cite{Grover:2006wy,Grover:2006ps}. This implies that preonic backgrounds are
necessarily quotients of maximally supersymmetric solutions. While M-theory
preons cannot arise by quotients \cite{Figueroa-O'Farrill:2007ic}, it remains to
be seen if 3/4 supersymmetric solutions to ${\cal N}=2$, $D=4$ or $D=5$ gauged supergravities
really do not exist. The only maximally supersymmetric backgrounds in these theories
are AdS$_4$ \cite{Caldarelli:2003pb} and AdS$_5$ \cite{Gauntlett:2003fk} respectively,
so the putative preonic configurations must be quotients of AdS.

Finally, it would be interesting to apply spinorial geometry techniques to
classify all supersymmetric solutions of four-dimensional ${\cal N}=2$ matter-coupled gauged
supergravity. Work in this direction is in progress \cite{cckmmor}.

\acknowledgments

We are grateful to Alessio Celi, Marcello Ortaggio and Christoph Sieg for useful discussions. This work was partially supported by INFN, MURST and
by the European Commission program MRTN-CT-2004-005104. D.R.~wishes to thank
the Universit\`a di Milano for hospitality. Part of this work was completed
while he was a post-doc at King's College
London, for which he would like to acknowledge the PPARC grant
PPA/G/O/2002/00475. In addition, he is presently supported by the European
EC-RTN project MRTN-CT-2004-005104, MCYT FPA 2004-04582-C02-01 and CIRIT GC
2005SGR-00564.

\normalsize

\appendix

\section{Spinors and forms}

In this appendix, we summarize the essential information needed to realize
the spinors of Spin(3,1) in terms of forms. For more details, we refer
to \cite{Lawson:1998yr}.
Let $V = \bR^{3,1}$ be a real vector space equipped with the Lorentzian inner
product $\langle\cdot,\cdot\rangle$. Introduce an orthonormal basis $e_1, e_2, e_3, e_0$,
where $e_0$ is along the time direction, and consider the subspace $U$
spanned by the first two basis vectors $e_1, e_2$. The space of Dirac spinors
is $\Delta_c = \Lambda^{\ast}(U\otimes \bC)$, with basis
$1, e_1, e_2, e_{12} = e_1 \wedge e_2$.
The gamma matrices are represented on $\Delta_c$ as
\eqn
\Gamma_{0}\eta&=&-e_2\wedge\eta+e_2\rfloor\eta\,, \qquad
\Gamma_{1}\eta=e_1\wedge\eta+e_1\rfloor\eta\,, \nonumber \\
\Gamma_{2}\eta&=&e_2\wedge\eta+e_2\rfloor\eta\,, \qquad
\Gamma_{3}\eta=ie_1\wedge\eta-ie_1\rfloor\eta\,,
\feqn
where
\begin{displaymath}
\eta = \frac 1{k!}\eta_{j_1\ldots j_k} e_{j_1}\wedge\ldots\wedge e_{j_k}
\end{displaymath}
is a $k$-form and
\begin{displaymath}
e_i \wedge \eta = \frac 1{(k-1)!}\eta_{ij_1\ldots j_{k-1}} e_{j_1}\wedge\ldots
                  \wedge e_{j_{k-1}}\,.
\end{displaymath}
One easily checks that this representation of the gamma matrices satisfies
the Clifford algebra relations $\{\Gamma_a, \Gamma_b\} = 2\eta_{ab}$.
The parity matrix is defined by $\Gamma_5 = i\Gamma_0\Gamma_1\Gamma_2\Gamma_3$,
and one finds that the even forms $1, e_{12}$ have positive chirality,
$\Gamma_5\eta = \eta$, while the odd forms $e_1, e_2$ have negative chirality,
$\Gamma_5\eta = -\eta$, so that $\Delta_c$ decomposes into two complex chiral
Weyl representations $\Delta_c^+ = \Lambda^{\mathrm{even}}(U\otimes \bC)$ and
$\Delta_c^- = \Lambda^{\mathrm{odd}}(U\otimes \bC)$.\\
Let us define the auxiliary inner product
\begin{equation}
\langle\sum_{i=1}^2 \alpha_i e_i, \sum_{j=1}^2 \beta_j e_j\rangle = \sum_{i=1}^2
\alpha_i^{\ast}\beta_i
\end{equation}
on $U\otimes \bC$, and then extend it to $\Delta_c$. The Spin(3,1) invariant
Dirac inner product is then given by
\begin{equation}
D(\eta, \theta) = \langle\Gamma_0\eta, \theta\rangle\,.
\end{equation}
In many applications it is convenient to use a basis in which the
gamma matrices act like creation and annihilation operators, given
by \eqn
\Gamma_{+}\eta\equiv\frac1{\sqrt2}\left(\Gamma_{2}+\Gamma_{0}\right)\eta
&=&\sqrt2\,e_{2}\rfloor\eta\,, \qquad
\Gamma_{-}\eta\equiv\frac1{\sqrt2}\left(\Gamma_{2}-\Gamma_{0}\right)\eta
=\sqrt2\,e_{2}\wedge\eta\,, \nonumber \\
\Gamma_{\bullet}\eta\equiv\frac1{\sqrt2}\left(\Gamma_{1}-i\Gamma_{3}\right)\eta
&=&\sqrt2\,e_{1}\wedge\eta\,, \qquad
\Gamma_{\bar\bullet}\eta\equiv\frac1{\sqrt2}\left(\Gamma_{1}+i\Gamma_{3}\right)\eta
=\sqrt2\,e_{1}\rfloor\eta\,.
\feqn
The Clifford algebra relations in this  basis are $\{\Gamma_A,\Gamma_B\} = 2\eta_{AB}$,
where $A,B,\ldots = +,-,\bullet,\bar\bullet$ and the nonvanishing components of
the tangent space metric read
$\eta_{+-} = \eta_{-+} = \eta_{\bullet\bar\bullet} = \eta_{\bar\bullet\bullet} = 1$.
The spinor 1 is a Clifford vacuum, $\Gamma_{+}1 = \Gamma_{\bar\bullet}1 = 0$,
and the representation $\Delta_c$ can be constructed by acting on 1 with the
creation operators $\Gamma^+ = \Gamma_-, \Gamma^{\bar\bullet} = \Gamma_{\bullet}$,
so that any spinor can be written as
\begin{displaymath}
\eta = \sum_{k=0}^2 \frac 1{k!}\phi_{{\bar a}_1\ldots {\bar a}_k}\Gamma^{{\bar a}_1
\ldots {\bar a}_k}1\,, \qquad \bar a = +,\bar\bullet\,.
\end{displaymath}
The action of the Gamma matrices and the Lorentz generators $\Gamma_{AB}$ is
summarized in the  table \ref{tab:gamma}.

\begin{table}[ht]
\begin{center}
\begin{tabular}{|c||c|c|c|c|}
\hline
& 1 & $e_{1}$ & $e_{2}$ & $e_{1}\wedge e_{2}$\\
\hline\hline
$\Gamma_{+}$ & 0 & 0 & $\sqrt2$ & $-\sqrt2e_{1}$\\
\hline
$\Gamma_{-}$ & $\sqrt2e_{2}$ & $-\sqrt2e_{1}\wedge e_{2}$ & 0 & 0\\
\hline
$\Gamma_{\bullet}$ & $\sqrt2e_{1}$ & $0$ & $\sqrt2e_{1}\wedge e_{2}$ & 0\\
\hline
$\Gamma_{\bar\bullet}$ & 0 & $\sqrt2$ & 0 & $\sqrt2e_{2}$\\
\hline\hline
$\Gamma_{+-}$ & 1 & $e_{1}$ & $-e_{2}$ & $-e_{1}\wedge e_{2}$\\
\hline
$\Gamma_{\bar\bullet\bullet}$ & 1 & $-e_{1}$ & $e_{2}$ & $-e_{1}\wedge e_{2}$\\
\hline
$\Gamma_{+\bullet}$ & 0 & 0 & $-2e_{1}$ & 0\\
\hline
$\Gamma_{+\bar\bullet}$ & 0 & 0 & 0 & 2\\
\hline
$\Gamma_{-\bullet}$ & $-2e_{1}\wedge e_{2}$ & 0 & 0 & 0\\
\hline
$\Gamma_{-\bar\bullet}$ & 0 & $2e_{2}$ & 0 & 0\\
\hline
\end{tabular}
\end{center}
\caption{The action of the Gamma matrices and the Lorentz generators
  $\Gamma_{AB}$ on the different basis elements. \label{tab:gamma}}
\end{table}

Note that $\Gamma_A = {U_A}^a\Gamma_a$, with
\begin{displaymath}
\left({U_A}^a\right) = \frac1{\sqrt2} \left(\begin{array}{cccc} 1 & 0 & 1 & 0 \\
-1 & 0 & 1 & 0 \\ 0 & 1 & 0 & -i \\ 0 & 1 & 0 & i
\end{array}\right) \in {\mathrm U}(4)\,,
\end{displaymath}
so that the new tetrad is given by $E^A = {(U^{\ast})^A}_a E^a$.

\section{Spinor bilinears}

Given a Killing spinor
\eq
\epsilon=c_0 1+c_1e_1+c_2e_2+c_{12}e_1\w e_2\,,
\feq
one can construct the bilinears
\begin{eqnarray}
\tilde f & = & -iD(\epsilon,\epsilon)=-i\lp c_0c_2^*-c_1c_3^*-c_2c_0^*+c_{12}c_1^*\rp\,, \\
\tilde g & = &-iD(\epsilon,\Gamma_5\epsilon)=c_0c_2^*+c_1c_3^*+c_2c_0^*+c_{12}c_1^*\,, \\
\tilde V & = &D(\ep,\Ga_{\mu}\ep)\,\dd x^{\mu}=
\frac12\left[\frac1{|b|^2}\lp |c_2|^2+|c_{12}|^2\rp
-|c_0|^2-|c_1|^2\right]\dd z\nonumber\\
&& -2\left[ |c_2|^2+|c_{12}|^2
+|b|^2\lp|c_0|^2+|c_1|^2\rp\right]\left(\dd t+\sigma\right)\nonumber\\
&&+\frac 1{|b|}e^\psi
\left[\lp c_2c_1^*-c_0c_{12}^*\rp\dd w
+\lp c_1c_2^*-c_{12}c_0^*\rp\dd\bar w\right]\,, \\
\tilde B & = &D(\ep,\Ga_5\Ga_{\mu}\ep)\,\dd x^{\mu}=
\frac12\left[\frac1{|b|^2}\lp |c_2|^2-|c_{12}|^2\rp
+ |c_0|^2-|c_1|^2\right] \dd z\nonumber\\
&&-2\left[ |c_2|^2-|c_{12}|^2
-|b|^2\lp |c_0|^2-|c_1|^2\rp\right] (\dd t+\sigma)\nonumber\\
&&+\frac 1{|b|}e^\psi\left[\lp c_2c_1^*+c_0c_{12}^*\rp \dd w
+\lp c_1c_2^*+c_{12}c_0^*\rp \dd\bar w\right]\,,
\end{eqnarray}
\begin{eqnarray}
\tilde\Phi=\frac 12 D(\ep,\Ga_{\mu\nu}\ep)\,\dd x^{\mu}\w\dd x^{\nu}=
-\left(c_0c_2^*-c_1c_{12}^*+c_2c_0^*-c_{12}c_1^*\rp \dd t\w\dd z
\nonumber\\
-\frac{2e^\psi}{|b|}\left(c_2c_{12}^*
+|b|^2c_0c_1^*\right)\dd t\w\dd w
-\frac{2e^\psi}{|b|}\left(c_{12}c_0^*
+4|b|^2c_1c_0^*\rp\dd t\w\dd\bar w
\nonumber\\
+\left[\left(c_0c_2^*-c_1c_{12}^*+c_2c_0^*-c_{12}c_1^*\rp\sigma_w
+\frac{e^\psi}{2|b|^3}c_2c_{12}^*
-\frac{e^\psi}{2|b|}c_0c_1^*\right]\dd z\w\dd w
\nonumber \\
+\left[\left(c_0c_2^*-c_1c_{12}^*+c_2c_0^*-c_{12}c_1^*\rp\sigma_{\bar w}
+\frac{e^\psi}{2|b|^3}c_{12}c_0^*
-\frac{e^\psi}{2|b|}c_1c_0^*\right]\dd z\w\dd\bar w
\nonumber
\end{eqnarray}
\begin{eqnarray}
+\frac{2e^\psi}{|b|}\left[c_2c_{12}^*\sigma_{\bar w}
-c_{12}c_0^*\sigma_w
+|b|^2\lp c_0c_1^*\sigma_{\bar w}
-c_1c_0^*\sigma_w\rp\right.\qquad\qquad\qquad\nonumber\\
\left.+\frac{e^{\psi}}{4|b|}\lp c_0c_2^*+c_1c_{12}^*-c_2c_0^*-c_{12}c_1^*\rp
\right]\dd w\w\dd\bar w\,.
\end{eqnarray}
Given the first Killing spinor of the form $\ep_1=1+be_2$ and the second Killing
spinor $\ep_2=c_0 1+c_1e_1+c_2e_2+c_{12}e_1\w e_2$, one can also construct mixed
bilinears of the type $D(\ep_1,\Ga_{\cdots}\ep_2)$, which verify the same
differential equations as the bilinears built from the original two Killing spinors:
\begin{align}
& \hat f=-i(\bar bc_0-c_2)\,,\qquad
\hat g=\bar bc_0+c_2\,, \\
& \hat V=\frac1{2b}\lp c_2+bc_0\rp\lp\dd t+\si\rp+\frac1{2b}\lp c_2-bc_0\rp\dd z
+\frac 1{|b|}e^\psi\lp\bar bc_1-c_{12}\rp\dd\bar w\,, \\
& \hat B=\frac1{2b}\lp c_2-bc_0\rp\lp\dd t+\si\rp+\frac1{2b}\lp c_2+bc_0\rp\dd z
+\frac 1{|b|}e^\psi\lp\bar bc_1+c_{12}\rp\dd\bar w\,.
\end{align}

\section{The case $P'=0$}
\label{app-P'=0}

In section \ref{halfBPS}, we simplified the equations for the second Killing
spinor under the assumption $P'\neq 0$, where $P=e^{-2\psi}b\partial b$.
Here we consider the case $P'=0$. To this end, we need the following subset of
the Killing spinor equations \eqref{linsys}:
\begin{eqnarray}
\partial_+\psi_2 - \sqrt 2 r^2\left(\frac{{\bar b}'}{\bar b} + \frac 1{\ell\bar b}
\right)\psi_2 - \sqrt 2 r^2\left(\frac{b'}b + \frac 1{\ell b}\right)\psi_{12} &=& 0\,,
\label{+2} \\
\partial_+\psi_{12} - r e^{-\psi}\partial_{\bar\bullet}\ln\bar b\,\psi_1 - \sqrt 2 r^2
\left(2\frac{r'}r + \frac 1{\ell\bar b}\right)\psi_{12} &=& 0\,, \label{+3} \\
\partial_-\psi_2 + \frac 1r e^{-\psi}\partial_{\bar\bullet}\ln b\,\psi_1 - \sqrt 2
\left(2\frac{r'}r + \frac 1{\ell b}\right)\psi_2 &=& 0\,, \label{-2} \\
\partial_-\psi_{12} - \sqrt 2\left(\frac{{\bar b}'}{\bar b} + \frac 1{\ell\bar b}
\right)\psi_2 - \sqrt 2\left(\frac{b'}b + \frac 1{\ell b}\right)\psi_{12} &=& 0\,,
\label{-3}
\end{eqnarray}
\begin{eqnarray}
r e^{-\psi}\partial_{\bullet}\left(\frac 1{r^2}e^{2\psi}\psi_2\right) - \sqrt 2
\left(\frac{b'}b + \frac 1{\ell b}\right)\psi_1 &=& 0\,, \label{bullet2} \\
r e^{-\psi}\partial_{\bullet}\left(\frac 1{r^2}e^{2\psi}\psi_{12}\right) + \sqrt 2
\left(\frac{{\bar b}'}{\bar b} + \frac 1{\ell\bar b}\right)\psi_1 &=& 0\,.
\label{bullet3}
\end{eqnarray}
If $P'=0$, \eqref{intcondP1-} implies $\psi_-=0$ or $\partial P=0$. Let us first
assume the former, i.~e.~, $\psi_2=\psi_{12}$. From \eqref{bullet3} -- \eqref{bullet2}
one obtains then $\psi_1=0$ or
\begin{equation}
\frac{b'}b + \frac{{\bar b}'}{\bar b} + \frac 1{\ell b} + \frac 1{\ell\bar b} = 0\,.
\label{b'/b}
\end{equation}
\begin{itemize}
\item If $\psi_1=0$, \eqref{-3} -- \eqref{-2} yields $\psi_2=0$, and thus there
exists no further Killing spinor.
\item If \eqref{b'/b} holds, one can use \eqref{+2} and \eqref{-3} to show that
$\partial_+\psi_2=\partial_-\psi_2=0$, or equivalently $\partial_t\psi_2=\psi_2'=0$.
Using this in \eqref{+3} and \eqref{-2} and deriving with respect to $t$, one gets
${\bar\partial}\bar b\,\partial_t\psi_1={\bar\partial}b\,\partial_t\psi_1=0$. When
$\partial_t\psi_1\neq 0$, this means that ${\bar\partial}b=\partial b=0$, so $b=b(z)$,
which is a case analyzed in section \ref{b(z)}. If instead $\partial_t\psi_1=0$, all
the $\psi_i$ are independent of $t$, and the Killing spinor equations reduce to
the system \eqref{KSE1.1} to \eqref{KSE4.3} with $G_0=0$.
\end{itemize}
In the case $\partial P = 0$, consider the integrability condition
\begin{equation}
\psi_1 Q' + \psi_-\partial Q = 0\,, \label{intcondQ1-}
\end{equation}
where $Q=e^{-2\psi}\bar b\partial\bar b$, following from the first line of
${\tilde N}_{wt}$. As long as $Q'\neq 0$, with the same reasoning as in section
\ref{halfBPS}, one obtains the system \eqref{KSE1.1} to \eqref{KSE4.3}. If
$Q'=0$, \eqref{intcondQ1-} implies $\psi_-=0$ or $\partial Q = 0$. The case
$\psi_-=0$ was already considered above, so the only remaining case is
$P'=\partial P=Q'=\partial Q=0$. For $P=Q=0$ we get again $b=b(z)$, so without
loss of generality we can assume $P\neq 0$ or $Q\neq 0$. Suppose that $Q=0$,
$P\neq 0$, so $b=b(w,z)$. Take the logarithm of $e^{-2\psi}b\partial b = P(\bar w)$,
derive with respect to $z$, use \eqref{Phi}, and apply $\bar\partial$. This leads
to $\partial b=0$, which is a contradiction to the assumption $P\neq 0$. In the
same way one shows that $P=0$, $Q\neq 0$ is not possible, so that both $P$ and $Q$
must be nonvanishing. Now use the third row of ${\tilde N}_{\bar w t}$, which leads
to $\bar Q\psi_2 =0$ and hence $\psi_2=0$. Finally, the last row of
${\tilde N}_{\bar w t}$ yields $\psi_-=0$, i.~e.~, the case already considered above.

Hence, the conclusion is that in the case $P' = 0$, the second Killing spinor either has $G_0$ time-dependence of the form \eqref{time-dep}, or leads to solutions with $b=b(z)$. The latter are treated separately in section 5.1. As can be found there, all 1/2-BPS solutions with $b=b(z)$ also have second Killing spinors with $G_0$ time-dependence of the form \eqref{time-dep}. Hence this time-dependence is a completely general result\footnote{The only counterexample is the third Killing spinor of AdS$_4$, see \eqref{AdS4-Ks}, but since this is maximally supersymmetric it does not contradict the result.} for second Killing spinors in the time-like case.

\section{Half-supersymmetric solutions with $G_0=0$}
\label{app-G0=0}

From the difference of equations (\ref{KSE1.2})$-$(\ref{KSE2.3}) and
(\ref{KSE4.2})$-$(\ref{KSE4.3}) one gets $\psi_-=\psi_-(w)$. Furthermore,
[(\ref{KSE2.2})$-$(\ref{KSE1.3})$+e^{-2\psi}$(\ref{KSE4.1})] and (\ref{KSE3.1}) yield
$\psi_1=\psi_1(z)$. Assuming $\psi_-\neq 0$, eqns.~(\ref{KSE1.1}) and (\ref{KSE2.1})
can be written in the form
\begin{equation}
\left(\frac{\beta}b\right)'+\partial\left(\frac1b\right)=0\,, \qquad
\left(\frac{\beta}{\bar b}\right)'+\partial\left(\frac1{\bar b}\right)=0\,, \label{beta}
\end{equation}
where $\beta=\psi_1/\psi_-$. Deriving \eqref{beta} with respect to $\bar w$ gives
\begin{displaymath}
\left[\bar\partial\left(\frac1b\right)\beta\right]'+\partial\bar\partial\left(
\frac1b\right)=0\,, \qquad
\left[\bar\partial\left(\frac1{\bar b}\right)\beta\right]'+\partial\bar\partial\left(
\frac1{\bar b}\right)=0\,.
\end{displaymath}
Now use \eqref{beta} in the difference between the first equation and the complex
conjugate of the second one to get
\begin{displaymath}
\left[\frac1b\left(\bar\beta\beta'-\bar\beta'\beta\right)\right]'=0\,.
\end{displaymath}
Observe that $\bar\beta\beta'-\bar\beta'\beta=|\psi_-|^{-2}\left(\bar\psi_1\psi_1'-
\bar\psi_1'\psi_1\right)(z)$, so that for $\bar\psi_1\psi_1'-\bar\psi_1'\psi_1\neq0$
there must exist a real function $B(z)$ and a generic function $h(w,\bar w)$ such that
\begin{displaymath}
b=B(z)h(w,\bar w)\,.
\end{displaymath}
Plugging this into \eqref{beta}, we conclude that
\begin{displaymath}
\partial\ln\left(\frac h{\bar h}\right)=0\,,
\end{displaymath}
so that the phase of the function $h$ is fixed, $h=h_R(w,\bar w)e^{i\varphi_0}$,
with $h_R$ real. Using \eqref{dzsigmanew}, the constancy of the phase of $b$ implies
that the shift vector $\sigma$ does not depend on $z$. \eqref{disigmanew} gives then
\begin{displaymath}
\partial_z\left(\frac{e^{2\psi}}{B^3}\right)=0\,,
\end{displaymath}
or, using \eqref{Phi},
\begin{displaymath}
\frac23\frac{\cos\varphi_0}{\ell h_R}+B'=0\,,
\end{displaymath}
and thus
\begin{displaymath}
B'=c\,,\quad\frac23\frac{\cos\varphi_0}{\ell h_R}=-c\,,
\end{displaymath}
where $c$ denotes a real constant. Now we have to distinguish to cases:
\begin{enumerate}
\item{$c\neq0$}:
In this case $b(z)=\left(B_0-\frac{2\cos\varphi_0}{3\ell}z\right)e^{i\varphi_0}$.
Plugging this into the first of eqns.~\eqref{beta} one gets
\begin{displaymath}
\left(\frac{\psi_1}b\right)'=0\,,
\end{displaymath}
which is solved by $\psi_1=\eta b$ where $\eta$ is a constant. But this yields
$\bar\psi_1\psi_1'-\bar\psi_1'\psi_1=0$, which contradicts our assumption.
\item{$c=0$}:
In this case $b(w,\bar w)=ih_R(w,\bar w)$. The combination
(\ref{KSE1.3})+(\ref{KSE2.2})$-$(\ref{KSE1.2})$-$\\
(\ref{KSE2.3}) leads to $\psi_-=0$,
which again contradicts one of our assumptions.
\end{enumerate}
We thus conclude that $\bar\psi_1\psi_1'-\bar\psi_1'\psi_1=0$, and hence
$\psi_1=\zeta(z)e^{i\theta_0}$ where $\theta_0$ is a constant and $\zeta(z)$ is a real
function. Sending $\psi_i\rightarrow e^{-i\theta_0}\psi_i$ we can take $\psi_1$ real
and non-negative without loss of generality. Let us now consider the case where
both $\psi_1$ and $\psi_-$ are non-vanishing. This allows to introduce new coordinates
$Z,W,\bar W$ such that
\begin{displaymath}
\dd Z=\frac1{\psi_1(z)}\dd z\,, \qquad \dd W=\frac{\dd w}{\psi_-(w)}\,, \qquad
\dd\bar W=\frac{\dd\bar w}{\bar\psi_{-}(\bar w)}\,.
\end{displaymath}
Note that one can set $\psi_-=1$ using the residual gauge invariance $w\mapsto W(w)$,
$\psi\mapsto\tilde\psi = \psi - \frac 12\ln(\dd W/\dd w) - \frac 12\ln(\dd\bar W/
\dd\bar w)$ leaving invariant the metric $e^{2\psi}\dd w\dd\bar w$. We can thus
take $W=w$ in the following. Equations (\ref{KSE1.1}) and (\ref{KSE2.1}) are then
equivalent to
\begin{displaymath}
\left(\partial_Z+\partial\right)\varphi = 0\,, \qquad
\partial_Z\ln\psi_1-\left(\partial_Z+\partial\right)\ln r = 0\,.
\end{displaymath}
From the real part of the first equation we have
\begin{displaymath}
\varphi=\varphi(Z-w-\bar w)\,.
\end{displaymath}
Using $\psi_1=\psi_1(Z)$, the second equation implies
\begin{displaymath}
\left(\partial_Z+\partial\right)\frac r{\psi_1}=0\,,
\end{displaymath}
and hence
\begin{displaymath}
\frac r{\psi_1}=\rho(Z-w-\bar w)\,.
\end{displaymath}
The function $b$ must thus have the form
\begin{displaymath}
b(Z,w+\bar w)=\psi_1(Z)B(Z-w-\bar w)\,,
\end{displaymath}
where $B(Z-w-\bar w)=\rho(Z-w-\bar w)e^{i\varphi(Z-w-\bar w)}$.
The difference between (\ref{KSE3.2}) and (\ref{KSE3.3}) yields
\begin{displaymath}
\left(\partial_Z+\partial\right)\left(\ln\psi_1-\psi\right)=0\,,
\end{displaymath}
so that $\ln\psi_1-\psi=-H(Z-w-\bar w)$ with $H$ real. This gives
\begin{displaymath}
e^{2\psi}=\psi_1(Z)^2 e^{2H}
\end{displaymath}
for the conformal factor. In terms of the new coordinate $Z$, \eqref{Phi} reads
\begin{displaymath}
\partial_Z\psi+\frac1{2\ell}\left(\frac1B+\frac1{\bar B}\right)=0\,.
\end{displaymath}
Using the definition of $H$ we get
\eq
\dot H + \partial_Z\ln\psi_1+\frac1{2\ell}\left(\frac1B+\frac1{\bar B}\right)=0\,,
\label{ReB}
\feq
where a dot denotes a derivative with respect to $Z-w-\bar w$.
We can thus conclude that $\partial_Z\ln\psi_1=\gamma/\ell$ for some constant $\gamma$,
i.~e.~,$\psi_1(Z)=\psi_1^{(0)}e^{\gamma Z/\ell}$. By shifting $Z$ one can set
$\psi_1^{(0)}=1$.
Calling $\chi=\psi_+/\psi_-$, the only remaining nontrivial Killing spinor equations read
\eqn
\partial_Z\chi-2\left(\frac{\dot \rho}\rho-\dot H\right)\chi+2i\dot{\varphi}+\frac1\ell
\left(\frac1B-\frac1{\bar B}\right)&=&0\,, \nonumber \\
\partial_Z\chi-\left(2\frac{\dot \rho}\rho-\dot H+\frac{\gamma}{\ell}\right)\chi-
2i e^{-2H}\dot{\varphi}-\frac1{2\ell}\left(\frac1B-\frac1{\bar B}\right)&=&0\,,
\nonumber \\
\partial\chi+2\left(\frac{\dot \rho}\rho-\dot H\right)\chi-2i\dot{\varphi}-\frac1\ell
\left(\frac1B-\frac1{\bar B}\right)&=&0\,, \nonumber \\
\bar\partial\chi+2\frac{\dot \rho}\rho\chi-2i\dot{\varphi}&=&0\,, \nonumber \\
\frac1{2\ell}\left(\frac1B-\frac1{\bar B}\right)\chi+2\left(1+e^{-2H}\right)
\frac{\dot \rho}\rho-\dot H+\frac{\gamma}{\ell}&=&0\,. \nonumber
\feqn
Summing the first and the third equation yields $\chi=\chi(Z-w-\bar w)$, so that
we are left with
\eqn
\dot{\chi}-2\left(\frac{\dot \rho}\rho-\dot H\right)\chi+2i\dot{\varphi}+\frac1\ell
\left(\frac1B-\frac1{\bar B}\right)&=&0\,, \\ \label{KSE-rem-1}
\dot{\chi}-\left(2\frac{\dot \rho}\rho-\dot H+\frac{\gamma}{\ell}\right)\chi-
2i e^{-2H}\dot{\varphi}-\frac1{2\ell}\left(\frac1B-\frac1{\bar B}\right)&=&0\,,\\
\label{KSE-rem-2}
-\dot{\chi}+2\frac{\dot \rho}\rho\chi-2i\dot{\varphi}&=&0\,,\\ \label{KSE-rem-3}
\frac1{2\ell}\left(\frac1B-\frac1{\bar B}\right)\chi+2\left(1+e^{-2H}\right)
\frac{\dot \rho}\rho-\dot H+\frac{\gamma}{\ell}&=&0\,. \label{KSE-rem-4}
\feqn
Adding \eqref{KSE-rem-1} and \eqref{KSE-rem-3} one gets
\eq
\dot H\chi+\frac1{2\ell}\left(\frac1B-\frac1{\bar B}\right)=0\,, \label{ImB}
\feq
which means that $\chi$ is purely imaginary. From \eqref{ReB} and
\eqref{ImB} we obtain then the function $B$,
\eq
\frac1{\ell B}+\frac{\gamma}{\ell}+\dot H(1+\chi)=0\,. \label{B}
\feq
Using this, the remaining Killing spinor equations reduce further to
\eqn
\left[\left(1+e^{2H}\right)\frac{\chi}{\rho^2}\right]^{\cdot}-\frac{\gamma}{\ell}
\left(e^{2H}\frac\chi{\rho^2}\right)&=&0\label{lastKSE1}\,,\\
\left(\frac\chi{\rho^2}\right)^{\cdot}+2i\frac{\dot{\varphi}}{\rho^2}&=&0\,,
\label{lastKSE2}\\
\dot H\left(1+\chi^2\right)-2\frac{\dot\rho}\rho\left(1+e^{-2H}\right)&=&
\frac{\gamma}{\ell}\,. \label{lastKSE3}
\feqn
Note that \eqref{lastKSE1} automatically implies the integrability condition
for the system \eqref{dzsigmanew}, \eqref{disigmanew}, which reduces to
\eq
\partial_Z\sigma_w = \frac1{4\psi_1}\left(\frac\chi{\rho^2}\right)^{\cdot}\,, \qquad
\partial\sigma_{\bar w}-\bar\partial\sigma_w = -\frac1{2\psi_1}\left(e^{2H}
\frac\chi{\rho^2}\right)^{\cdot}\,. \label{dsigma-G0=0}
\feq
Thus, also equation \eqref{maxwell} is satisfied, whereas \eqref{bianchi} reads
\eq
\left(1+e^{-2H}\right)2\ddot H+{\dot H}^2\left(1+3\chi^2\right)=\frac{\gamma^2}{\ell^2}\,.
\label{lastEOM}
\feq
From \eqref{B} we obtain the phase $\varphi$ and the modulus $\rho$ of $B$,
\begin{displaymath}
\tan\varphi=i\frac{\dot H\chi}{\frac{\gamma}{\ell}+\dot H}\,, \qquad
\frac1{\ell^2\rho^2}=\left(\frac{\gamma}{\ell}+\dot H\right)^2-{\dot H}^2\chi^2\,.
\end{displaymath}
Plugging this into equation (\ref{lastKSE2}) yields
\begin{displaymath}
2\ddot H\dot H\chi\left(1-\chi^2\right)-{\dot H}^2\dot{\chi}\left(1+3\chi^{2}\right)+
\frac{\gamma^2}{\ell^2}\dot{\chi}=0\,.
\end{displaymath}
Using (\ref{lastEOM}), this can be rewritten as
\begin{displaymath}
2\ddot H\chi\left[\dot H\chi\left(1-\chi^2\right)+\left(1+e^{-2H}\right)\dot{\chi}
\right]=0\,,
\end{displaymath}
so that either $\ddot H=0$ or $\dot H\chi\left(1-\chi^2\right)+\left(1+e^{-2H}\right)
\dot{\chi}=0$. It is straightforward to show that the first case leads to AdS$_4$,
whereas the second one implies
\eq
\left(e^{2H}+1\right)\frac{\chi^2}{1-\chi^2}=-\alpha^2\,, \label{secondcase}
\feq
where $\alpha$ is a real integration constant. Equations (\ref{lastKSE1}) and
(\ref{lastKSE3}) are then identically satisfied. Solving \eqref{secondcase} for
$\chi$ and plugging into \eqref{lastEOM} yields finally the ordinary differential
equation \eqref{equ_H}, which determines half-supersymmetric solutions with $G_0=0$.
Putting together all our results, we obtain \eqref{metric-G0=0} for the metric.
Note that in the case $\gamma\neq 0$ one can always set $\gamma=1$ by rescaling
the coordinates.\\
The second Killing spinor for these backgrounds is given by
\begin{displaymath}
\alpha^T = (\alpha_0, \rho^{-2}e^{-\gamma Z/\ell}, \frac{\chi+1}{2\rho}e^H,
\frac{\chi-1}{2\rho}e^H)\,,
\end{displaymath}
where
\begin{displaymath}
\alpha_0 = -\frac{2\gamma t}{\ell} + {\hat{\alpha}}_0(Z,w,\bar w)\,,
\end{displaymath}
and ${\hat{\alpha}}_0$ is a solution of the system
\eqn
\partial_Z{\hat{\alpha}}_0 &=& \frac 1{\psi_1\rho^2}\left[\frac{\dot\rho}{\rho}
-i\dot{\varphi} + \frac{\gamma}{2\ell}\right]\,, \nonumber \\
\partial{\hat{\alpha}}_0 &=& -\frac{2\gamma}{\ell}\sigma_w + \frac 1{\psi_1\rho^2}
\left[-\frac{\dot\rho}{\rho} + i\dot{\varphi}\right]\,, \\
\bar\partial{\hat{\alpha}}_0 &=& -\frac{2\gamma}{\ell}\sigma_{\bar w} + \frac 1{\psi_1
\rho^2}\left[-\frac{\dot\rho}{\rho} + i\dot{\varphi}\right] + \frac{\gamma\chi\,e^{2H}}
{\ell\psi_1\rho^2}\,. \nonumber
\feqn
It is straightforward to verify that the integrability conditions for this system are
already implied by \eqref{lastKSE1}, \eqref{lastKSE2} and \eqref{dsigma-G0=0}.

Consider now the case $\psi_-=0$. From the difference of equations (\ref{KSE1.1}) and (\ref{KSE2.1}) it follows that $b'/b$ is real. Then (\ref{KSE1.1}) and (\ref{KSE3.1}) imply that $\psi$ is a real function, depending only on $z$, $\psi_1=\psi_1(z)$. Moreover, since $\psi_{12}=\psi_{2}$, the difference of equations (\ref{KSE3.2}) and (\ref{KSE3.3}) imply that $b'/b+1/\ell b$ is imaginary.

The conditions $b'/b$ real and $b'/b+1/\ell b$ imaginary can be satisfied simultaneously in three different ways:
\begin{itemize}
\item $b'/b=0$ hence $b=b(w,\bar w)$ is an imaginary function independent of $z$. This case is solved completely in section~\ref{im-b}.
\item $b'/b+1/\ell b=0$ implies $b=-z/\ell+c$ and corresponds to AdS$_2\times\HH^2$, analyzed in section~\ref{ssan}. It is also a subcase of the following, general case,
\item if we are not in one of the previous special cases, the function $b$ must take the form
\eq
b=-\frac1\ell\frac{z}{1-iY(w,\bar w)}\,,
\label{bY}\feq
where $Y(w,\bar w)$ is some real function to be
determined.
\end{itemize}
We thus have to solve just for the ansatz (\ref{bY}). Equation~(\ref{KSE1.1}) implies $\psi_1'/\psi_1=b'/b$ than is solved by $\psi_1=z$, where we have reabsorbed the integrability constant in the scale of $z$. Equation
(\ref{KSE1.2}) (or equivalently (\ref{KSE2.3})) tells us that
$\psi_2=\psi_2(w,\bar w)$, so that the remaining independent equations read
\eqn
iz^{2}e^{-2\psi}\frac{\bar\partial Y}{1+Y^{2}}-\psi_{2}&=&0\,,
\nonumber \\
\partial\psi_2+\partial\left[\log\left(1+Y^{2}\right)+2\psi\right]\psi_{2}-iY&=&0\,,
\nonumber \\
\bar\partial\psi_2+\bar\partial\log\left(1+Y^{2}\right)\psi_{2}&=&0\,. \nonumber
\feqn
The first equation allows us to define a function $H(w,\bar w)$ such that
\eq
e^{\psi}=ze^{H(w,\bar w)}\,,
\feq
while the last one implies that there must exist a holomorphic function $C(w)$ such that
\eq
\psi_{2}=\frac{C(w)}{1+Y^{2}}\,.
\feq
Thus we are left with
\eqn
e^{2H}C(w)&=&i\bar\partial Y\,,\label{psi-01}\\
\partial\left[e^{2H}C(w)\right]&=&ie^{2H}Y\left(1+Y^{2}\right)\label{psi-02}\,.
\feqn
This set of equations automatically implies the integrability condition for the system \eqref{dzsigmanew}, \eqref{disigmanew}, which reduces to
\eqn
\partial_{z}\sigma&=&i\frac{\ell^{2}}2\frac{\partial Y}{z^{2}}\,,\\
\partial\bar\sigma-\bar\partial\sigma&=&i\ell^{2}e^{2H}Y\left(1+Y^{2}\right)\frac1z\,.
\feqn
Thus also \eqref{maxwell}, which reads
\eq
\partial\bar\partial Y-e^{2H}Y\left(1+Y^{2}\right)=0\,,
\feq
is satisfied and it turns out that also the Bianchi identity
\eqref{bianchi}, namely
\eq
\partial\bar\partial2H-e^{2H}\left(1+3Y^{2}\right)=0\,,
\feq
holds. We conclude that a solution to the system \eqref{psi-01}, \eqref{psi-02} describes a 1/2-BPS configuration of the ``gravitational Chern-Simons'' system discussed in \cite{Cacciatori:2004rt}. If $C(w)=0$ then necessarily also $Y=0$ so that we are left with AdS. If $C(w)\neq0$ then we can define new variables $W$ and $\bar W$ such that
\eq
\partial_W=C(w)\partial\,,\qquad \partial_{\bar W}=\bar C(\bar w)\bar\partial\,,
\feq
so that we have
\eqn
e^{2H}C\bar C&=&i\partial_{\bar W} Y\,, \nonumber \\
\partial_W\left[e^{2H}C\bar C\right]&=&ie^{2H}C\bar
CY\left(1+Y^{2}\right)\,. \nonumber
\feqn
As what we did in the previous case, we can set $C(w)=1$ using the residual gauge invariance $w\mapsto W(w)$, $\psi\mapsto\tilde\psi = \psi - \frac 12\ln(\dd W/\dd w) - \frac 12\ln(\dd\bar W/\dd\bar w)$ leaving invariant the metric $e^{2\psi}\dd w\dd\bar w$. We can thus
take $W=w$ without loss of generality, and get
\eqn
e^{2H}&=&i\bar\partial Y\,,\label{e2H} \\
2\partial H&=&iY\left(1+Y^{2}\right)\,. \nonumber
\feqn
\eqref{e2H} implies $Y=Y[i(w-\bar w)]$ and hence $H=H[i(w-\bar w)]$. Denoting with a dot the derivative with respect to the combination $i(w-\bar w)$ we have
\eqn
e^{2H}&=&\dot Y\,,\label{psi-01mod}\\
2\dot H&=&Y\left(1+Y^{2}\right)\,.\label{psi-02mod}
\feqn
The equations for the shift form can now be integrated, giving
\eq
\sigma=\frac{\ell^{2}}{2z}\dot Y\dd\left(w+\bar w\right)
\feq
Plugging \eqref{psi-01mod} into \eqref{psi-02mod} leads to
\eq
\ddot Y=\dot YY(1+Y^{2})\,,
\feq
which, integrated once, gives
\eq
\dot Y= \frac{L}{\ell^4} -\frac{k}2Y^{2}+\frac14Y^{4}\equiv P(Y)\,,
\feq
where $L$ is a real constant and $k=-1$\footnote{The link with the notation of \cite{Grumiller:2003ad}, where $\cal C$ and $k$ are the Casimirs of the Poisson sigma model equivalent to the dimensionally reduced gravitational Chern-Simons model in 2D, is given by $2 \cal C = L / \ell^4$.}. We can thus use $Y$ as a new coordinate, instead of $i(w-\bar w)$. Call $X=w+\bar w$, so that the solution reads
\eqn
\dd s^{2}&=&-\frac4{\ell^{2}}\frac{z^{2}}{1+Y^{2}}\left[\dd t+\frac{\ell^{2}}{2z}P_{{\cal C}}(Y)\dd X\right]^{2}+\frac{\ell^{2}}4\frac{1+Y^{2}}{z^{2}}\left[\dd z^{2}+z^{2}\left(P_{\cal C}(Y)\dd X^{2}+\frac{\dd Y^{2}}{P_{\cal C}(Y)}\right)\right]\,,\nonumber\\
\A&=&\frac2\ell z\frac{Y}{1+Y^{2}}\dd t+\ell Y\left[\frac{P_{\cal C}(Y)}{1+Y^{2}}-\frac14\left(1+Y^{2}\right)\right]\dd X+\frac\ell2\frac{\dd Y}{1+Y^{2}}\,. \label{metric}
\feqn
We can thus finally compute the second Killing spinor, with the result
\eqn
\epsilon_{2}&=&-\left[\frac{\ell^{2}}{2z}\left(1+Y^{2}\right)+2t\right]1-\frac{\ell^{2}}z\sqrt{P_{\cal C}(Y)}\sqrt{\frac{1+iY}{1-iY}}e^{1}+\nonumber\\
&&+\left[\frac{\ell^{2}}z\left(1+Y^{2}\right)+\frac\ell2\left(1+iY\right)+\frac2\ell\frac{zt}{1-iY}\right]e^{2}+\ell\sqrt{\frac{P_{\cal C}(Y)}{1+Y^{2}}}e^{1}\wedge e^{2}\,. \label{Ks}
\feqn

\end{document}